\documentclass[acmsmall]{acmart}

\AtBeginDocument{%
  \providecommand\BibTeX{{%
    \normalfont B\kern-0.5em{\scshape i\kern-0.25em b}\kern-0.8em\TeX}}}

\usepackage{tikz,pgfplots}
\usepackage{standalone}
\usepackage[listings,skins,breakable]{tcolorbox}
\usepackage{multirow}
\usepackage{subcaption}
\usepackage{makecell}
\usepackage{hyperref}
\usepackage{pifont}
\usepackage{balance}
\usepackage{booktabs}
\usepackage{pgfplots}
\usepackage{amsmath}
\usepackage[ruled, linesnumbered, vlined, commentsnumbered]{algorithm2e}
\usepackage{xcolor,colortbl}
\usepackage{adjustbox}

\usepackage{csquotes}
\usepackage{pgf-pie}
\usepackage{amsmath}
\usepackage{url}
\usepackage[tikz]{bclogo}
\usetikzlibrary{arrows.meta}
\usetikzlibrary{pgfplots.statistics}
\usepackage{fancyhdr}
\usepackage{threeparttable}
\usepackage{subcaption}
\usepackage{makecell}
\usepackage{hyperref}
\usepackage{pifont}
\usepackage{balance}

\usepackage{newtxmath}
\pgfplotsset{compat=newest}

\newtcolorbox{quotebox}{colback=teal!10,boxrule=0.4pt,colframe=black,fonttitle=\bfseries,top=2pt,bottom=2pt}
\newtcolorbox{expbox}{colback=red!10,boxrule=0.4pt,colframe=black,fonttitle=\bfseries,top=2pt,bottom=2pt}
\definecolor{steel}{rgb}{0, 0.2, 0.9} 

\definecolor{mycolor}{rgb}{0.122, 0.435, 0.698}

\newtcbox{\mytag}{nobeforeafter,colframe=mycolor,colback=mycolor!30!white,boxrule=0.7pt,arc=0pt,
 boxsep=-3pt,left=6pt,right=6pt,top=4pt,bottom=5pt,tcbox raise base}

\DeclareMathAlphabet\mathbfcal{OMS}{cmsy}{b}{n}

\setcopyright{cc}
\setcctype{by}
\acmJournal{TOSEM}
\acmYear{2025} \acmVolume{1} \acmNumber{1} \acmArticle{1} \acmMonth{1} 
\acmDOI{10.1145/3750040}

\pgfplotsset{compat=1.11,
   /pgfplots/xbar legend/.style={
   /pgfplots/legend image code/.code={%
      \draw[##1,/tikz/.cd,yshift=-0.3em]
       (0cm,0cm) rectangle (3pt,0.8em);},
  },
}

\acmJournal{TOSEM}
\acmVolume{37}
\acmNumber{4}
\acmArticle{111}
\acmMonth{8}

\begin{document}
% % Start of cover letter
% \pagenumbering{gobble} % Suppress page numbers
% \input{cover-letter}
% \pagebreak

% \input{novelty}
% \pagebreak

% % Reset page numbering for main document
% \pagenumbering{arabic}
% \setcounter{page}{1}

%\title[Bug Report Identification with Effort-aware and Pseudo-labeled Active Learning]{Boosting Human Machine Teaming: Bug Report Identification with Effort-aware and Pseudo-labeled Active Learning}

%\title{Bug Report Identification with Mutualistic Neural Active Learning}
%\title{Learning Bug Reports for Software Projects: A Systematic Literature Review}
\title{Learning Software Bug Reports: A Systematic Literature Review}

\author{Guoming Long}
\authornote{Guoming Long is also supervised in the IDEAS Lab.}
\affiliation{%
  \institution{University of Electronic Science and Technology of China}
  \city{Chengdu}
  \country{China}}
\email{logan31413@outlook.com}

\author{Jingzhi Gong}
\email{j.gong@leeds.ac.uk}
\orcid{0000-0003-4551-0701}
\affiliation{%
  \institution{University of Leeds}
  \city{Leeds}
  \country{United Kingdom}
}

\author{Hui Fang}
\affiliation{%
  \institution{Loughborough University}
  \city{Loughborough}
  \country{United Kingdom}}
\email{h.fang@lboro.ac.uk}

\author{Tao Chen}
\authornote{Corresponding author: Tao Chen, t.chen@bham.ac.uk.}
\affiliation{%
  \institution{IDEAS Lab, University of Birmingham}
  \city{Birmingham}
  \country{United Kingdom}}
\email{t.chen@bham.ac.uk}

\begin{abstract}
The recent advancement of artificial intelligence, in particular machine learning (ML), has witnessed its significant growth in various software engineering research fields. Among them, bug report analysis is one of such examples as it aims to automatically understand, extract and correlate information from the reports with the help of ML approaches. Despite the importance of ML in automating and enhancing bug report analysis, a comprehensive review that systematically examines the state-of-the-art in this area is still lacking. In this paper, we provide a systematic literature review on this promising research topic. Our review covers 1,825 papers, from which we extract 204 most relevant studies for detailed analysis. Based on the statistics and trends observed in these reviewed studies, we obtained seven key findings summarized as follows: 1) the extensive use of Convolutional Neural Network (CNN), Long Short-Term Memory (LSTM) and $k$-Nearest Neighbor ($k$NN) for bug report analysis, noting the underutilization of more advanced models like BERT due to their complexity and computational demands. 2) Word2Vec and TF-IDF are the most common methods for feature representation, with a notable increase in deep learning-based methods in recent years. 3) Stop word removal is the most common preprocessing method, followed by tokenization and stemming. Structural methods surged post-2020. 4) Eclipse and Mozilla are the most frequently evaluated software projects, reflecting their prominence in the field. 5) Bug categorization is the most popular task, followed by bug localization, assignment, and severity/priority prediction, with a growing interest in bug report summarization driven by advancements in NLP. 6) Most studies focus on general bug types, but there is increasing attention on specific bugs such as non-functional and performance bugs. 7) Common evaluation metrics include F1-score, Recall, Precision, and Accuracy, but bug report related evaluation metrics have not received significant attention. The majority of studies prefer $k$-fold cross-validation for model evaluation. and 8) many studies lack robust statistical tests or effect size measurements. Finally, based on the key findings, we discover six promising future research directions, by which we hope, together with the findings, can offer useful insights to practitioners of this particular research direction.  

%This provides a perfect place for machine learning, which was designed precisely for these tasks.
%It also has the best Scott-Knott rank over four SOTA approaches. We have made our dataset and code publicly available to facilitate further research and reproducibility at \textcolor{blue}{\texttt{\url{???}}}.

%optimizing the selection of informative reports for the model while reducing the developers' efforts in labeling. 

%The model not only queries the most informative and representative bug reports, enhanced by selected pseudo-labeled ones to enhance performance, but also considers the developers' labeling efforts in the selection process, thereby promoting effective collaboration between humans and machines. 

\end{abstract}





\maketitle

\section{Introduction}
\label{sec:intro}

Bugs are prevalent in the process of software development, implementation, and maintenance. These software bugs, if left untreated, can often cause serious consequences. For example, Knight Capital Group, one of the biggest American market makers, lost 440 million dollars in just 30 minutes due to a software bug\footnote{\url{https://raygun.com/blog/costly-software-errors-history/}}. A common practice for bug fixing is to leverage bug reports, which document software issues encountered by developers and users. However, as the complexity and scale of software systems increase, handling bug reports becomes profoundly complex and time-consuming. As such, automated bug report analysis (or simply bug report analysis) is a vital way to free developers from tedious manual and repetitive tasks involved, achieving more efficient and effective bug fixing during software maintenance.

During the past decade, automatic analysis of bug reports has been an effective way to prioritize, track, and understand bugs, improving bug-fixing efficiency and productivity~\cite{Zhang2015,DBLP:conf/ease/LongC22}. However, it is a challenging task to achieve such automation due to the following two reasons: (1) The sheer volume of bug reports continues to grow exponentially. For example, in addition to the unresolved ones, the number of newly reported issues for the PyTorch project in 2023 has reached 8,352\footnote{\url{https://tinyurl.com/294mrax5}}, which is a 44$\times$ increment from what was reported in 2016; and (2) Bug reports are inherently unstructured, consisting of free-form text written by various reporters with different levels of detail and technical expertise. 

Consequently, in the last decade, there has been an increasing trend to exploit machine learning (ML) algorithms for automatic bug report analysis. ML has proven to be a crucial tool in this domain by enabling automated classification, deduplication, prioritization, and summarization of bug reports. Traditional rule-based approaches struggle to handle the complexity and scale of modern bug reports, whereas ML methods can learn patterns from historical data and make informed decisions with minimal human intervention. Moreover, recent advances in deep learning (DL) and natural language processing (NLP) have further boosted the performance of bug report analysis, making ML an indispensable approach for modern software maintenance. For example, Long Short-Term Memory (LSTM) networks have been effectively applied to bug report analysis, capturing contextual dependencies and retaining long-term information to improve tasks such as bug report classification and summarization~\cite{DBLP:conf/issta/LiLZZ19}. Further, Transformer-based models like BERT have demonstrated strong performance in bug report prioritization, leveraging contextual information to rank reports based on severity and urgency, enabling faster issue resolution~\cite{DBLP:conf/dsa/ZhengCWFSC21}.

% Therefore, practitioners in the field demand a survey that covers state-of-the-art work for research of learning bug report analysis and providing a taxonomy of the various concepts therein. 

\subsection{Motivation}

Despite the rapid evolution of ML methods and their growing application in software maintenance, the literature remains fragmented---many studies concentrate on specific tasks, such as duplicate detection~\cite{jayarajah2016duplicate, lin2016enhancements} or triaging~\cite{banerjee2017automated, DBLP:conf/comad/ManiSA19}, without providing a comprehensive view of how these techniques connect and complement each other.

Moreover, while several reviews exist in adjacent areas of software engineering, few have specifically addressed the integration of ML in bug report analysis~\cite{DBLP:journals/air/UddinGDNS17, DBLP:conf/icetc/TararAB19, DBLP:journals/infsof/GomesTC19}, and those that do are often outdated given the swift progress in deep learning and natural language processing~\cite{DBLP:journals/tr/StrateL13, Zhang2015}. This lack of a unified, up-to-date synthesis hampers both researchers and practitioners, who need to understand the current state of the art, identify recurring challenges, and uncover emerging trends.

A systematic survey in this domain is therefore essential to bridge the gap between isolated findings and to provide a comprehensive roadmap that can guide future research and practical tool development. First, this survey is anticipated to offer fresh insights by formulating a unified taxonomy that integrates traditional ML techniques with recent advancements in deep learning and natural language processing. Readers will be introduced to emerging trends---such as the increasing adoption of deep learning algorithms, innovative feature representation methods, and evolving preprocessing strategies---as well as to critical research gaps like the underutilization of bug report-specific evaluation metrics and the lack of rigorous statistical testing, thereby fostering innovation and direct future investigations toward more effective, data-driven approaches to bug report analysis.

Beyond academic contributions, this work also seeks to support practical advancements in software engineering (SE). For example, the integrative perspective provided by this survey will serve as a guide to explore underinvestigated areas and refine existing methodologies, ultimately advancing the state of the art in automated bug report analysis. Additionally, the synthesis of best practices and emerging trends will offer actionable strategies for enhancing bug-fixing processes, reducing maintenance efforts, and improving overall software quality.

\subsection{Contributions}

Driven by these motivations, we conduct a systematic literature review following the widely used systematic literature review (SLR) methodology in software engineering~\cite{DBLP:journals/infsof/KitchenhamBBTBL09}. Specifically, we reviewed 1,825 papers from 46 venues and three repositories, from which we extracted 585 highly relevant studies for detailed analysis. In a nutshell, our key contributions can be summarized as follows:

\subsubsection{New Insights and Findings} 
Our study presents a novel taxonomy that bridges traditional ML techniques with modern deep learning and NLP approaches for bug report analysis, providing a cohesive view of the field's evolution. In doing so, we uncover several important findings and insights, for example:

\begin{itemize}
    \item \textbf{Emergence of deep learning:}
    The following findings indicate a shift from classical methods to deep learning, which reflects a growing sophistication in handling textual data for bug report analysis.
    \begin{itemize}
        \item \textbf{Finding 1:} CNN, LSTM, and $k$NN have become the leading algorithms, with a marked increase in the use of deep learning techniques from 2020 to 2023.
        \item \textbf{Finding 2:} While traditional methods like Word2Vec and TF-IDF remain popular, there is a clear trend towards adopting deep learning-based feature representations.
        \item \textbf{Finding 3:} Standard preprocessing techniques such as stop word removal, tokenization, and stemming continue to be widely applied, accompanied by a notable rise in the use of structural preprocessing methods.
    \end{itemize}

    \item \textbf{Attention on evaluation datasets and bug task analysis:} 
    Our findings reveal a dual focus: reliance on well-established datasets and an emerging interest in diversifying the types of bug-related tasks addressed. This indicates a maturing field that is beginning to tackle more complex and varied challenges.
    \begin{itemize}
        \item \textbf{Finding 4:} Benchmark projects like Eclipse and Mozilla Core are frequently used for evaluation, with a growing emphasis on projects that include unstructured bug reports.
        \item \textbf{Finding 5:} Bug categorization remains the dominant task, yet bug report summarization has seen significant growth, expanding from one primary study to nine over the past four years.
        \item \textbf{Finding 6:} Although most studies focus on general bug analysis, there is an increasing trend towards examining specific bug types, reflecting a deeper inquiry into nuanced software issues.
    \end{itemize}

    \item \textbf{Gaps in methodology and evaluation practices:}
    We also uncover limitations in the evaluation frameworks, particularly the underutilization of bug report-specific metrics and rigorous statistical methods. This gap highlights opportunities for refining the analytical rigor in future studies.
    \begin{itemize}
        \item \textbf{Finding 7:} Conventional metrics such as precision, F1-score, accuracy, and recall are predominantly used, with most studies relying on $k$-fold cross-validation, while bug report-specific evaluation measures remain scarce.
        \item \textbf{Finding 8:} Although statistical tests like the Wilcoxon signed-rank test and Cliff's delta are sometimes applied, many studies overlook rigorous statistical validation and fail to report effect sizes.
    \end{itemize}
\end{itemize}

\subsubsection{Impacts on the Software Engineering Community} 
The insights and recommendations distilled from this survey are expected to make a significant impact on the software engineering community---(1) for researchers, our findings serve as a guide to identify promising directions and methodological improvements for advancing automated bug report analysis; and (2) for practitioners, the synthesized best practices and emerging trends offer actionable strategies to enhance automated bug-fixing processes, thereby reducing maintenance efforts and improving software quality.

In addition to these immediate impacts, our work outlines several actionable future research directions that can further advance the field. Specifically, we propose that future work should:
\begin{itemize}
    \item Leverage Transformer-based architectures and Large Language Models (e.g., BERT, GPT-4) to improve the precision and efficiency of tasks such as bug triaging, duplicate detection, and report summarization.
    \item Develop advanced, integrated tools for automated bug report analysis on platforms like GitHub by combining traditional ML, deep learning, and hybrid methods to address scalability and contextual challenges.
    \item Investigate specialized approaches for analyzing specific types of bugs by developing domain-specific models and dedicated evaluation metrics tailored to capture the unique characteristics of bug reports.
    \item Incorporate explainable deep learning techniques, such as attention mechanisms and hybrid models, to provide transparent and actionable insights, fostering stronger collaboration between researchers and practitioners.
\end{itemize}

By consolidating disparate studies into a unified framework, this work not only advances theoretical understanding but also paves the way for more effective, data-driven solutions in real-world software maintenance scenarios.

\subsection{Organization}

This paper is organized as follows: Section~\ref{section:bg} provides background information on bug report analysis. Following it, section~\ref{section:method} presents the systematic literature review protocol. Section~\ref{sec:results} analyzes the results of the research questions. Subsequently, future research directions are discussed in section~\ref{sec:future}. Finally, the threats to validity is discussed in sections~\ref{sec:threat} and conclusions are drawn in~\ref{sec:conclusion}, respectively.

\section{Background} 
\label{section:bg}
\subsection{Bug Report}
\begin{figure}[t!]
  \centering
  \includegraphics[width=0.6\columnwidth]{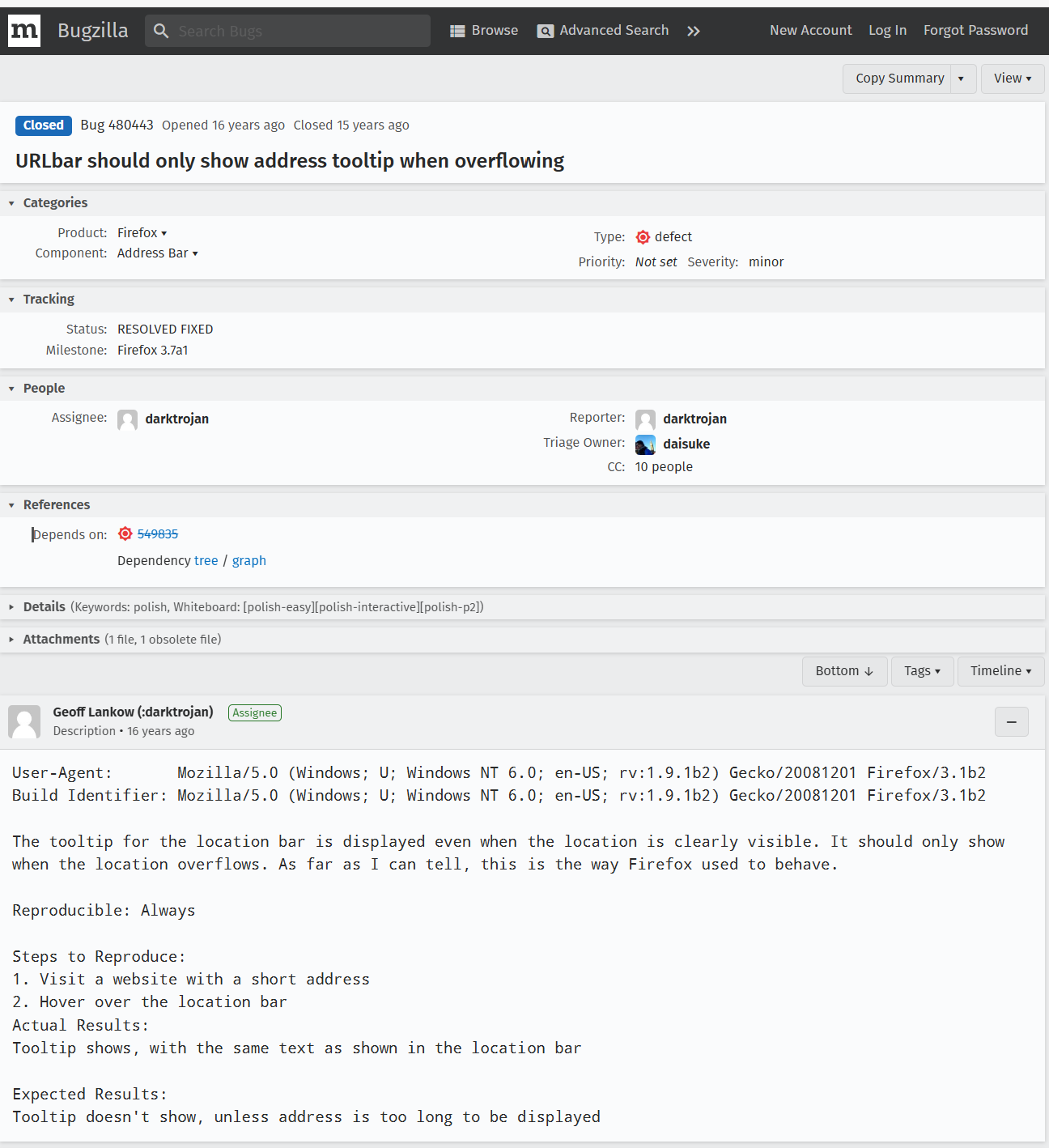}
  \caption{A bug report from the Mozilla project.}
  \label{fig:report-example}
\end{figure}

Software bugs refer to the defects in the code or the software configurations that generate unexpected behaviors \cite{Zhang2015}. An example of a Mozilla bug report\footnote{\url{https://bugzilla.mozilla.org/show_bug.cgi?id=480443}} has been shown in Figure~\ref{fig:report-example}. As can be seen, a typical bug report contains various attributes and fields, among which the most common ones include:

\begin{itemize}
    \item \textbf{Title:} a short summary of a bug report.
    \item \textbf{Priority and severity:} indication of the importance and severity level of the bug.
    \item \textbf{Status:} stage of the bug in the life cycle of bug fixing, e.g., \textit{closed}, \textit{resolved} and \textit{reopened}.
    \item \textbf{Assignee:} The person responsible for fixing the bug.
    \item \textbf{Description:} detailed textual information, such as what the bug is, its affected modules, and steps to reproduce.
\end{itemize}

\begin{figure}[t!]
  \centering
  \includegraphics[width=0.7\columnwidth]{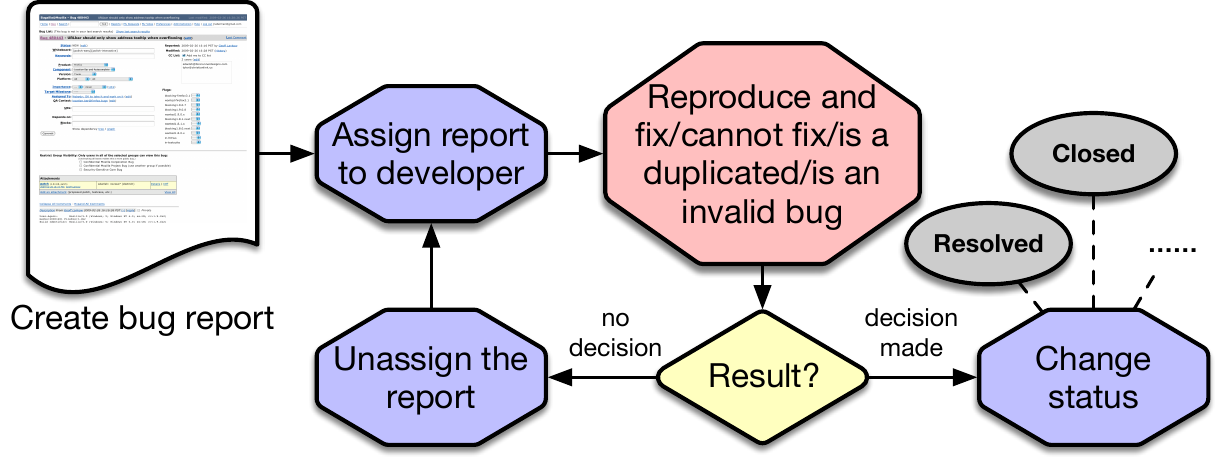}
  \caption{Typical bug report handling protocol.}
  \label{fig:example}
\end{figure}

A bug report is produced by either a developer, a tester, or an end-user describing their identified bug, its stack trace, and the steps to reproduce it, etc. ~\cite{DBLP:journals/air/UddinGDNS17}. Therefore, bug reports provide useful information for the developers to fix bugs. Large-scale software projects, such as Mozilla, Eclipse and Netbeans, have their own repositories to store bug reports. Figure~\ref{fig:example} illustrates how the bug reports are processed in a standard procedure.
%An example of how bug reports are handled can be seen from .

\subsection{Machine Learning Algorithms}

The application of machine learning (ML) algorithms has surged with the advancement of computational capabilities. Fundamental paradigms like statistical learning and deep learning are at the core of this technological evolution:

\textbf{Statistical Learning} methods involve models that infer patterns from data based on statistical principles. In the context of bug report analysis, statistical learning can utilize features such as the frequency of specific keywords, the length of the bug report, the time of submission, the severity of the bug, and the historical data of similar bugs. These attributes help in identifying patterns and making predictions about new bug reports.

\textbf{Deep Learning} techniques, employing complex neural networks, are particularly adept at understanding patterns in large volumes of data. These techniques excel in tasks that require the analysis of unstructured data, such as the textual content of bug reports. By automatically extracting relevant context features from the bug report, deep learning can enhance the accuracy of both discriminative and generative models. \textbf{Large Language Models} (LLMs) are a transformative subfield within deep learning, leveraging transformer-based architectures and self-attention mechanisms to efficiently process sequential data at significant scale. LLMs can be categorized into various architectural types, including encoder-only, decoder-only, and encoder-decoder models. Encoder-only models specialize in understanding contextual relationships within texts but are typically not classified as LLMs due to their smaller size and limited generative capabilities. In contrast, decoder-only models, like OpenAI's GPT-4\footnote{https://openai.com/index/gpt-4/} (closed-source) and Meta's open-source LLaMA 3\footnote{https://llama.meta.com/llama3/}, are optimized for generating coherent and contextually relevant text. Unlike traditional neural networks, LLMs undergo extensive pre-training on vast and diverse textual corpora, allowing them to capture intricate linguistic patterns, contextual dependencies, and semantic relationships. These models demonstrate superior capabilities in natural language understanding and generation, making them highly effective across a wide array of complex text-based applications.

\subsection{Related Surveys}

Despite the growing body of work on bug report analysis, to the best of our knowledge, there is no comprehensive, up-to-date review focusing exclusively on ML-based approaches. One of the earliest surveys related to bug report analysis was conducted by~\citeauthor{DBLP:journals/tr/StrateL13}\cite{DBLP:journals/tr/StrateL13}, covering 104 studies. That survey reviews the connections among various domains related to bugs/defects, such as bug report analysis, bug detection, and bug repair, rather than solely concentrating on bug report analysis. Similarly, in 2015, \citeauthor{Zhang2015}\cite{Zhang2015} reviewed work specifically on analyzing bug reports, including papers published before or in 2014. Their focus has been on providing a comprehensive overview of various tasks by using bug report analysis. Interestingly, the above reviews have not applied a systematic protocol as we use in this work.

There also exist surveys on specific problems in bug report analysis, e.g., bug prioritization~\cite{DBLP:journals/air/UddinGDNS17}, bug report summarization~\cite{DBLP:conf/icetc/TararAB19}, and severity prediction~\cite{DBLP:journals/infsof/GomesTC19}, as opposed to the entire bug report analysis domain. A more recent survey on bug report analysis was conducted by \citeauthor{8466000}~\cite{8466000}. Their purpose is to understand whether the problems in bug report analysis are important from the perspective of industrial practitioners. As part of the process, they carried out a systematic review on 121 studies with the sole aim of identifying all the problems that are being investigated. The result of that review forms the basis for designing a questionnaire for developers to rank their importance.

While prior surveys have examined bug reports across different platforms, they have largely overlooked the differences between desktop software and mobile applications. \citeauthor{DBLP:journals/software/ZhangCLL19}~\cite{DBLP:journals/software/ZhangCLL19} systematically analyzed bug reports in GitHub and found that mobile app reports tend to be shorter but contain more debugging elements, such as stack traces and code examples, which contribute to faster bug resolution. In contrast, desktop software reports are longer yet often lack these crucial elements, leading to longer fixing times. This distinction highlights the importance of tailored approaches in bug report analysis based on software type.

In comparison to other surveys, we seek to systematically understand what, how, why, and where machine learning algorithms are used and assessed in various tasks related to bug report analysis. Thus our work is in fact complementary to the work by~\citeauthor{8466000}~\cite{8466000}, as well as to those problem-specific surveys~\cite{DBLP:journals/air/UddinGDNS17,DBLP:conf/icetc/TararAB19,DBLP:journals/infsof/GomesTC19}. Although machine learning-based methods have been briefly discussed by~\citeauthor{DBLP:journals/tr/StrateL13}~\cite{DBLP:journals/tr/StrateL13} and~\citeauthor{Zhang2015}~\cite{Zhang2015}, our work provides a more comprehensive review of this specialized topic, i.e., bug report analysis, by summarizing key findings and capturing the latest trends, particularly on their designed algorithms, feature representations, and evaluations, which have not been covered in previous surveys.
 \section{Review Methodology}
\label{section:method}

In this article, we conduct an incremental systematic literature review on bug report analysis with machine learning, for which most of the processes follow the best practice on systematic literature review for software engineering~\cite{DBLP:journals/infsof/KitchenhamBBTBL09}. 

\subsection{Stage 1: Automatic search}

\begin{figure}[t]
  \centering
  \includegraphics[width=0.95\columnwidth]{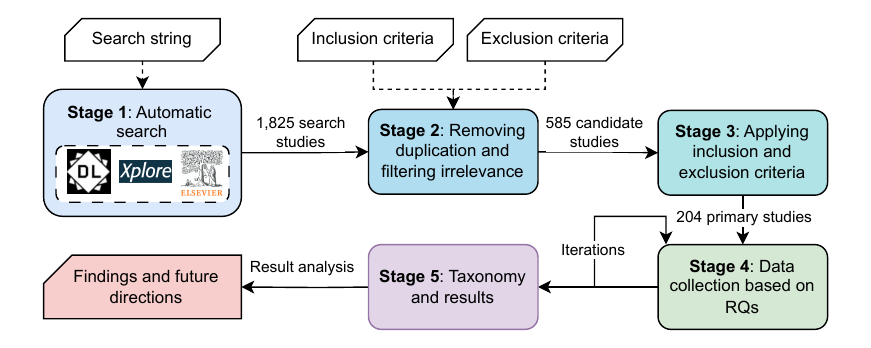}
  \caption{Overview of the systematic literature review protocol. }
  \label{fig:protocol}
\end{figure}

As shown in Figure~\ref{fig:protocol}, starting from \textbf{Stage 1}, from 18th Jan to 27th Dec 2023, we conducted an automatic search over three well-known indexing services, i.e., ACM Library, IEEE Xplore and Google Scholar from 2014 to 2023 by using the search string below:

\begin{tcolorbox}[breakable,left=2pt,right=2pt,top=2pt,bottom=2pt] 
\emph{("bug report" OR "issue report" OR "fault report") AND  ("machine learning" OR "artificial intelligence" OR "data mining" OR "information retrieval")}
\end{tcolorbox}

This manual search is enhanced with a focused automated search to increase the reliability of the findings obtained manually and to enhance reproducibility. Even though manual searches require significant effort, they often yield more comprehensive results compared to automated searches. Automated searches utilizing standard search strings on indexing services may overlook relevant literature, especially those addressing interdisciplinary topics that lack established terminology. Moreover, focused searches conducted at specific reputable venues can ensure the retrieval quality. %Literature lacking peer review, which would likely be identified accurately in a purely automated search, may indeed be of low quality.

\subsection{Stage 2: Removing duplication and filtering irrelevant studies}

In the second stage, our goal is to ensure that the reviewed studies are both unique and highly pertinent. To this end, we remove duplicates and carefully check if the articles are relevant. This step guarantees that every study we include offers distinct contributions to the knowledge. Following this, we perform a preliminary assessment to exclude studies that are clearly not related to bug report analysis. Through these filtering steps, we have selected 585 candidate studies that are highly relevant to our investigation.

\subsection{Stage 3: Applying inclusion and exclusion criteria}
Inclusion and exclusion criteria help us to identify studies directly related to our research questions. We apply inclusion and exclusion criteria during the initial search as well as the later stages to those studies identified through the initial search. A paper is selected as a primary study if it meets all of the inclusion criteria below:

\begin{itemize}
\item The paper pertains specifically to the software domain. Our focus is only on bug report analysis in software engineering. Investigations concerning hardware or alternative system types are out of the scope of this survey.
\item The paper seeks to address a problem in bug report analysis using machine learning algorithms.
\item The paper clearly states which machine learning algorithm(s) has been applied in its work.
\item The paper discusses how to design bug report analysis solutions.
%if there is no details of the algorithm(s).
\item The paper clarifies its experimental setup and presents quantitative experimental results. %and has an explanation of experiment setup. 
\item The paper performs evaluations based on real-world software projects.
\end{itemize}

For each paper selected after the initial search, we apply the following exclusion criteria. A study is removed if it meets any of the following criteria: 
\begin{itemize}
%\item The paper focus on bugs related analysis based on source code rather than analyzing the bug report.
\item The paper type is either an editorial, abstract, opinion paper, short paper, poster abstract, keynote, opinion, summary of a tutorial, summary of a conference (or an introduction to conference proceedings), summary of a workshop, or a summary of a panel discussion. %Formats like editorials, abstracts, surveys, or tutorials, among others, typically do not offer a substantial amount of information.
\item The paper is not software engineering related. 
\item The paper is not published in a peer-reviewed venue.
%\item The paper is a survey or tutorial type of study.
%\item The paper is a short publication, i.e., it is shorter than 8 pages for double-column or 12 pages for single-column.
\end{itemize}

The above process stops when we finish reviewing all candidate studies. Finally, 204 primary studies are selected for further data collection. 

\subsection{Stage 4: Data collection based on RQs}
In this stage, we formulate the research questions (RQs) for this literature review outlined below. Precisely, we have identified six questions suitable for investigation. 

\subsubsection{Research Questions}

Initially, considering the potential utilization of diverse machine learning algorithms in bug report analysis, the primary area involves exploring machine learning algorithms, feature representation and preprocessing techniques. In order to grasp this concept, we pose the first research question:

\begin{tcolorbox}[breakable,left=2pt,right=2pt,top=2pt,bottom=2pt] 
\emph{RQ1: What machine learning algorithms, feature representation and preprocessing methods are used?}
\end{tcolorbox}

Subsequently, we observed that the bug reports used in bug report analysis as datasets are sourced from a multitude of software projects. In order to investigate these software projects, we aim to comprehend:

\begin{tcolorbox}[breakable,left=2pt,right=2pt,top=2pt,bottom=2pt] 
\emph{RQ2: What software projects are evaluated?}
\end{tcolorbox}

Furthermore, we explore the multifaceted landscape of the associated tasks of bug report analysis using machine learning techniques. Our aim is to explain the detailed challenges and issues encountered within this domain. Through a comprehensive examination, we seek to clarify:

\begin{tcolorbox}[breakable,left=2pt,right=2pt,top=2pt,bottom=2pt] 
\emph{RQ3: What tasks in bug report analysis are tackled by machine learning?}
\end{tcolorbox}

Moreover, an integral aspect of our investigation relates to the categorization of bug types within the selected studies. To understand the breadth and scope of bug types used in bug report analysis, we pose the question: 

\begin{tcolorbox}[breakable,left=2pt,right=2pt,top=2pt,bottom=2pt] 
\emph{RQ4: What types of bugs are analyzed?}
\end{tcolorbox}

Additionally, to measure the effectiveness and reliability of bug report analysis techniques, we carefully examine the evaluation metrics and methodologies employed across various studies. Hence, we inquire:

\begin{tcolorbox}[breakable,left=2pt,right=2pt,top=2pt,bottom=2pt] 
\emph{RQ5: Which evaluation metrics and procedures are leveraged in the evaluation?}
\end{tcolorbox}

Lastly, to ensure the robustness and validity of collected bug report analysis studies, we investigate the statistical tests and standardized effect size measurements utilized by researchers to mitigate potential biases in bug report analysis studies. Therefore, we ask:

\begin{tcolorbox}[breakable,left=2pt,right=2pt,top=2pt,bottom=2pt] 
\emph{RQ6: Which statistical tests and standardized effect size measurements are applied to mitigate bias?}
\end{tcolorbox}

\subsubsection{Data Items}
We have devised an extensive list of data items during the review process to answer the research questions. The data elements, listed in Table~\ref{tab:collect}, comprise a total of 12 items, each serving a specific purpose pertaining to its corresponding research question. In this section, we explain the rationale behind their design and explain the procedure for extracting and categorizing data from these items. As shown in Table~\ref{tab:collect}, the data items F1 to F3, including primary study ID, title, author, and publisher, reflect the meta-information of the reviewed studies. This meta-information facilitates accurate referencing and organization of the studies, laying a solid groundwork for subsequent analysis and interpretation.

To address RQ1, we aim to explore machine learning algorithms, feature representations and preprocessing methods used in these primary studies. For this purpose, data items F4 to F6 are crafted to extract information on these methods. For RQ2, we use data item F7 to investigate which software projects are evaluated in their bug report analysis. We examine F8 to explore what particular tasks are tackled by using machine learning in RQ3. After it, we explore what types of bugs are analyzed by examining F9 for RQ4. Furthermore, we discuss the evaluation metrics and procedures leveraged in the evaluation using F10 and F11 for RQ5, respectively. Finally, to answer RQ6, F12 is designed to identify which statistical tests are used in the studies, whereas F13 is for further clarifying the effect size measurement that examines the magnitude of difference in the statistical tests. 

\begin{table}[t!]
	\centering
	\caption{Data items for collection.}
	%\begin{threeparttable}
    \footnotesize
	\begin{tabular}{lll}\toprule
	\textbf{Field ID \#}&\textbf{Field}&\textbf{Research question}\\\midrule
    F1&Primary ID&n/a\\
    F2&Title&n/a\\
    F3&Publication site&n/a\\
    F4&Algorithm&RQ1\\
    F5&Representation&RQ1\\
    F6&Preprocessing&RQ1\\
    F7&Project&RQ2\\
    F8&Task&RQ3\\
    F9&Bug Type&RQ4\\
    F10&Metrics&RQ5\\
    F11&Procedure&RQ5\\
    F12&Statistical test&RQ6\\
    F13&Effect size&RQ6\\

    \bottomrule 
	\end{tabular}
	%\end{threeparttable}
	\label{tab:collect}%
\end{table}%

\subsubsection{Data Collection}

The process of data collection aims to obtain thorough information from the primary studies, as outlined in Table~\ref{tab:collect}. The collection protocol follows standard practices commonly employed in Software Engineering literature reviews~\cite{DBLP:journals/infsof/KitchenhamBBTBL09}. This involves a series of steps conducted by all authors in three iterations.

During the initial iteration, each author independently conducts data collection from the 204 primary studies. They meticulously review each study, extract data according to predefined criteria, summarize it into a table with 12 columns and 204 rows, and conduct initial classification. Some challenges arise where certain data items are missing, poorly defined, or require inference from context. Such instances are flagged for further investigation in subsequent iterations.

In the second iteration, authors cross-reference and merge their individual data summaries into an integrated table. Unclear data items are revisited and resolved through discussions, further research, or consultation with external experts. For instance, a category labeled "unspecified" is added for the evaluation procedure due to a lack of justification in many studies, requiring additional research to determine the subject system domains.

The third iteration aims to summarize the integrated table statistics, count the number of studies for each technique associated with data items, and group similar techniques into broader categories. For example, studies focusing on bug report title generation are grouped under "bug report summarization" due to shared characteristics. All authors participate in comprehensive discussions to finalize data classifications for the study.

Finally, this process has resulted in the data used in this study\footnote{Details of the primary studies and data can be found at our repository: \url{https://tinyurl.com/4eztvr7n}.}, which undergoes thorough discussion among all authors to achieve comprehensive classifications.

\subsection{Stage 5: The taxonomy and result}

Using the data collected in the previous stage, we develop a taxonomy and obtain results that enhance our understanding of bug report analysis using machine learning techniques. 

As illustrated in Figure~\ref{fig:content}, we construct a taxonomy of bug report analysis with machine learning techniques by connecting Research Questions (RQs) and their corresponding data items. The main content flow and categorization of the survey are effectively summarized, focusing on various aspects of bug report analysis using machine learning techniques. The key elements are broken down into distinct sections, each addressing a specific area of interest. Here’s what each category specifically means for the research questions (RQs):

\begin{figure}[t!]
  \centering
  \includegraphics[width=0.8\columnwidth]{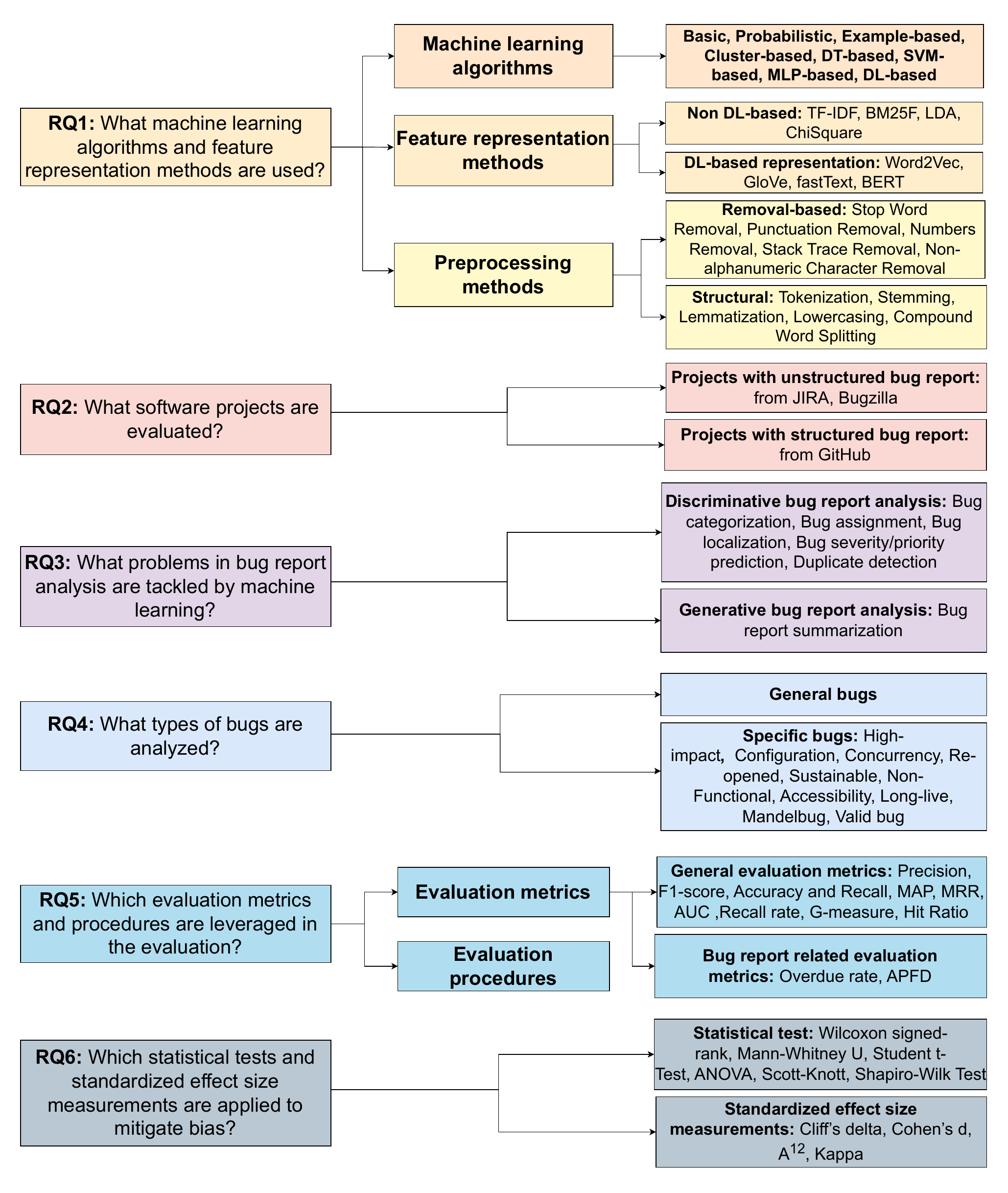}
  \caption{The taxonomy of learning-based bug report analysis. We use color to indicate different sections.}
  \label{fig:content}
\end{figure}

\begin{itemize}
    \item \textbf{RQ1: What machine learning algorithms, feature representation and preprocessing methods are used?}
    \begin{itemize}
        \item \textbf{Machine Learning Algorithms:} This involves various types of algorithms used for bug report analysis, including basic algorithms, probabilistic methods, example-based techniques, decision tree-based methods, support vector machines, and neural network-based approaches.
        \item \textbf{Feature Representation Methods:} These are divided into non-deep learning and deep learning-based methods, suggesting an exploration of how bug report data is structured and utilized by different machine learning algorithms.
        \item \textbf{Preprocessing Methods:} These techniques are essential for refining raw bug report data before applying models. They can be broadly categorized into removal-based methods, which eliminate unnecessary elements, and structural methods, which transform text while preserving all information. The selection of preprocessing methods plays a crucial role in improving data quality and model performance.
    \end{itemize}
    
    \item \textbf{RQ2: What software projects are evaluated?}
    \begin{itemize}
        \item \textbf{Projects with Unstructured Bug Report:} This focuses on studies that deal with projects where bug reports are not formatted according to a predefined structure.
        \item \textbf{Projects with Structured Bug Report:} This focuses on studies involving projects that use a specific, predefined format for bug reports.
    \end{itemize}
    
    \item \textbf{RQ3: What problems in bug report analysis are tackled by machine learning?}
    \begin{itemize}
        \item \textbf{Discriminative Bug Report Analysis:} Machine learning techniques that classify or differentiate between types or qualities of bug reports.
        \item \textbf{Generative Bug Report Analysis:} Machine learning techniques that generate new content or predictions related to bug reports.
    \end{itemize}
    
    \item \textbf{RQ4: What types of bugs are analyzed?}
    \begin{itemize}
        \item \textbf{General Bugs:} Studies that focus on a wide range of bugs without specific limitations or definitions.
        \item \textbf{Specific Bugs:} Studies that focus specifically on certain types of bugs, possibly those that are more critical or common in certain contexts.
    \end{itemize}
    
    \item \textbf{RQ5: Which evaluation metrics and procedures are leveraged in the evaluation?}
    \begin{itemize}
        \item \textbf{General Evaluation Metrics:} Metrics that are widely used across various studies and contexts.
        \item \textbf{Bug Report Related Evaluation Metrics:} Specific metrics that are tailored to evaluate bug report analyses.
        \item \textbf{Evaluation Procedures:} Methods and protocols used to assess the effectiveness of machine learning techniques in analyzing bug reports.
    \end{itemize}

    \item \textbf{RQ6: Which statistical tests and standardized effect size measurements are applied to mitigate bias?}
    \begin{itemize}
        \item \textbf{Statistical Test:} Refers to the statistical methods used to analyze data and test hypotheses in the studies.
        \item \textbf{Standardized Effect Size Measurements:} Measures used to quantify the size of an effect or difference seen in the studies, helping to address the magnitude of impacts in statistical terms.
    \end{itemize}
\end{itemize}

In the next section, we will present the results analysis, where a thorough examination of the comprehensive data statistics gathered from the primary studies will be discussed. This analysis will reveal key findings and outline potential future directions for research. 

%Using the compiled statistics derived from the Research Questions (RQs), data items, and data collection methodology outlined earlier, we construct a taxonomy illustrating bug report analysis with machine learning techniques, as depicted in Figure~\ref{fig:content}. Our analysis reveals numerous categories for the RQs and associated data items. In the following sections, we will provide comprehensive data statistics and detailed explanations of each aspect related to bug report analysis.

%\subsection{Overview of the Results}

%After the execution of the search process and removing redundancy, we have identified a total of 339 searched studies. Then, we rule out dozens of studies by reviewing their abstracts, leading to a more precise set of 112 candidate studies. Finally, following the inclusion and exclusion criteria, we extract 57 primary studies from 36 different venues\footnote{The comprehensive list of primary studies and other summarized data can be accessed via our anonymous repository: \url{http://tiny.cc/phy7jz}}. The primary studies from the top 10 most common venues have been shown in Table~\ref{tab:freq}.

%From Figure~\ref{fig:figure2}, we see that the number of primary studies increase over years: from 2014 to 2017 there is a steeper increment, but it slows down between 2017 and 2018, after which the number of primary studies increase again at a considerably high pace by the time of writing this paper. 

\section{Results Analysis}
\label{sec:results}
% In what follows, we elaborate on the purpose of the RQs and the corresponding results analysis.

\subsection{RQ1: What machine learning algorithms, feature representation and preprocessing methods are used?}
\label{sec:RQ1}

\subsubsection{Machine Learning Algorithm}
\label{sec:RQ1-1}

\begin{figure}[!t]
\centering
\begin{subfigure}[t]{0.46\columnwidth}
\includestandalone[width=\columnwidth]{tikz/RQ1-1-1}
    \subcaption{Top 10 machine learning algorithms used.}
  \label{fig:RQ1-1}
    \end{subfigure}
\begin{subfigure}[t]{0.53\columnwidth}
\begin{tikzpicture}[scale=0.4]
\tikzstyle{every node}=[font=\scriptsize]
    \pie[
        /tikz/every pin/.style={align=center},
        text=label,
        rotate=-40,
        explode=0.1,
        color={black!10,black!30},
        ]
        {24/DL-based,
        76/Non DL-based
        }
    \end{tikzpicture}
   \subcaption{\% of machine learning algorithm type.}
 \label{fig:RQ1-3}
    \end{subfigure}
    \caption{Statistics of machine learning algorithms.}
  \end{figure}
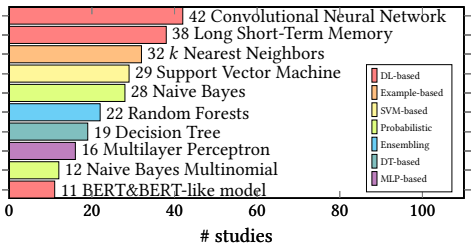
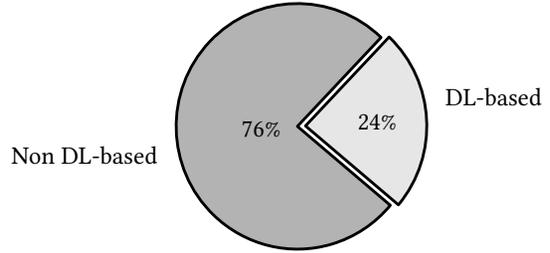

\begin{figure}[!t]
\centering
\begin{subfigure}[t]{0.46\columnwidth}
\begin{tikzpicture}
\begin{axis}[
    xbar=0.05cm,
    width=10cm,
    height=5.5cm,
    xmax=60,
    bar width=0.35cm,
 enlarge y limits=0.08,
 enlarge x limits=0,
    legend style={at={(0.91,0.29)},anchor=north,legend columns=1,font=\tiny},
    legend cell align={left},
     nodes near coords,
          point meta=explicit symbolic,
          every node near coord/.append style={anchor=west,font=\large},
       symbolic y coords={Emotion score,fastText,ChiSquare,BERT’s Word represenntation,TF,GloVe,BM25F,LDA,TF-IDF,Word2Vec},
      yticklabels={,,},
         y tick label style={font=\large},
         ytick=data,
         xtick=\empty
    ]
\addplot [fill=orange!50,draw=black] coordinates {(43,Word2Vec)[43 Word2Vec](0,TF-IDF)[] (0,LDA)[] (0,BM25F)[](0,GloVe)[](0,TF)[](0,BERT’s Word represenntation)[](0,ChiSquare)[](0,fastText)[](5,Emotion score)[]};
\end{axis}

\begin{axis}[
    xbar=0.05cm,
    width=10cm,
    height=5.5cm,
    xmax=60,
    bar width=0.35cm,
 enlarge y limits=0.08,
 enlarge x limits=0,
    legend style={at={(0.89,0.15)},anchor=north,legend columns=1,font=\tiny},
    legend cell align={left},
     nodes near coords,
          point meta=explicit symbolic,
          every node near coord/.append style={anchor=west,font=\large},
       symbolic y coords={Emotion score,fastText,ChiSquare,BERT’s Word represenntation,TF,GloVe,BM25F,LDA,TF-IDF,Word2Vec},
      yticklabels={,,},
         y tick label style={font=\large},
         ytick=data,
         xtick=\empty
    ]
\addplot [fill=cyan!50,draw=black] coordinates {(0,Word2Vec)[](39,TF-IDF)[39 TF-IDF] (16,LDA)[] (12,BM25F)[](9,GloVe)[](9,TF)[](8,BERT’s Word represenntation)[](6,ChiSquare)[](6,fastText)[](5,Emotion score)[]};
\end{axis}

\begin{axis}[
    xbar=0.05cm,
    width=10cm,
    height=5.5cm,
    xmax=60,
    bar width=0.35cm,
 enlarge y limits=0.08,
 enlarge x limits=0,
    legend style={at={(1.05,0.7)},anchor=north,legend columns=1,font=\tiny},
    legend cell align={left},
     nodes near coords,
          point meta=explicit symbolic,
          every node near coord/.append style={anchor=west,font=\large},
       symbolic y coords={Emotion score,fastText,ChiSquare,BERT’s Word represenntation,TF,GloVe,BM25F,LDA,TF-IDF,Word2Vec},
      yticklabels={,,},
         y tick label style={font=\large},
          x tick label style={font=\large},
         ytick=data,
         xtick=\empty
    ]
\addplot [fill=cyan!50,draw=black] coordinates {(0,Word2Vec)[](0,TF-IDF)[] (16,LDA)[16 Latent Dirichlet Allocation] (12,BM25F)[](9,GloVe)[](9,TF)[](8,BERT’s Word represenntation)[](6,ChiSquare)[](6,fastText)[](5,Emotion score)[]};
\end{axis}

\begin{axis}[
    xbar=0.05cm,
    width=10cm,
    height=5.5cm,
    xmax=60,
    bar width=0.35cm,
 enlarge y limits=0.08,
 enlarge x limits=0,
    legend style={at={(1.05,0.7)},anchor=north,legend columns=1,font=\tiny},
    legend cell align={left},
     nodes near coords,
          point meta=explicit symbolic,
          every node near coord/.append style={anchor=west,font=\large},
       symbolic y coords={Emotion score,fastText,ChiSquare,BERT’s Word represenntation,TF,GloVe,BM25F,LDA,TF-IDF,Word2Vec},
      yticklabels={,,},
         y tick label style={font=\large},
         ytick=data,
         xtick=\empty
    ]
\addplot [fill=cyan!50,draw=black] coordinates {(0,Word2Vec)[](0,TF-IDF)[] (16,LDA)[] (12,BM25F)[12 BM25F](9,GloVe)[](9,TF)[](8,BERT’s Word represenntation)[](6,ChiSquare)[](6,fastText)[](5,Emotion score)[]};
\end{axis}

\begin{axis}[
    xbar=0.05cm,
    width=10cm,
    height=5.5cm,
    xmax=60,
    bar width=0.35cm,
 enlarge y limits=0.08,
 enlarge x limits=0,
    legend style={at={(1.05,0.7)},anchor=north,legend columns=1,font=\tiny},
    legend cell align={left},
     nodes near coords,
          point meta=explicit symbolic,
          every node near coord/.append style={anchor=west,font=\large},
       symbolic y coords={Emotion score,fastText,ChiSquare,BERT’s Word represenntation,TF,GloVe,BM25F,LDA,TF-IDF,Word2Vec},
      yticklabels={,,},
         y tick label style={font=\large},
         ytick=data,
         xtick=\empty
    ]
\addplot [fill=orange!50,draw=black] coordinates {(0,Word2Vec)[](0,TF-IDF)[] (0,LDA)[] (0,BM25F)[](9,GloVe)[9 GloVe](9,TF)[](8,BERT’s Word represenntation)[](6,ChiSquare)[](6,fastText)[](5,Emotion score)[]};
\end{axis}

\begin{axis}[
    xbar=0.05cm,
    width=10cm,
    height=5.5cm,
    xmax=60,
    bar width=0.35cm,
 enlarge y limits=0.08,
 enlarge x limits=0,
    legend style={at={(1.05,0.7)},anchor=north,legend columns=1,font=\tiny},
    legend cell align={left},
     nodes near coords,
          point meta=explicit symbolic,
          every node near coord/.append style={anchor=west,font=\large},
       symbolic y coords={Emotion score,fastText,ChiSquare,BERT’s Word represenntation,TF,GloVe,BM25F,LDA,TF-IDF,Word2Vec},
      yticklabels={,,},
         y tick label style={font=\large},
         ytick=data,
         xtick=\empty
    ]
\addplot [fill=cyan!50,draw=black] coordinates {(0,Word2Vec)[](0,TF-IDF)[] (0,LDA)[] (0,BM25F)[](0,GloVe)[](9,TF)[9 Term Frequency](8,BERT’s Word represenntation)[](6,ChiSquare)[](6,fastText)[](5,Emotion score)[]};
\end{axis}

\begin{axis}[
    xbar=0.05cm,
    width=10cm,
    height=5.5cm,
    xmax=60,
    bar width=0.35cm,
 enlarge y limits=0.08,
 enlarge x limits=0,
    legend style={at={(1.05,0.7)},anchor=north,legend columns=1,font=\tiny},
    legend cell align={left},
     nodes near coords,
          point meta=explicit symbolic,
          every node near coord/.append style={anchor=west,font=\large},
       symbolic y coords={Emotion score,fastText,ChiSquare,BERT’s Word represenntation,TF,GloVe,BM25F,LDA,TF-IDF,Word2Vec},
      yticklabels={,,},
         y tick label style={font=\large},
         ytick=data,
         xtick=\empty
    ]
\addplot [fill=orange!50,draw=black] coordinates {(0,Word2Vec)[](0,TF-IDF)[] (0,LDA)[] (0,BM25F)[](0,GloVe)[](0,TF)[](8,BERT’s Word represenntation)[8 BERT’s Word represenntation](6,ChiSquare)[](6,fastText)[](5,Emotion score)[]};
\end{axis}

\begin{axis}[
    xbar=0.05cm,
    width=10cm,
    height=5.5cm,
    xmax=60,
    bar width=0.35cm,
 enlarge y limits=0.08,
 enlarge x limits=0,
    legend style={at={(1.05,0.7)},anchor=north,legend columns=1,font=\tiny},
    legend cell align={left},
     nodes near coords,
          point meta=explicit symbolic,
          every node near coord/.append style={anchor=west,font=\large},
       symbolic y coords={Emotion score,fastText,ChiSquare,BERT’s Word represenntation,TF,GloVe,BM25F,LDA,TF-IDF,Word2Vec},
      yticklabels={,,},
         y tick label style={font=\large},
         ytick=data,
         xtick=\empty
    ]
\addplot [fill=cyan!50,draw=black] coordinates {(0,Word2Vec)[](0,TF-IDF)[] (0,LDA)[] (0,BM25F)[](0,GloVe)[](0,TF)[](0,BERT’s Word represenntation)[](6,ChiSquare)[6 ChiSquare](6,fastText)[](5,Emotion score)[]};
\end{axis}

\begin{axis}[
    xbar=0.05cm,
    width=10cm,
    height=5.5cm,
    xmax=60,
    bar width=0.35cm,
 enlarge y limits=0.08,
 enlarge x limits=0,
    legend style={at={(1.05,0.7)},anchor=north,legend columns=1,font=\tiny},
    legend cell align={left},
     nodes near coords,
          point meta=explicit symbolic,
          every node near coord/.append style={anchor=west,font=\large},
       symbolic y coords={Emotion score,fastText,ChiSquare,BERT’s Word represenntation,TF,GloVe,BM25F,LDA,TF-IDF,Word2Vec},
      yticklabels={,,},
         y tick label style={font=\large},
         ytick=data,
         xtick=\empty
    ]
\addplot [fill=orange!50,draw=black] coordinates {(0,Word2Vec)[](0,TF-IDF)[] (0,LDA)[] (0,BM25F)[](0,GloVe)[](0,TF)[](0,BERT’s Word represenntation)[](0,ChiSquare)[](6,fastText)[6 fastText](5,Emotion score)[]};
\end{axis}

\begin{axis}[
    xbar=0.05cm,
    width=10cm,
    height=5.5cm,
    xmax=60,
    bar width=0.35cm,
 enlarge y limits=0.08,
 enlarge x limits=0,
    legend style={at={(1.05,0.7)},anchor=north,legend columns=1,font=\tiny},
    legend cell align={left},
   xlabel={\# studies},
     x label style={font=\Large},
     nodes near coords,
          point meta=explicit symbolic,
          every node near coord/.append style={anchor=west,font=\large},
       symbolic y coords={Emotion score,fastText,ChiSquare,BERT’s Word represenntation,TF,GloVe,BM25F,LDA,TF-IDF,Word2Vec},
      yticklabels={,,},
         y tick label style={font=\large},
          x tick label style={font=\large},
         ytick=data
    ]
\addplot [fill=cyan!50,draw=black] coordinates {(0,Word2Vec)[](0,TF-IDF)[] (0,LDA)[] (0,BM25F)[](0,GloVe)[](0,TF)[](0,BERT’s Word represenntation)[](0,ChiSquare)[](0,fastText)[](5,Emotion score)[5 Emotion score]};
\end{axis}

\begin{axis}[
    xbar=0.05cm,
    width=10cm,
    height=5.5cm,
    xmax=60,
    bar width=0.35cm,
 enlarge y limits=0.08,
 enlarge x limits=0,
    legend style={at={(0.86,0.25)},anchor=north,legend columns=1,font=\tiny},
    legend cell align={left},
   xlabel={\# studies},
     x label style={font=\Large},
     nodes near coords,
          point meta=explicit symbolic,
          every node near coord/.append style={anchor=west,font=\large},
       symbolic y coords={Emotion score,fastText,ChiSquare,BERT’s Word represenntation,TF,GloVe,BM25F,LDA,TF-IDF,Word2Vec},
      yticklabels={,,},
         y tick label style={font=\large},
          x tick label style={font=\large},
         ytick=data
    ]
\addplot [fill=cyan!50,draw=black] coordinates {(0,Word2Vec)[](0,TF-IDF)[] (0,LDA)[] (0,BM25F)[](0,GloVe)[](0,TF)[](0,BERT’s Word represenntation)[](0,ChiSquare)[](0,fastText)[](0,Emotion score)[]};
\addplot [fill=orange!50,draw=black] coordinates {(0,Word2Vec)[](0,TF-IDF)[] (0,LDA)[] (0,BM25F)[](0,GloVe)[](0,TF)[](0,BERT’s Word represenntation)[](0,ChiSquare)[](0,fastText)[](0,Emotion score)[]};
\legend{Non DL-based, DL-based}
\end{axis}

\end{tikzpicture}
   \subcaption{Top 10 feature representation methods (TF-IDF denotes Term Frequency-Inverse Document Frequency).}
 \label{fig:RQ1-2}
    \end{subfigure}
\begin{subfigure}[t]{0.53\columnwidth}
\begin{tikzpicture}[scale=0.4]
\tikzstyle{every node}=[font=\scriptsize]
    \pie[
        /tikz/every pin/.style={align=center},
        text=label,
        rotate=-40,
        explode=0.1,
        color={black!10,black!30},
        ]
        {22.1/DL-based,
        77.9/Non DL-based
        }
    \end{tikzpicture}
   \subcaption{\% of feature representation method type.}
 \label{fig:RQ1-4}
    \end{subfigure}
    \caption{Statistics of feature representation methods.}
  \end{figure}
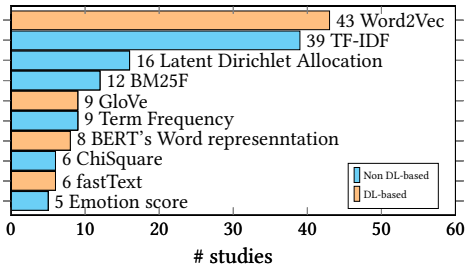
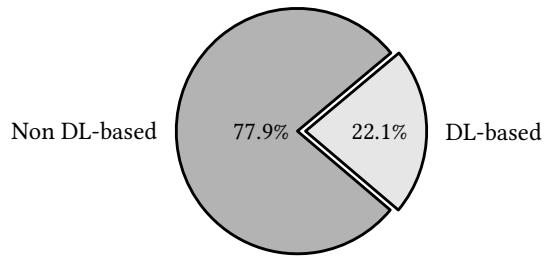
  
Since our focus is on bug report analysis with machine learning, it is essential to understand what machine learning algorithms are of the most interest. In this section, we particularly review the machine learning algorithms that are commonly utilized for bug report analysis within the community. Figure~\ref{fig:RQ1-1} shows the Top 10 machine learning algorithms used in primary studies. It is observed that Convolutional Neural Network (CNN), Long Short-Term Memory (LSTM) and $k$-Nearest Neighbor ($k$NN) are the most commonly selected methods compared with the others. Here, we categorize the used machine learning methods into: basic methods, probabilistic methods, example-based methods, ensemble methods, cluster-based methods, decision tree (DT)-based methods, support vector machine (SVM)-based methods, and neural network methods. %It's important to note that these categories include algorithms that did not make the top 10. However, due to their representativeness, we have also included them. These include Linear Regression, Logistic Regression, Voting, Bagging, Boosting, Stacking, Recurrent Neural Networks (RNNs) with Gated Recurrent Units (GRUs), and Transformers.
%To ensure fair coverage, for all primary studies, we summarize all algorithms that are either used as the core that underpins the proposed approach or those that are used in the comparative evaluation. 

Between 2020 and 2023, there has been a notable surge in the adoption of deep learning-based models for bug report analysis, with a total of 58 studies utilizing these advanced computational techniques, compared to just 18 studies between 2014 and 2019. This significant increase underscores the growing recognition and reliance on deep learning methodologies within this domain. The enhanced capabilities of deep learning models to handle complex patterns and large datasets effectively have made them increasingly favored for analyzing and extracting meaningful insights from bug reports.

Here, we go through each type of machine learning in detail:

\paragraph{Basic methods. }
\textbf{Linear Regression}, modeling the relationship between a dependent variable and its independent variables, has been applied to predict severity or priority levels in bug report analysis \cite{DBLP:journals/ese/TianLXS15, DBLP:journals/access/UmerLS18}. However, its effectiveness may be limited by the linear assumption, whereas bug report analysis involves complex and non-linear patterns. In contrast, \textbf{Logistic Regression}, estimating probabilities with a logistic function for binary classification task, is useful in bug report identification~\cite{DBLP:conf/issre/WenYH16,DBLP:journals/tse/PetersTYN19} and duplicate bug report detection~\cite{DBLP:journals/ese/HindleAS16,DBLP:conf/wcre/AggarwalRTHGS15}. It can also be extended to multi-class tasks using a one-vs-all strategy for severity~\cite{DBLP:conf/fedcsis/KumarKNMKP21,DBLP:conf/qrs/SuHCQM22} and priority level prediction~\cite{DBLP:conf/ccs/ParkKS19,DBLP:conf/qrs/SuHCQM22,DBLP:conf/internetware/HuangSFYYZ22}. %Logistic regression is typically used in comparison with other models to find the best one, and its performance is often not the highest among other models.

\paragraph{Probabilistic methods.}
Probabilistic classifiers predict the probability that a document belongs to a class using Bayes’ theorem~\cite{DBLP:conf/icetc/TararAB19,DBLP:conf/ecml/Lewis98}. \textbf{Naive Bayes}~\cite{rish2001empirical}, which assumes independence between features, is used in bug report analysis due to its simplicity and effectiveness with textual data~\cite{DBLP:conf/apsec/XuLXSL15,DBLP:journals/tse/PetersTYN19,DBLP:journals/access/UmerLS18,DBLP:journals/ese/HindleAS16,DBLP:conf/bdccf/GaoL0GG18}. However, its assumption of feature independence often does not hold true in real-world text data, where word occurrences can be highly dependent on each other.

\paragraph{Example-based methods.}
\textbf{$k$NN} is a simple, non-parametric lazy learning algorithm for classification and regression tasks that identifies the top $k$ objects similar to a given instance. In bug report analysis, $k$NN efficiently handles feature-rich datasets by measuring the proximity of bug reports in feature space, aiding in pattern and similarity identification~\cite{DBLP:conf/compsac/YangZL14,DBLP:journals/tr/XiaLSW16,DBLP:conf/qrs/SuHCQM22,DBLP:conf/internetware/HuangSFYYZ22}. However, $k$NN can be computationally intensive, as computing distances between a query point and all training examples becomes inefficient with larger datasets.

\paragraph{Cluster-based methods.}
Clustering techniques group similar bug reports to identify common patterns and improve resolution strategies. Two commonly used models are: \textbf{$k$-means} selects centroids for each cluster and associates each feature vector with the closest centroid. It is known for its simple implementation and computational efficiency, making it effective for categorizing large volumes of bug reports~\cite{DBLP:conf/esem/WangCWW16,DBLP:conf/iwpc/ThungLL15}. However, it faces challenges like determining the optimal number of clusters and sensitivity to initial centroids and outliers. \textbf{Fuzzy $c$-means} allows each data point to have a degree of membership in each cluster, addressing $k$-means' limitations in handling datasets with unclear cluster boundaries~\cite{DBLP:journals/access/JindalK20}.

\paragraph{DT-based methods.}
\textbf{Decision Trees} use a tree-like model to make decisions by segmenting reports based on their attributes. They are effective for extracting rules and patterns from bug reports, aiding in various analysis tasks~\cite{DBLP:journals/access/JainKSNT19,DBLP:conf/ijcnn/CiancariniPRS17,DBLP:journals/access/Guo0W0L18,DBLP:journals/jss/GarciaSN18,DBLP:conf/ease/ZhouNG15,DBLP:conf/w4a/AljedaaniML0J22}. Decision trees are noted for their interpretability and ability to handle non-linear relationships. However, they can overfit by creating overly complex trees and struggle with imbalanced data, potentially biasing the algorithm towards predominant classes~\cite{DBLP:conf/w4a/AljedaaniML0J22}.

\paragraph{Ensembling methods.}
Ensemble Learning combines multiple classifiers to improve performance, accuracy, and robustness. Common methods include \textbf{voting}~\cite{DBLP:journals/jss/WuZCWM20} and \textbf{bagging}~\cite{le2017will,DBLP:journals/saem/KaurJ19}. \textbf{Random Forest}, a bagging method, constructs multiple decision trees and averages their predictions, reducing overfitting. However, bagging can struggle with imbalanced datasets, addressed by techniques like constructing classifiers on disjoint subsets~\cite{DBLP:journals/infsof/XiaLSWY15}. \textbf{Boosting} trains models sequentially, correcting errors of previous models, with algorithms like AdaBoost~\cite{freund1995desicion}, Gradient Boosting~\cite{friedman2001greedy}, and XGBoost~\cite{DBLP:journals/corr/ChenG16}. Boosting is robust against unbalanced data but can be sensitive to noise and computationally demanding. \textbf{Stacking} combines model predictions using another model to find optimal weights, potentially outperforming bagging but at a higher computational cost~\cite{DBLP:journals/ese/JonssonBBSER16}.

\paragraph{SVM-based methods.}
\textbf{SVM} classifies data by finding the hyperplane that best separates classes with maximum margin. In bug report analysis, SVM is favored for handling complex, high-dimensional textual data, making it effective for various tasks~\cite{DBLP:journals/array/HirschH22,DBLP:conf/iccsa/SharmaKSS14,DBLP:conf/apsec/TsurudaMA15,DBLP:conf/icaisc/FloreaAA17}. Despite its strengths, SVM can be time-consuming and sensitive to data scale and dimensionality, requiring dimensionality reduction or feature selection to maintain accuracy and efficiency~\cite{zhang2008improved}.

\paragraph{MLP-based methods.}
\textbf{Multilayer Perceptron (MLP)} is a feedforward neural network with an input layer, hidden layers, and an output layer, interconnected through weights adjusted via the backpropagation algorithm. This process helps minimize the difference between actual and predicted outputs and addresses overfitting through regularization. Studies often compare MLPs with other algorithms, but MLPs rarely outperform alternatives like SVM and Naive Bayes in bug report analysis~\cite{DBLP:conf/iceccs/KochharTL14,DBLP:journals/access/JainKSNT19,DBLP:conf/crisis/Gawron0M17,DBLP:conf/ccs/ParkKS19,DBLP:conf/fedcsis/KumarKNMKP21}. Despite their complexity and lengthy training times, MLPs do not consistently show superior performance.

\paragraph{DL-based methods.}
Deep Learning (DL) refers to a subset of machine learning that employs neural networks with many layers to model complex patterns in data. Unlike traditional machine learning algorithms that require manual feature extraction, DL methods automatically learn hierarchical feature representations, making them particularly powerful for tasks involving large and complex datasets. It's important to distinguish between DL and simpler neural network architectures like MLPs. MLPs are a type of neural network with one or more layers of nodes, typically used for tasks requiring fixed-size input and output. However, MLPs are not considered part of DL in this work because they lack the depth and complexity of architectures, which can automatically learn and extract features from data at multiple levels of abstraction.

\textbf{Convolutional Neural Networks (CNNs)} learn spatial hierarchies of features from input data, originally designed for images but also effective for text processing in bug report analysis. They automatically extract features from text represented as numerical vectors, identifying patterns like n-grams without extensive preprocessing. This makes CNNs suitable for various bug report analysis tasks~\cite{DBLP:conf/apsec/AnhL21,DBLP:conf/apsec/WangL21,DBLP:conf/issre/HeXF0YL20}. However, while CNNs excel in spatial feature extraction, they are less effective than RNNs for tasks requiring long-term temporal dependency processing, such as time-series analysis or natural language processing.

\textbf{Recurrent Neural Networks (RNNs)}, including \textbf{Long Short-Term Memory (LSTM)} and \textbf{Gated Recurrent Unit (GRU)} architectures, are essential in bug report analysis for processing sequential data and retaining past information. RNNs capture contextual dependencies in bug reports, enhancing understanding of issues. Traditional RNNs face vanishing and exploding gradient problems, but LSTM and GRU networks address these by retaining long-term dependencies. LSTM networks are notable for their gated mechanism that maintains effective learning over long sequences, making them invaluable for bug report analysis tasks~\cite{DBLP:conf/comad/ManiSA19,DBLP:conf/icmla/0003FWBL18,DBLP:conf/issta/LiLZZ19,DBLP:journals/access/KimY22d,DBLP:conf/esem/LiCHWWWW22}. GRUs balance effectiveness and computational efficiency by simplifying gating mechanisms, suitable for large-scale bug report analysis~\cite{DBLP:conf/internetware/Xi0XX018}. However, GRUs' simplified structure can struggle with capturing complex dependencies.

\textbf{Transformer} models, such as \textbf{BERT} and its variants, have revolutionized bug report analysis by capturing bidirectional contexts and global dependencies within textual data. They excel at tasks such as bug report summarization, sentiment analysis, and identifying similarities between bug reports and code repositories. For instance, the research identified as~\citeauthor{DBLP:conf/dsa/ZhengCWFSC21}~\cite{DBLP:conf/dsa/ZhengCWFSC21} introduces an approach for predicting bug severity by integrating Transformer models with domain-specific pre-training strategies. Meanwhile, study~\citeauthor{DBLP:conf/noms/FukudaWHT22}~\cite{DBLP:conf/noms/FukudaWHT22} leverages the model to generate readable fault reports from diverse data, demonstrating its capability in fault detection. Additionally, \citeauthor{DBLP:journals/access/WeiZR23}~\cite{DBLP:journals/access/WeiZR23} employs the Transformer to refine the extraction of security domain knowledge, enhancing keyword quality. These examples highlight the model's potential in enhancing software reliability and security through sophisticated analysis of bug reports.

Despite their effectiveness, transformer models face several adoption barriers beyond theoretical limitations. The main reasons include:

\begin{itemize}
\item \textbf{High Computational Requirements:} Training BERT from scratch requires massive datasets, such as the 3.3 billion words used for the original BERT model, and significant GPU resources, which are often inaccessible to many research teams. A likely solution could be leveraging transfer learning, such as pre-trained models like CodeBERT~\cite{DBLP:conf/emnlp/FengGTDFGS0LJZ20}, fine-tuned on domain-specific datasets to significantly reduce computational demands.

\item \textbf{Data Scarcity:} Labeled bug report datasets are often small and imbalanced, posing practical barriers for training transformer models effectively. A possible solution is synthetic data generation. For example, \cite{DBLP:journals/tr/MaKYZZL23} employs a BERT-based encoder-decoder framework combined with a transformer decoder and copy mechanism to generate high-quality, concise bug report titles from lengthy descriptions, thus effectively mitigating data inadequacy in low-resource settings.

\item \textbf{Lack of Interpretability:} The "black-box" nature of transformer-based deep learning models presents significant challenges, particularly for trust-sensitive tasks like severity prioritization. One way to address this challenge could be the integration of attention visualization into BERT-based systems. For instance, \cite{DBLP:conf/iccel/HirakawaTN20} introduces attention heatmaps within their fine-tuned model, achieving high performance (F1 score of 0.87) on real-world OSS defect reports, while transparently indicating how category-specific keywords influence classification decisions.

\end{itemize}

\begin{table}[t!]
  \caption{Summaries of the strengths, weaknesses, and best-suited usage scenarios for the ML approaches.}
% \vspace{-0.3cm}
\centering
\footnotesize
\begin{adjustbox}{width=\linewidth,center}
\begin{tabular}{p{2cm}p{4cm}p{3.5cm}p{3.5cm}}
\toprule

\textbf{Category} & \textbf{Strength} & \textbf{Weakness} & \textbf{Best-Suited Scenario} \\ \hline
Basic methods & Simple and interpretable models that are computationally efficient. & Limited by linear assumptions and sensitive to outliers. & Predicting severity or priority levels in bug reports with linear relationships. \\ \hline
Probabilistic methods & Simple and fast, effective with large datasets and handles missing data well. & Assumes feature independence, which often does not hold true in real-world data. & Text classification tasks and bug report analysis where computational efficiency is crucial. \\ \hline
Example-based methods & Simple and intuitive, effective with multi-modal data and no training phase required. & Computationally intensive with large datasets and sensitive to irrelevant features. & Pattern and similarity identification in feature-rich bug report datasets. \\ \hline
Cluster-based methods & Simple implementation, computationally efficient, and handles large datasets well. & Requires pre-specifying the number of clusters and sensitive to initial centroids and outliers. & Categorizing large volumes of bug reports with well-separated clusters. \\ \hline
DT-based methods & Interpretable and easy to visualize, handles non-linear relationships well. & Prone to overfitting and struggles with imbalanced data. & Extracting rules and patterns from bug reports where interpretability is crucial. \\ \hline
Ensembling methods & Combines multiple classifiers for improved performance, accuracy, and robustness. & Computationally intensive and less interpretable than single models. & High accuracy tasks and bug report analysis with imbalanced datasets. \\ \hline
SVM-based methods & Effective in high-dimensional spaces and robust to overfitting. & Computationally intensive and sensitive to the choice of kernel parameters. & Classification tasks with clear margin of separation and high-dimensional data. \\ \hline
MLP-based methods & Capable of learning complex patterns and relationships in data. & Requires large amounts of data and computational resources for training. & Tasks involving complex, non-linear relationships and large datasets. \\ \hline
CNN-based methods & Excellent for image and spatial data analysis, capturing local dependencies. & Requires large datasets and computational resources, sensitive to overfitting. & Image-based bug report analysis and tasks involving spatial data. \\ \hline
RNN-based methods & Effective for sequential data and capturing temporal dependencies. & Computationally intensive and prone to vanishing gradient problem. & Analyzing sequential bug report data and tasks involving time-series data. \\ \hline
Transformer-based methods & Handles long-range dependencies well and highly parallelizable. & Requires substantial computational resources and large datasets. & Natural language processing tasks and analyzing complex bug report texts. 
\\ \bottomrule
\end{tabular}
\end{adjustbox}
\label{tb:pro_con_ml_algorithm}
% \vspace{-0.3cm}
\end{table}

\paragraph{\textbf{Discussions}}
Through our investigation, as the increasing trend to embrace deep learning methods, BERT \& its variants, along with other transformer models, have been highlighted for their superior performance in bug report analysis tasks. However, within the period from 2020 to 2023, only 14 studies have employed BERT \& BERT-like or transformer-based models. As illustrated in Figure~\ref{fig:RQ1-3}, there is a modest 24\% of all the deep learning-based approaches, which suggests that their adoption is not as widespread as the general category of deep learning models, although their merit for bug report analysis is highly recognized.

The relatively lower adoption rate of BERT \& BERT-like and transformer-based models may be attributed to several factors, including the requirement for substantial computational resources and the necessity for large datasets to train these models effectively. Nevertheless, the demonstrated efficacy of these models in handling bug report analysis tasks presents a compelling case for their increased utilization, suggesting that future research and application in this area could benefit significantly from a deeper exploration and broader adoption of transformer-based deep learning models.

% There are also other algorithms used, including the fairly complex variants of neural network model, which form the remaining minority. 

Based on our review, we have compiled Table~\ref{tb:pro_con_ml_algorithm} to highlight the strengths, weaknesses, and ideal use cases for various categories of ML methods. This table serves as a valuable resource for researchers and practitioners, enabling them to choose the most appropriate ML method for different scenarios.

In a nutshell, our findings can be summarized as follows:

\begin{tcolorbox}[breakable,left=5pt,right=5pt,top=5pt,bottom=5pt] 
\textbf{\underline{Finding 1}}: CNN, LSTM and $k$NN are the top three commonly used machine learning algorithms for bug report analysis. It also notes a significant increase in the use of deep learning algorithms from 2020 to 2023.

% CNN, LSTM and $k$NN are the top three commonly used machine learning algorithms for bug report analysis. Despite BERT \& BERT-like models along with other transformer-based architectures are proven superiority in handling bug report analysis, they have been underutilized, possibly due to their high computational demands and necessity for large datasets.
\end{tcolorbox}

%Given the recent development of AI, deep learning algorithms, particularly deep Recurrent Neural Networks (RNN) and its kind, have been becoming very successful in natural language processing (NLP)~\cite{young2018recent}. Indeed, it remains lack of evidence about the effectiveness of deep learning outside the fields of NLP and Computer Vision, and therefore suggesting it as the alternative needs to take this into account. However, bug report analysis actually bear many similarity to the NLP tasks, e.g., they all tends to deal with classification; the documents/reports are both nonstructural and involve the naturalness of human language; there are a large number of documents/bug reports to be analyzed. This provides a perfect avenue for the adoption of deep learning.

%Yet, despite that there are a few primary studies have investigated deep learning, majority of them remain rely on the classic machine learning algorithms, making no breakthrough but staying in the `comfort zone'. In fact, on those primary studies that adopt $k$NN, NB and SVM, we found no comparison to other deep learning algorithms. This rises the concern about the bias of using classic machine learning algorithms have not been well-justified.

\subsubsection{Feature Representation}
\label{sec:RQ1-2}

The employment of appropriate feature representation is important for extracting meaningful insights from textual data. Typical methods encompass a wide spectrum ranging from traditional non-deep learning approaches to the more advanced DL-based techniques. Here, we outline the significance and utility of both categories of feature representation methods in facilitating comprehensive bug report analysis.

As illustrated in Figure~\ref{fig:RQ1-2}, Word2Vec and TF-IDF are the most common feature representation methods, accounting for 43 and 39 studies respectively. They each have three times the number of studies compared to Latent Dirichlet Allocation, which is the third most prevalent method. 

Moreover, our analysis revealed an intriguing trend: compared to the period from 2014 to 2019 (with 16 studies), the years from 2020 to 2023 witnessed a substantial increase in the adoption of DL-based feature representation methods, with 48 studies. 

Here, we present an overview of these methods, categorized into non-deep learning (DL) based representation and DL-based representation, each serving distinct approaches in representing the intricacies of bug reports:

\paragraph{Non DL-based Method.}
\textbf{Term Frequency-Inverse Document Frequency (TF-IDF)} is widely used in bug report analysis to capture term importance by weighing terms based on their document frequency relative to the corpus~\cite{DBLP:journals/ese/JonssonBBSER16,DBLP:conf/iwpc/TianWLG16,DBLP:conf/iwpc/LamNNN17,DBLP:journals/ese/TianLXS15,DBLP:conf/iceccs/KochharTL14}. This highlights significant terms within bug reports but ignores word order, context, and semantics, and may bias towards longer documents. Some studies use simple \textbf{Term Frequency} to count term occurrences~\cite{DBLP:conf/iwpc/LiJLRL18,DBLP:journals/jss/GarciaSN18,DBLP:conf/esem/WangZW14,DBLP:conf/compsac/JiPCM18,DBLP:conf/esem/FanYYWW17,DBLP:conf/icse/WangCWW17}, offering quick insights but lacking sophistication for complex tasks.

\textbf{BM25F} extends BM25 with field-based weighting, making it suitable for bug report analysis where different fields have varying importance. It requires careful parameter tuning for optimal performance~\cite{DBLP:journals/asc/ShiKBZ18,DBLP:journals/ese/HindleAS16,DBLP:journals/jss/ZhangCYLL16,DBLP:journals/jss/LinYLC16,DBLP:journals/tosem/ZhangHVIXTLJ23}. \textbf{Latent Dirichlet Allocation (LDA)} uncovers latent topics within bug reports, aiding in text representation by identifying topics based on word distributions~\cite{DBLP:conf/qrs/JonssonBMSVE16,jayarajah2016duplicate,DBLP:conf/icaisc/FloreaAA17,DBLP:journals/smr/XiaLWZ15,DBLP:conf/iwpc/ZhangCJLX17}. However, LDA struggles with large vocabularies. \textbf{ChiSquare} identifies significant terms for issue classification in bug report analysis, essential for feature selection. Higher Chi-Square values indicate stronger term-topic associations~\cite{DBLP:conf/apsec/XuLXSL15,DBLP:conf/issre/WenYH16,sharma2015novel,roy2014towards,DBLP:conf/icaisc/FloreaAA17,DBLP:conf/compsac/XiaLQWZ14}.

\paragraph{DL-based Method. }

\textbf{Word2Vec} captures semantic similarities between terms by representing words as dense vectors in a high-dimensional space, enhancing bug report analysis by identifying semantically similar terms~\cite{DBLP:journals/corr/abs-1301-3781}. Some researchers use pre-trained Word2Vec models, while others train their own, like the CBOW model developed by \citeauthor{DBLP:conf/iwpc/Zhang0YZZ20}~\cite{DBLP:conf/iwpc/Zhang0YZZ20}. However, Word2Vec may struggle with global corpus statistics and rare words. \textbf{GloVe (Global Vectors for Word Representation)} uses global co-occurrence statistics for deeper contextual understanding of terms in bug reports, valuable for comprehensive analysis~\cite{DBLP:conf/apsec/AnhL21,DBLP:conf/sac/KimKL22a,DBLP:conf/apsec/LongCC22,DBLP:conf/msr/RodriguesAFD20,DBLP:conf/kbse/ChenXYJCX20}. GloVe, however, requires more memory and computational resources due to its large co-occurrence matrix.

\textbf{fastText} is used in bug report analysis~\cite{DBLP:journals/compsec/SunOZLWZ23,DBLP:journals/scp/KallisSCP21,DBLP:conf/qrs/SuHCQM22,DBLP:conf/issre/ZhengZTCCWS21,DBLP:journals/jksucis/GuptaIF22} for its ability to incorporate subword details, managing out-of-vocabulary terms and morphologically complex languages. By summing character n-grams to represent words, fastText provides robust embeddings that handle unseen terms and capture morphological variations, enhancing the vector representations for bug summaries and keywords.

\textbf{BERT's word representation}, derived from Bidirectional Encoder Representations from Transformers (BERT), is highly effective in capturing contextual information within bug reports. By considering the surrounding context of each word, BERT's embeddings provide rich representations that account for the nuanced meanings of terms within their specific contexts, yielding rich representations that encapsulate nuanced term meanings within their specific contexts. This methodology has proven to be crucial in recent bug report analysis research~\cite{DBLP:journals/access/KimY22f,DBLP:conf/qrs/WangZLYZ22,DBLP:journals/access/WangL22d,DBLP:journals/tr/MessaoudMJMG23,DBLP:journals/access/MohsenHWMM23,DBLP:journals/tr/MaKYZZL23,DBLP:conf/icse/HaeringSM21,DBLP:conf/icsm/IsotaniWFNOS21}, significantly contributing to the high accuracy achieved by the BERT model.

% fastText is utilized in bug report analysis for its ability to capture subword information, particularly useful for handling out-of-vocabulary terms and morphologically rich languages. By representing words as the sum of their character n-grams, fastText embeddings provide robust representations that can handle unseen terms and capture morphological variations effectively. This method is essential for bug report analysis in diverse linguistic contexts, ensuring that relevant information is captured accurately even for terms with limited occurrences or variations.

% BERT's word representation, derived from Bidirectional Encoder Representations from Transformers (BERT), is highly effective in capturing contextual information within bug reports. By considering the surrounding context of each word, BERT embeddings provide rich representations that account for the nuanced meanings of terms within their specific contexts. This approach is critical for bug report analysis as it allows for more accurate understanding of the relationships between terms, improving the precision of issue identification and resolution.

\begin{table}[t!]
  \caption{Summaries of the strengths, weaknesses, and best-suited usage scenarios for the feature representation methods.}
% \vspace{-0.3cm}
\centering
\footnotesize
\begin{adjustbox}{width=\linewidth,center}
\begin{tabular}{p{2cm}p{4cm}p{3.5cm}p{3.5cm}}
\toprule

\textbf{Category} & \textbf{Strength} & \textbf{Weakness} & \textbf{Best-Suited Scenario} \\ \hline
TF-IDF & Captures term importance   by weighing terms based on their document frequency relative to the corpus. & Ignores word order,   context, and semantics; may bias towards longer documents. & Highlighting significant   terms within bug reports. \\ \hline
Term Frequency & Simple and quick insights   by counting term occurrences. & Lacks sophistication for   complex tasks. & Basic analysis of term   occurrences in bug reports. \\ \hline
BM25F & Extends BM25 with   field-based weighting, suitable for varying field importance. & Requires careful   parameter tuning for optimal performance. & Analyzing bug reports   with different field importances. \\ \hline
LDA & Uncovers latent topics   within bug reports by identifying topics based on word distributions. & Struggles with large   vocabularies. & Topic modeling and text   representation in bug reports. \\ \hline
ChiSquare & Identifies significant   terms for issue classification, essential for feature selection. & May not capture complex   term relationships. & Feature selection for   issue classification in bug reports. \\ \hline
Word2Vec & Captures semantic   similarities between terms by representing words as dense vectors. & May struggle with global   corpus statistics and rare words. & Identifying semantically   similar terms in bug reports. \\ \hline
GloVe & Uses global co-occurrence   statistics for deeper contextual understanding of terms. & Requires more memory and   computational resources due to its large co-occurrence matrix. & Comprehensive analysis of   terms in bug reports. \\ \hline
fastText & Incorporates subword   details, managing out-of-vocabulary terms and morphologically complex   languages. & Requires substantial   computational resources. & Enhancing vector   representations for bug summaries and keywords. \\ \hline
BERT & Captures contextual   information within bug reports, providing rich representations. & Requires substantial   computational resources and large datasets. & High accuracy tasks in   bug report analysis, such as summarization and sentiment analysis. 
\\ \bottomrule
\end{tabular}
\end{adjustbox}
\label{tb:pro_con_feat_repre}
% \vspace{-0.3cm}
\end{table}

\paragraph{\textbf{Discussions}}
We observe that Word2Vec has become the most frequently used feature representation due to the growing recognition of the importance of DL-based models, consistent with our findings in Section~\ref{sec:RQ1-2}, which noted the increasing application of DL-based machine learning models. While DL-based machine learning algorithms can be combined with non-DL-based feature representation methods, such as Word2Vec with Naive Bayes or SVM, such combinations often necessitate dimensionality reduction operations like pooling, which can lead to information loss and limited improvements in accuracy. Therefore, as shown in Figure~\ref{fig:RQ1-4} , most studies opt to combine non-DL-based machine learning algorithms with non-DL-based feature representation methods, or DL-based machine learning algorithms with DL-based feature representation methods.

In recent years, we have observed a new trend emerging in some studies, wherein researchers directly utilize the output of BERT as feature representation. We believe that future research directions should pay more attention to this approach, as BERT possesses superior capabilities in capturing contextual information and semantic nuances within textual data, potentially leading to enhanced bug report analysis performance.

Based on our review, we have created Table~\ref{tb:pro_con_feat_repre}, which details the strengths, weaknesses, and optimal use cases for each feature representation method. This table serves as a practical guide for researchers and practitioners, aiding them in selecting the most appropriate feature representation method for different scenarios.

In summary, we obtain findings shown below:

\begin{tcolorbox}[breakable,left=5pt,right=5pt,top=5pt,bottom=5pt] 
\textbf{\underline{Finding 2}}: Word2Vec is the most popular method for feature representation. TF-IDF ranked the second. There has been a notable increase in the adoption of DL-based feature representation methods from 2020 to 2023, outpacing previous years.
\end{tcolorbox}

%\footnote{We consider all variants of TF-IDF used as the same kind.}

%\subsection{Questionable Choice of the Popular Feature Representation}

%Bug reports, being the documents that are written and submitted by human reporters, would inevitably involve strong semantic naturalness of human language. Therefore, ignoring the semantics in bug report analysis is an unwise simplification of the problem's nature. For example, in the duplicated bug reports detection problem, the summaries of two duplicated bugs reports of OpenOffice are shown as below:

%\begin{itemize}
 %   \item \textbf{Bug ID 85502:} \texttt{Alt+<letter> does not work in dialogs}
  %  \item \textbf{Bug ID 85819:} \texttt{Alt-<key> no longer works as expected}
%\end{itemize}

%As a human developer, one would know easily that they are highly likely to be duplicated reports. However, when representing feature using TF-IDF metric, the summaries of these two bug reports would have rather low similarity, since there are nearly no words that are identical. This clearly does not match with the fact. 

%Indeed, there is existing work which shows that word embedding methods, such as Word2Vec, can provide better feature representation when combined with TF-IDF ~\cite{lilleberg2015support}\cite{ge2017improving}. However, the predominated usage of sole TF-IDF still rise a concern about whether it is a reasonable choice for representing features in bug report analysis, which is often semantics related.

\subsubsection{Preprocessing method}
\label{sec:RQ1-3}

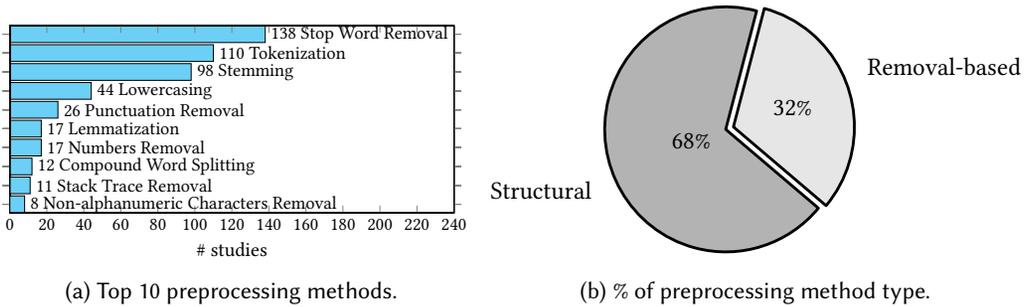
\begin{figure}[!t]
\centering
\begin{subfigure}[t]{0.46\columnwidth}
\includestandalone[width=\columnwidth]{tikz/RQ1-3-1}
   \subcaption{Top 10 preprocessing methods.}
 \label{fig:RQ1-3-1}
    \end{subfigure}
\begin{subfigure}[t]{0.53\columnwidth}
\begin{tikzpicture}[scale=0.4]
\tikzstyle{every node}=[font=\scriptsize]
    \pie[
        /tikz/every pin/.style={align=center},
        text=label,
        rotate=-40,
        explode=0.1,
        color={black!10,black!30},
        ]
        {32/Removal-based,
        68/Structural
        }
    \end{tikzpicture}
   \subcaption{\% of preprocessing method type.}
 \label{fig:RQ1-3-2}
    \end{subfigure}
    \caption{Statistics of preprocessing methods.}
  \end{figure}

In bug report analysis, preprocessing methods are essential for refining raw textual data to improve the effectiveness of analysis tasks. These methods help reduce noise, standardize text formats, and enhance the quality of extracted features. Figure~\ref{fig:RQ1-3-1} shows the top 10 preprocessing methods used in primary studies. It is observed that stop word removal is the most frequently used method, appearing in 138 papers. Tokenization and stemming rank second and third, with 110 and 98 papers, respectively. These three methods are dominant, as there is a significant drop starting from the fourth-ranked method, lowercasing, which appears in only 44 papers.

Preprocessing techniques can be broadly categorized based on whether they involve the removal of certain elements from the text. We classify preprocessing methods into two groups: Removal-based methods and structural methods. Removal-based methods eliminate specific components from the text, such as stop words, punctuation, or stack traces, to reduce redundancy and focus on meaningful content. In contrast, structural methods transform or normalize text without discarding any information, including techniques like tokenization, stemming, and lowercasing. As shown in Figure~\ref{fig:RQ1-3-2}, our analysis reveals that 32\% of the preprocessing methods used in primary studies are removal-based, while structural methods account for 68\%. The following provides an overview of each category, highlighting their benefits and limitations:

\paragraph{Removal-based methods.}
\textbf{Stop Word Removal}~\cite{le2014predicting,jonsson2016automated,yang2017high,pandey2017automated,kochhar2014automatic} eliminates common but uninformative words (e.g., "the," "is") to reduce noise in text analysis. Its advantage is improved efficiency and relevance in text processing, but it may lead to loss of contextual meaning. \textbf{Punctuation Removal}~\cite{peters2017text,chaparro2017detecting,DBLP:journals/infsof/JahanshahiC22,pandey2017automated,ardimento2017knowledge} deletes symbols such as commas and periods to standardize text input. This enhances consistency but can remove valuable syntactic cues. \textbf{Numbers Removal}~\cite{chaparro2017detecting,DBLP:journals/infsof/JahanshahiC22,yang2017high,jain2019particle,jahanshahi2021dabt} eliminates numerical values to prevent them from skewing textual analysis. While it reduces dimensionality, it may also discard meaningful identifiers or version numbers. \textbf{Stack Trace Removal}~\cite{DBLP:conf/cascon/SaborHH16,DBLP:journals/infsof/SaborHH20,DBLP:journals/infsof/GeFQGQ22,jiang2020ltrwes,hirsch2022using} filters out automatically generated error traces, reducing redundancy in bug reports. This helps focus on descriptive content but may remove useful debugging information. \textbf{Non-alphanumeric Character Removal}~\cite{deshmukh2017towards,peters2017text,DBLP:conf/msr/RodriguesAFD20,DBLP:journals/access/AfricVSD23,kukkar2020duplicate} strips out special characters to simplify text representation. It increases uniformity but can result in the loss of structured information, such as code snippets or log patterns.

\paragraph{Structural methods.}
\textbf{Tokenization}~\cite{DBLP:conf/msr/RodriguesAFD20,DBLP:conf/esem/LiCHWWWW22,DBLP:conf/issre/HeXF0YL20,li2022deeplabel,he2020duplicate} splits text into individual words or subwords to facilitate further processing. It improves feature extraction but can increase data sparsity if done improperly. \textbf{Stemming}~\cite{kim2021novel,DBLP:conf/issre/ZhengZTCCWS21,DBLP:journals/tse/HuoTLLS21,DBLP:conf/kbse/ChenXYJCX20,kukkar2020duplicate} reduces words to their root form to normalize variations of the same term. While it helps unify vocabulary, it may lead to over-stemming, reducing interpretability. \textbf{Lemmatization}~\cite{roy2014towards,ramay2019deep,umer2019cnn,DBLP:journals/tosem/AssiHGZ23,DBLP:journals/access/KimY22d} converts words to their base dictionary form, improving consistency in word representations and reducing vocabulary size. While it enhances text normalization, it requires linguistic knowledge and can be computationally expensive. \textbf{Lowercasing}~\cite{xiao2018bug,DBLP:conf/internetware/Xi0XX018,polisetty2019usefulness,DBLP:conf/noms/FukudaWHT22,DBLP:conf/apsec/WangL21} converts all text to lowercase to ensure uniformity in text matching. This reduces case sensitivity errors but may discard case-specific information, such as proper nouns. \textbf{Compound Word Splitting}~\cite{xiao2017improving,shi2018comparing,huo2019deep,hu2014effective,DBLP:journals/access/LiangSWY19} breaks down multi-word expressions into smaller components, improving granularity in text analysis. Although it enhances feature representation, it can introduce errors if not context-aware. In modern deep learning-based bug report analysis models, stop word removal is less necessary because these models learn contextual embeddings rather than relying on traditional frequency-based methods like TF-IDF or bag-of-words. Models such as transformers (e.g., BERT) capture semantic meaning and understand the importance of words in context, allowing them to assign appropriate weights to stop words rather than discarding them outright. Additionally, some stop words can be important for understanding the relationships between key terms in bug reports. 

Based on the analysis above, we summarize the strengths, weaknesses, and best-suited scenario for each preprocessing method in Table~\ref{tab:rq1-3-1}  This table provides a useful resource, helping researchers and practitioners choose the most suitable methods according to their specific needs and circumstances.

\begin{table}[t!]
  \caption{Summaries of the strengths, weaknesses, and best-suited usage scenarios for the preprocessing methods.}
% \vspace{-0.3cm}
\centering
\footnotesize
\begin{adjustbox}{width=\linewidth,center}
\begin{tabular}{p{1.5cm}p{4cm}p{3.5cm}p{4cm}}
\toprule

\textbf{Category} & \textbf{Strength} & \textbf{Weakness} & \textbf{Best-Suited Scenario} \\ \hline

Stop Word Removal & Reduces noise and improves efficiency by removing common but uninformative words. & May lead to loss of contextual meaning. & Traditional NLP models like TF-IDF or bag-of-words where reducing dimensionality. \\ \hline

Tokenization & Splits text into meaningful units, facilitating feature extraction and downstream processing. & Improper tokenization may increase data sparsity and negatively impact word relationships. & Preprocessing step for word-based or subword-based models. \\ \hline

Stemming & Normalizes words by reducing them to their root form, improving vocabulary consistency. & Over-stemming can lead to loss of meaning and reduced readability. & Search engines or retrieval-based systems where word variations need to be matched. \\ \hline

Lowercasing & Ensures uniformity by converting text to lowercase, reducing case sensitivity errors. & May discard important case-sensitive information, such as proper nouns and acronyms. & Applications where case differences do not alter meaning, such as basic text classification. \\ \hline

Punctuation Removal & Standardizes text input by removing symbols like commas and periods, improving consistency. & Removes valuable syntactic cues, which may impact grammatical analysis. & When punctuation is not crucial, such as when using bag-of-words models. \\ \hline

Lemmatization & Converts words to their base dictionary form, improving consistency in word representations. & Requires linguistic knowledge and can be computationally expensive. & Applications needing semantic accuracy, such as deep learning-based text models. \\ \hline

Numbers Removal & Reduces dimensionality by eliminating numerical values that might skew analysis. & May remove important numeric information, such as version numbers or error codes. & Sentiment analysis or topic modeling where numbers are not semantically important. \\ \hline

Compound Word Splitting & Breaks down multi-word expressions into smaller components, enhancing granularity. & Can introduce errors if not context-aware, leading to unnatural splits. & Text mining applications where fine-grained word representations are required. \\ \hline

Stack Trace Removal & Removes automatically generated error traces, reducing redundancy in bug reports. & May remove useful debugging information that could aid developers. & When focusing on descriptive aspects of bug reports rather than technical traces. \\ \hline

Non-alphanumeric Characters Removal & Simplifies text representation by stripping special characters. & Can result in loss of structured information, such as code snippets or logs. & When standardizing text for simpler NLP models where symbols do not contribute meaning. \\ 

\\ \bottomrule
\end{tabular}
\end{adjustbox}
\label{tab:rq1-3-1}
% \vspace{-0.3cm}
\end{table}

\paragraph{\textbf{Discussions}}
Our analysis reveals that stop word removal is the most frequently used preprocessing method. Surprisingly, from 2020 to 2023, only 27 papers employed stop word removal, accounting for just 19.6\% of the total. We attribute this decline to the widespread adoption of deep learning (DL)-based models in recent years.

Similarly, from 2020 to 2023, 64 papers utilized tokenization, representing 58.1\% of the total, compared to just 41.9\% between 2014 and 2019. We believe this increase is primarily due to the reliance on modern DL-based bug report analysis models on tokenization. These models require tokenization because they rely on subword embeddings (e.g., WordPiece, Byte Pair Encoding) to handle out-of-vocabulary words, typos, and rare terms effectively. Tokenization helps break down complex words into meaningful units, ensuring that the model can learn better representations and handle diverse linguistic patterns in bug reports. 

Based on these observations, we believe that future research should place greater emphasis on structural preprocessing methods. Since DL-based bug report analysis models can inherently handle stop words, out-of-vocabulary words, typos, and rare terms without needing to remove them, adopting transformation-based preprocessing techniques can further enhance bug report analysis performance.

In summary, we obtain the findings shown as below:

\begin{tcolorbox}[breakable,left=5pt,right=5pt,top=5pt,bottom=5pt] 
\textbf{\underline{Finding 3}}: Stop word removal remains the most widely used preprocessing method, followed by tokenization and stemming in second and third place, respectively. From 2020 to 2023, the adoption of structural preprocessing methods has seen a significant rise, surpassing their usage in earlier years.
\end{tcolorbox}

\subsection{RQ2: What software projects are evaluated?}
\label{sec:RQ2}

% \subsection{Purpose}

Given the fact that bug reports are commonly used in real-world scenarios, it is important to know in which open-source software projects machine learning based approaches have been widely applied for automating the analysis task. Thus, in the following,  we explore the landscape of bug reporting practices within various software projects. 

Overall, we found ML-based bug report analysis has been used in 172 different projects\footnote{To ensure fairness, we count open-sourced projects only.}. Figure~\ref{fig:RQ2} illustrates the top 10 software projects considered and their number of appearances in the primary studies. As can be seen, Eclipse\textemdash a major IDE for Java development\textemdash and Mozilla Core are the most predominantly used projects in the evaluation, appearing in nearly half of the primary studies. Java Development Toolkit ranks the third, which has significantly outnumbered others in the primary studies. Note that all projects were created before 2014, which is the starting year of this survey. 

Throughout our investigation, we encounter two prominent categories of projects: those with structured bug reports and those with unstructured bug reports. The distinction between these categories lies in the methodologies and tools utilized for bug tracking and management. Projects falling into the structured category, such as Eclipse, Mozilla Core, and Java Development Toolkit, often employ platforms like Bugzilla or JIRA, which enforce a structured approach to bug reporting. On the other hand, projects like those hosted on GitHub typically feature unstructured bug reports, characterized by a more flexible and less standardized format. We found that although most projects in primary studies use structured bug reports, a considerable number (13) of primary studies utilized unstructured bug reports during 2020-2023. In contrast, only one paper from 2014-2019 did so.

In this context, we explore the differences and implications of both structured and unstructured bug reporting methodologies. By analyzing a diverse range of software projects, we aim to explain the strengths, weaknesses, and implications of each approach. 

\subsubsection{Projects with Structured Bug Reports}
Structured bug reports help developers and testers communicate more efficiently with predefined bug description items. These reports typically include specific and standardized fields that provide all the information to understand, replicate, and address bugs. The main components of a structured bug report usually contain: title, description, environment, severity, priority, and assignee, etc. These projects predominantly utilize Bugzilla or JIRA to issue tracking and management. Bugzilla stands out as a leading free Issue Tracking System (ITS) globally, with major projects of Eclipse, firefox, Mozilla core, JDT, SWT, Aspectj and Firefox relying on its robust features. A typical bug reporting process in Bugzilla involves comprehensive textual and categorical details, resembling a form-filling activity. Similarly, JIRA has witnessed a surge in adoption over the past decade, with over 65,000 organizations leveraging its capabilities\footnote{https://tinyurl.com/y8vhxujv}. JIRA, akin to Bugzilla, follows a form-filling design for bug report submission, ensuring structured information gathering, which is used by the projects of Netbeans, Tomcat, and OpenOffice. Chromium, however, utilizes a unique platform for its structured issue tracking system, distinct from either Bugzilla or JIRA. This approach allows Chromium to tailor the issue tracking process to meet specific project needs while maintaining the benefits of structured data collection. 

\subsubsection{Projects with Unstructured Bug Reports}
GitHub serves as a prominent Git repository hosting service, offering diverse functionalities including issue tracking. GitHub issues, however, tend to contain unstructured information compared to Bugzilla and JIRA. While textual details are prevalent, categorical fields are often customizable per repository and may not be mandatory or well-defined. Consequently, GitHub lacks the structured nature inherent in Bugzilla and JIRA. Nonetheless, GitHub repositories can utilize labels to categorize issues, albeit in a less systematic manner. The absence of rigid categorical structures in GitHub offers both advantages and disadvantages. On one hand, it simplifies issue reporting by providing flexibility; on the other hand, it complicates the systematic extraction of useful categorical information. Reflecting this trend, Netbeans has also adopted GitHub's issue tracking system since 2022.

 \begin{figure}[!t]
\centering
\begin{subfigure}[t]{0.43\columnwidth}
\begin{tikzpicture}
\begin{axis}[
    xbar=0.05cm,
    width=8cm,
    height=5.5cm,
    xmax=110,
    bar width=0.35cm,
 enlarge y limits=0.08,
 enlarge x limits=0,
    legend style={at={(1.05,0.7)},anchor=north,legend columns=1,font=\tiny},
    legend cell align={left},
   xlabel={\# studies},
     x label style={font=\Large},
     nodes near coords,
          point meta=explicit symbolic,
          every node near coord/.append style={anchor=west,font=\large},
       symbolic y coords={Chromium,OpenOffice,Tomcat,Firefox,AspectJ,Netbeans,SWT,JDT,Mozilla,Eclipse},
      yticklabels={,,},
         y tick label style={font=\large},
          x tick label style={font=\large},
         ytick=data
    ]
   
\addplot [fill=cyan!50,draw=black] coordinates {(79,Eclipse)[79 Eclipse] (64,Mozilla)[64 Mozilla Core] (34,JDT)[34 Java Development Toolkit] (25,SWT)[25 Standard Widget Toolkit](23,Netbeans)[23 Netbeans] (23,AspectJ)[23 AspectJ] (22,Firefox)[22 Firefox] (19,Tomcat)[19 Tomcat](18,OpenOffice)[18 OpenOffice] (12,Chromium)[12 Chromium]};

\end{axis}
\end{tikzpicture}
    \subcaption{Top 10 projects. All belong to projects with structured bug reports.}
  \label{fig:RQ2}
    \end{subfigure}
~\hspace{-0.1cm}
\begin{subfigure}[t]{0.56\columnwidth}
\begin{tikzpicture}
\begin{axis}[
 xbar stacked,
    width=10cm,
    height=5.5cm,
    xmax=120,
    bar width=0.35cm,
 enlarge y limits=0.05,
 enlarge x limits=0,
    	nodes near coords, 
	nodes near coords={\pgfmathfloatifflags{\pgfplotspointmeta}{0}{}{\pgfmathprintnumber{\pgfplotspointmeta}}},
  enlargelimits=0,
 enlarge y limits=0.05,
    legend style={at={(0.66,0.39)},
      anchor=north,legend columns=1},
    xlabel={\# studies},
        x label style={font=\Large},
    symbolic y coords={$>$10 projects,nine projects,eight projects,seven projects,six projects,five projects,four projects,three projects,two projects,one project},
     yticklabels={,,},
         y tick label style={font=\large},
          x tick label style={font=\large},
         ytick=data
    ]
\addplot[xbar, fill=white, draw=black] coordinates {(1,one project) (9,two projects) (21,three projects) (17,four projects) (10,five projects) (6,six projects) (0,seven projects) (0,eight projects) (0,nine projects)(0,$>$10 projects)};

\addplot[xbar, fill=yellow!30, draw=black] coordinates {(5,one project) (13,two projects) (23,three projects) (19,four projects) (11,five projects) (5,six projects) (2,seven projects) (0,eight projects) (1,nine projects)(0,$>$10 projects)};

\addplot[xbar, fill=cyan!50,, draw=black] coordinates {(24,one project) (14,two projects) (33,three projects) (26,four projects) (23,five projects) (17,six projects) (7,seven projects) (2,eight projects) (2,nine projects)(13,$>$10 projects)};

\legend{\# Studies use Mozilla,\# Studies use Eclipse,Total \# studies use any project}
\end{axis}

\coordinate (A) at (3.2,3.75);
\coordinate (B) at (3.8,3.35);
\coordinate (C) at (6.8,2.95);
\coordinate (D) at (5.6,2.55);
\coordinate (E) at (4.3,2.15);
\coordinate (F) at (3.1,1.75);
\coordinate (G) at (1.6,1.35);
\coordinate (H) at (1.2,0.95);
\coordinate (I) at (1.3,0.55);
\coordinate (J) at (1.8,0.15);
\node[rotate=0] at (A) {1 project};
\node[rotate=0] at (B) {2 projects};
\node[rotate=0] at (C) {3 projects};
\node[rotate=0] at (D) {4 projects};
\node[rotate=0] at (E) {5 projects};
\node[rotate=0] at (F) {6 projects};
\node[rotate=0] at (G) {7 projects};
\node[rotate=0] at (H) {8 projects};
\node[rotate=0] at (I) {9 projects};
\node[rotate=0] at (J) {$>$10 projects};

\end{tikzpicture}
   \subcaption{\# projects used in a study.}
 \label{fig:proj-no}
    \end{subfigure}
\begin{subfigure}[t]{0.4\columnwidth}
\begin{tikzpicture}[scale=0.4]
\tikzstyle{every node}=[font=\scriptsize]
    \pie[
        /tikz/every pin/.style={align=center},
        text=label,
        rotate=-40,
        explode=0.1,
        color={black!10,black!30},
        ]
        {
        6.9/Unstructured,
        93.1/Structured
        }
    \end{tikzpicture}
   \subcaption{\% of project type.}
 \label{fig:RQ2-3}
    \end{subfigure}
    
    \caption{Statistics of projects used to evaluate the machine learning algorithms.}
  \end{figure}
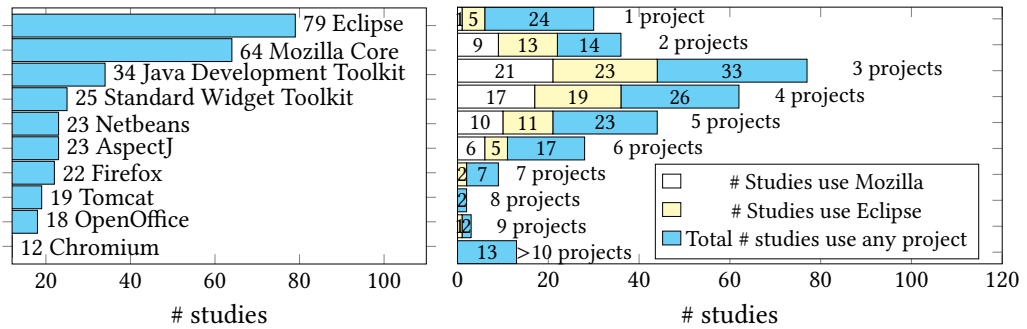
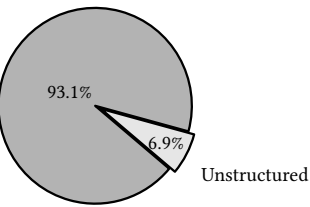

Some GitHub repositories, such as the \textbf{VSCode} project, attempt to bridge this gap by offering templates to embed categorical information within textual descriptions. For instance, the VSCode template prompts users to provide details like VSCode version and operating system, enhancing the categorization process. Despite this, adherence to such templates is not obligatory and may be disregarded by issue reporters. 

This unique characteristic of GitHub's bug reporting system may facilitate the differentiation between bug reports, feature requests, and documentation updates. However, it also poses challenges in maintaining consistency and extracting structured information systematically.

\begin{table}[t!]
  \caption{Summaries of the strengths, weaknesses, and best-suited usage scenarios for the projects.}
% \vspace{-0.3cm}
\centering
\footnotesize
\begin{adjustbox}{width=\linewidth,center}
\begin{tabular}{p{2cm}p{4cm}p{3.5cm}p{3.5cm}}
\toprule

\textbf{Category} & \textbf{Strength} & \textbf{Weakness} & \textbf{Best-Suited Scenario} \\ \hline
Structured & Facilitates efficient   communication with predefined bug description items, ensuring all necessary   information is provided. & May be rigid and less   flexible, potentially discouraging detailed reporting. & Projects requiring   comprehensive and standardized bug reporting, such as Eclipse and Mozilla   Core. \\ \hline
Unstructured & Offers flexibility in   issue reporting, allowing for customizable and less standardized formats. & Complicates systematic   extraction of useful categorical information and maintaining consistency. & Projects benefiting from   flexible reporting, such as those hosted on GitHub, including VSCode.
\\ \bottomrule
\end{tabular}
\end{adjustbox}
\label{tb:pro_con_projects}
% \vspace{-0.3cm}
\end{table}

\paragraph{\textbf{Discussions}}
Figure~\ref{fig:proj-no} shows the number of projects considered in each study. The majority of primary studies utilized datasets comprising three projects, followed by those with four and five projects. In particular, Eclipse and Mozilla are often considered in those cases, and up to 90\% of the time they are considered together. In particular, there are 94 studies with Eclipse/Mozilla and they cover 38 projects. For the other 110 without these two, 132 projects are covered. This implies that, current studies tend to believe Eclipse and Mozilla are good representatives of the others, especially when only three projects are used.

Although, as presented in Figure~\ref{fig:RQ2-3}, most papers over the past decade have adopted projects with structured bug reports. However, the application of projects with unstructured bug reports in the recent four years, which indicates an increasing focus on projects employing unstructured bug reports, suggests that more powerful language models should be used in current bug report analysis, as they are better at handling complex unstructured data.

Following a comprehensive review, the advantages, limitations, and best-suited scenarios for each project type have been identified and summarized in Table~\ref{tb:pro_con_projects}. This table serves as a valuable reference, enabling researchers and practitioners to select the most appropriate projects based on specific situations and requirements.
 
Overall, we conclude that:

%The reason behind this is mainly because (i) Eclipse has a mature procedure to report bugs, and it owns one of the largest repositories of bug reports, which is publicly accessible; (ii) it is based on Java and support for Java development, which itself is a language that has a large number of active developers. Instead of a single software product, Mozilla Core refers to a collection of shared components for all Mozilla's software, and this is why it may involve much higher number of bug reports than other software projects.

\begin{tcolorbox}[breakable,left=5pt,right=5pt,top=5pt,bottom=5pt] 
\textbf{\underline{Finding 4}}: For projects with unstructured bug reports, Eclipse is the predominantly used software project in evaluation. Mozilla Core comes the second. There has been a growing trend in focusing on projects with unstructured bug reports in the past four years.
\end{tcolorbox}

\subsection{RQ3: What tasks in bug report analysis are tackled by machine learning?}
\label{sec:RQ3}

% \subsection{Purpose}

In software engineering, bug report analysis is a broad topic that involves various tasks. In our \textbf{RQ3}, we seek to understand what tasks in bug report analysis have been specifically investigated and addressed by using machine learning. Table~\ref{tab:2} shows the tasks of bug report analysis tackled by machine learning, sorted based on their popularity using the number of studies found. As we can see, out of the seven task types, bug categorization tends to be of the highest popularity (28.2\% of the primary studies). In general, the bug categorization task may be binary or multi-labeled. The former concerns with identifying whether a given bug reports have indeed contain bugs or not; the latter, in contrast, is about deciding which categories of bug types (e.g., performance bugs, security bugs, \emph{etc}.) that a bug report should be assigned to, and thereby improving the productivity of the developers. The actual categories of bug types used vary widely across different studies and software projects~\cite{inproceedings}. The popularity of bug categorization tasks may be attributed to their clear problem structure and the availability of labeled datasets, which align well with supervised learning approaches in machine learning. Bug localization, bug assignment and bug severity/priority prediction account for 21.8\%, 18.0\%, and 14.1\% respectively. In theory, again, all of the above are part of the broad classification task, but they require different sources of features and different types of output form in order to fulfill their purposes. Additionally, a trend has been observed in bug report summarization: from 2014-2019, only one primary study addressed this task, but from 2020-2023, there were nine.

\begin{table}
\setlength{\tabcolsep}{1mm}
\centering
  \caption{Bug report analysis tasks and their popularity in studies from the academia.}
  \label{tab:2}
  \footnotesize
  \begin{tabular}{p{2.2cm}p{0.5cm}p{3cm}p{5.4cm}}
    \toprule
    \textbf{Task}&\textbf{\%}&\textbf{Type}&\textbf{Description}\\
    \midrule
    Bug categorization &28.2&Discriminative& To distinguish whether a bug report describes a bug or not; or to identify its suitability to a bug type.\\
    Bug localization &21.8&Discriminative&  To locate potentially buggy files in a software project given a bug report.\\
    Bug assignment &18.0&Discriminative& To assign bug reports to suitable developers.\\
    Bug severity/priority prediction &14.1&Discriminative& To predict the severity and priority for the bug reports.\\
    Duplicate detection &9.7&Discriminative&  To detect bug reports that are duplicates.\\
    Summarization&4.9&Generative&To summarize bug reports in brief.\\
    Others&3.3&Discriminative/Generative&Other bug report tasks.\\
    %Bug-commit linking&2&-\\
    %Bug report completion/refinement&1&-\\
    %Bug fixing time prediction&1&-\\
    %Bug report summarization&1&-\\
    %Re-opened bug prediction&1&-\\
    %Story point estimating&1&-\\
  \bottomrule
\end{tabular}
\end{table}

Based on the nature of the tasks and their respective objectives, we have categorized these tasks into two major groups: discriminative bug report analysis and generative bug report analysis.

% \subsection{Results}

\subsubsection{Discriminative Bug Report Analysis}

Discriminative bug report analysis refers to the use of machine learning techniques to classify, sort, or make decisions about bug reports in software engineering. This approach primarily involves identifying distinct characteristics or categories within bug reports and using this information to make informed decisions on handling them. The key tasks typically associated with discriminative bug report analysis include:

\textbf{Bug categorization}~\cite{DBLP:journals/infsof/GomesTC21,DBLP:journals/access/AfricVSD23,DBLP:conf/icse/FangZTJXS23,DBLP:journals/tr/DuZXZT22,DBLP:journals/tse/WuZXL22} is a process in software engineering where reports submitted by users or developers are analyzed to identify and categorize bugs in software. This process is foundational to maintaining and improving software quality. It requires a thorough understanding and analysis of the structure and content of bug reports to diagnose software issues efficiently. Automatic efficient bug categorization facilitates the rapid identification of software bugs, which, in turn, contributes to enhanced software quality and reliability. By accurately classifying bug reports, development teams can more effectively allocate their resources to address the types of bugs they wish to focus on. 

\textbf{Bug assignment}~\cite{DBLP:journals/access/DipongkorIHYM23,DBLP:journals/access/ZaidiALWL20,DBLP:journals/kbs/WuMX0H22,DBLP:journals/jss/DaiLXLWZ23,DBLP:journals/asc/AlkhaziDAAYM20} refers to the process of assigning bug reports to appropriate developers or teams responsible for their resolution. Effective bug assignment helps ensure timely bug resolution by directing reports to individuals with the requisite expertise and availability. Automating this task reduces manual effort and accelerates the bug fixing process, thereby improving software quality and developer productivity.

\textbf{Bug localization}~\cite{DBLP:journals/sigsoft/RaoMK15,DBLP:conf/models/ArcegaFC18,DBLP:conf/iwpc/KhatiwadaTM20,DBLP:journals/kbs/ZhuTWL22,DBLP:journals/jss/MiryeganehHH21} involves identifying the specific code components or modules responsible for causing reported software bugs. This task is crucial for developers to isolate and fix defects efficiently. Automated bug localization techniques leverage information from bug reports, source code, and program execution traces to pinpoint the root causes of bugs, facilitating rapid debugging and resolution. Knowing where bugs are likely to be found can help in better allocating testing and debugging resources, focusing efforts where they are most needed.

\textbf{Bug Severity/Priority Prediction}~\cite{DBLP:journals/infsof/WangWH23,DBLP:journals/infsof/TongZ21,DBLP:journals/compsec/SunOZLWZ23,DBLP:journals/ese/TianLXS15,DBLP:journals/jikm/SinghMS17} is important for allocating resources effectively and addressing critical issues promptly. Automated prediction models analyze various features of bug reports, such as descriptions, stack traces, and historical data, to assess the severity and prioritize bugs accordingly. By accurately prioritizing bugs, development teams can focus their efforts on resolving critical issues and delivering high-quality software products.

Duplicate bug reports often arise when multiple users encounter the same underlying software defect and submit separate reports. \textbf{Duplicate detection}~\cite{DBLP:journals/ieicet/ZouXYZZH16,DBLP:journals/jss/LinYLC16,jayarajah2016duplicate,DBLP:journals/tr/MessaoudMJMG23,DBLP:conf/icse/CooperBCMP21} and its management is crucial for avoiding redundancy and optimizing resource utilization in bug resolution efforts. Automated duplicate detection systems compare incoming bug reports with existing reports, leveraging textual similarity and metadata attributes to identify duplicates and streamline the bug triaging process.

These discriminative bug report analysis tasks play a vital role in improving the efficiency, accuracy, and scalability of bug management processes, ultimately enhancing software quality and developer productivity.

\subsubsection{Generative Bug Report Analysis}

Generative Bug Report Analysis in software engineering focuses on the creation of new content based on existing bug reports. Unlike discriminative tasks that involve classification and sorting, generative tasks aim to synthesize and produce informative outputs that help in understanding and communicating the essence of bug reports more effectively. This category encompasses tasks such as bug report title generation and bug report summarization, which involves condensing the information contained in bug reports into concise and informative summaries. In this review, we collectively refer to these tasks as \textbf{bug report summarization}, recognizing their shared objective of distilling key insights from bug reports.

One notable example of generative bug report analysis is presented in a paper proposing a method for generating human-readable fault reports by leveraging time-series data and textual metadata of metrics for detecting faults in a target system. The authors integrated heterogeneous sources of monitoring data and demonstrated effectiveness through experiments on an e-commerce service implemented in a Kubernetes cluster~\cite{DBLP:conf/noms/FukudaWHT22}. Another noteworthy contribution comes from a study introducing a method named iTAPE, which automatically generates titles for bug reports to aid project practitioners in quickly grasping the core idea of reported bugs. This approach utilizes a Seq2Seq-based model and addresses challenges such as lacking off-the-shelf datasets and handling low-frequency human-named tokens to produce concise and precise bug report titles~\cite{DBLP:conf/kbse/ChenXYJCX20}. Additionally, researchers proposed a deep attention-based summarization model for generating high-quality bug report titles. By employing a robustly optimized bidirectional-encoder-representations-from-transformers approach and a stacked transformer decoder, this model effectively captures contextual semantic information and improves the accuracy of bug report title generation compared to LSTM-based models~\cite{DBLP:journals/tr/MaKYZZL23}.

\paragraph{\textbf{Discussion}}

From the trends over the last four years, it is evident that in the realm of generative bug report analysis, recent advancements in natural language processing, particularly with models like Meta LLaMA 3\footnote{https://llama.meta.com/llama3/} and OpenAI GPT-4\footnote{https://openai.com/index/gpt-4/}, offer promising capabilities for automatically generating high-quality bug report summaries. These models leverage sophisticated language understanding and generation techniques to distill the essential information from bug reports into coherent and informative summaries, empowering developers with actionable insights while minimizing manual effort.

While both LLaMA 3 and GPT-4 possess remarkable text generation capabilities, the choice between them may depend on factors such as accessibility, cost, and specific user requirements. LLaMA 3's open-source nature provides users with flexibility and affordability, making it a favorable option for experimentation and integration into various software development workflows. On the other hand, GPT-4, despite potentially offering higher accuracy and performance, entails the use of API services, which may incur additional costs but could be justified by its superior capabilities and support. We think generative bug report analysis, facilitated by advanced language models like LLaMA 3 and GPT-4, presents a promising approach to automating bug report summarization and enhancing the efficiency of software development processes. By harnessing the power of natural language processing, developers can extract actionable insights from bug reports more effectively, leading to improved software quality and developer productivity. Indeed, some of them may belong to the same general category of task from the machine learning perspective, i.e., classification tasks, such as bug categorization and bug assignment. However, their purposes are different, and thus the features and data labeling involved are also different.

% Here, we divide these tasks related to bug report analysis into two main categories and provide detailed explanations for each of them.

Therefore, we can conclude that:

%in bug assignment, the aim is to recommend suitable developers who is capable to fix the bug in a bug report. As a result, apart from the target bug report, additional features that represent the expertise of the developers are required, e.g., the historical bug reports that have been fixed by a developers. The bug severity/priority prediction, on the other hand, is more similar to the bug categorization problem. However, instead of classifying the bug reports into different types of bug, it aims to categorize them into different levels of severity, within each of which the bug reports should all come from the same bug type.

%\begin{figure}[t!]
 % \centering
%\includestandalone[width=0.9\columnwidth]{tikz/RQ3}
 %  \caption{Ratio of bug report analysis problems solved by machine learning}
 %\label{fig:RQ3}
%\end{figure}

\begin{tcolorbox}[breakable,left=5pt,right=5pt,top=5pt,bottom=5pt] 
\textbf{\underline{Finding 5}}: Bug categorization is the most popular bug analysis task. In contrast, bug report summarization is still niche. However, it has attracted more attention, evidenced by the significantly increasing number from only one primary study four years ago to nine recently. 
\end{tcolorbox}

\subsection{RQ4: What types of bugs are analyzed?}
\label{sec:RQ4}

%There are different bug types in software engineering, and therefore the bug reports may be associated with one or more bug types. 

% \subsection{Purpose}

% Understanding what bug types have been covered in bug report analysis with machine learning would have vast benefits, for example, enabling better study on the different characterises of bug types, identifying their origins, and thus developing better specialized machine learning approach.

% \subsection{Results}

Indeed, there are many specific and detailed categorizations of bug types from the literature~\cite{inproceedings}. In this work, we classify the main bug types addressed in each study based on what has been stated in the papers. Note that when studies specifically stated that certain types of bugs that possess unique characteristics and exert varying impacts on software quality and user experience are focused, we put them into \textit{Specific bug} category. When no bug type has been specifically stated, we put the study into \textit{General bug} category.

In this review, we found that there are 188 studies investigating general bugs with no clear statement on the bug types. Figure~\ref{fig:RQ4} shows the summary of the results in specific bug types. As we can see, 13 studies have focused on high-impact bug reports. High-impact bugs are composed of an aggregation of various bug types. Our statistical breakdown reveals the presence of one performance bug, two blocking bugs, eight security bugs, one surprise bug, and one breakage bug~\cite{DBLP:conf/icsm/KashiwaYKO14}. Additionally, nine other bug types, including three configuration bugs, two concurrency bugs, and other bug types, each with only one instance, form the minority of non-general bug type studies.  

Moreover, the comparative analysis of bug report studies between two distinct periods, 2014-2019 and 2020-2023, reveals notable trends and shifts in the focus of research. During the period from 2014 to 2019, there were 11 primary studies dedicated to specific bug types. In contrast, the period from 2020 to 2023 saw an increase in the number of studies focusing on specific bug types, with 13 primary studies reported. 

Next, we will describe these types of bugs in detail:

\subsubsection{General Bug Types}
Different from specific bug categories, bug report analysis studies encompass a broad spectrum of general bug types. While these types may not be explicitly defined or categorized, they collectively contribute to the understanding of software defects and the improvement of bug management processes. General bug types serve as a baseline for assessing overall software quality, identifying trends, and guiding quality assurance efforts.

\subsubsection{Specific Bug Types}
Bug reports often encompass a diverse array of issues, each with its unique characteristics and impacts on software quality and user experience. Within the realm of specific bug types, several categories emerge, each addressing distinct aspects of software functionality and behavior. Next, we will introduce each specific bug type we have discovered:

\textbf{High-impact bugs}~\cite{DBLP:conf/icsm/KashiwaYKO14} significantly affect software usability, stability, or security. These include surprise bugs, dormant bugs, blocking bugs, security bugs, performance bugs, and breakage bugs. Analyzing these types is crucial for user satisfaction, system reliability, and organizational reputation.

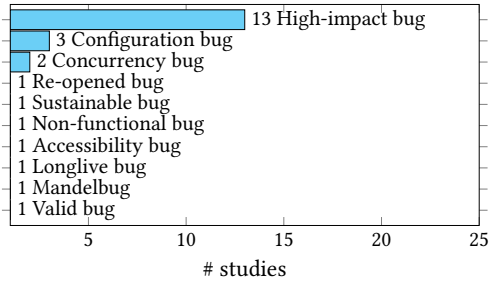
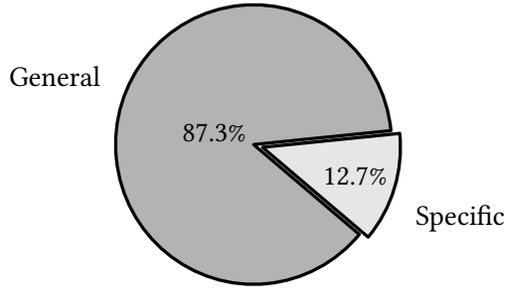
\begin{figure}[!t]
\centering
\begin{subfigure}[t]{0.49\columnwidth}
\newcommand\shift{1.5}
\begin{tikzpicture}[scale=1]
\begin{axis}[
    xbar=0.05cm,
    width=10cm,
    height=5.5cm,
    xmax=25,
    bar width=0.35cm,
 enlarge y limits=0.08,
 enlarge x limits=0,
    legend style={at={(1.05,0.7)},anchor=north,legend columns=1,font=\tiny},
    legend cell align={left},
   xlabel={\# studies},
     x label style={font=\Large},
     nodes near coords,
          point meta=explicit symbolic,
          every node near coord/.append style={anchor=west,font=\large},
       symbolic y coords={Valid,Mandelbug,Longlive,Accessibility,Non-functional,Sustainable,Re-opened,Concurrency,Configuration,High-impact},
      yticklabels={,,},
         y tick label style={font=\large},
          x tick label style={font=\large},
         ytick=data
    ]

\addplot [fill=cyan!50,draw=black] coordinates {(1,Valid)[1 Valid bug](1,Mandelbug)[1 Mandelbug](1,Longlive)[1 Longlive bug](1,Accessibility)[1 Accessibility bug](1,Non-functional)[1 Non-functional bug](1,Sustainable)[1 Sustainable bug](1,Re-opened)[1 Re-opened bug] (2,Concurrency)[2 Concurrency bug] (3,Configuration)[3 Configuration bug](13,High-impact)[13 High-impact bug]};

 \end{axis}
 \end{tikzpicture}
 
   \subcaption{Specific bug types involved in the analyzed reports.}
 \label{fig:RQ4}
    \end{subfigure}
~\hspace{-0.2cm}
\begin{subfigure}[t]{0.5\columnwidth}
\begin{tikzpicture}[scale=0.4]
\tikzstyle{every node}=[font=\scriptsize]
    \pie[
        /tikz/every pin/.style={align=center},
        text=label,
        rotate=-40,
        explode=0.1,
        color={black!10,black!30},
        ]
        {
        12.7/Specific,
        87.3/General
        }
    \end{tikzpicture}
   \subcaption{\% of bug types.}
 \label{fig:RQ4-2}
    \end{subfigure}
    \caption{Statistics of bug types.}
  \end{figure}

\textbf{Configuration bugs} arise from incorrect or incompatible software configurations, studied in three works~\cite{DBLP:conf/apsec/XuLXSL15,DBLP:conf/issre/WenYH16,DBLP:conf/compsac/XiaLQWZ14}. \textbf{Concurrency bugs}, appearing in two studies~\cite{DBLP:conf/ijcnn/CiancariniPRS17,DBLP:conf/ease/ZhouNG15}, result from improper handling of concurrent processes, leading to race conditions, deadlocks, or inconsistent states. \textbf{Re-opened bugs} refer to issues previously resolved but resurfacing due to incomplete fixes or regressions, studied by~\citeauthor{DBLP:journals/ase/XiaLSWZ15}~\cite{DBLP:journals/ase/XiaLSWZ15}.

\textbf{Sustainable bugs} persist across multiple versions or releases due to design flaws or complex dependencies. For example, \citeauthor{DBLP:conf/webi/ZhangZW21}\cite{DBLP:conf/webi/ZhangZW21} propose a method to balance accuracy and sustainability in developer recommendations using deep learning. \textbf{Non-functional bugs} affect attributes like usability, performance, reliability, and maintainability.  To address these, \citeauthor{DBLP:conf/apsec/LongCC22}\cite{DBLP:conf/apsec/LongCC22} develop a tool to identify non-functional bug reports in deep learning frameworks. \textbf{Accessibility bugs} hinder access for users with disabilities, requiring solutions for inclusive design. \citeauthor{DBLP:conf/w4a/AljedaaniML0J22}~\cite{DBLP:conf/w4a/AljedaaniML0J22} tackle this as a binary classification problem.

\textbf{Long-lived bugs} evade resolution over extended periods. \citeauthor{DBLP:journals/infsof/GomesTC21}\cite{DBLP:journals/infsof/GomesTC21} compare machine learning algorithms for predicting long-lived bugs. \textbf{Mandelbugs} are complex and unpredictable, analyzed using a combined semantic model and deep learning approach\cite{DBLP:journals/tr/DuZXZT22}. \textbf{Valid bugs} are genuine issues requiring immediate attention. \citeauthor{DBLP:journals/tse/FanXLH20}~\cite{DBLP:journals/tse/FanXLH20} propose a method to automatically verify bug report authenticity.

\paragraph{\textbf{Discussion}}
Our study categorizes bugs in machine learning-assisted bug report analysis into general and specific types, covering a wide range of issues from high-impact bugs to configuration and concurrency bugs. This broad scope demonstrates an evolving awareness of various software quality aspects, including sustainable, non-functional, and accessibility bugs. However, gaps remain, such as the underrepresentation of mandelbugs and long-lived bugs, indicating areas for future research. Identifying common or impactful bug types helps direct analytical efforts and resources more effectively.

Although most papers focus on general bug types as shown in Figure~\ref{fig:RQ4-2}, the last four years have seen a shift towards analyzing specific bug types, reflecting growing interest and sophistication in this area. Researchers are focusing on the nuances of non-functional, performance, and valid bugs, driven by advancements in machine learning and more available data. Despite this progress, specific bug types still represent a small proportion of research, highlighting an opportunity for further exploration. Expanding machine learning applications to include more specific bug types could improve bug detection precision and intervention effectiveness, enhancing software quality and user experience. Detailed bug categorization could also foster the development of specialized machine learning algorithms tailored to diverse issues.

%However, it is difficult, if not impossible, to utilize them as most of the primary studies have not clearly stated which specific bug types that they seek to cover. Therefore, in this work, we classify the bug types studied by primary studies on two categories: performance bugs and non-performance bugs. This is because they are the most general concepts to describe the quality/bug of software~\cite{DBLP:conf/msr/NistorJT13}\cite{DBLP:conf/msr/ZamanAH12}\cite{DBLP:conf/msr/ZamanAH11}\cite{DBLP:conf/sosp/YinMZZBP11}, and can often be inferred via directly looking into the case and datasets in a primary study. 

%For those primary studies whose bug types have not been specifically stated, we simply classify them as either performance bugs and non-performance bugs.

%This is because their categorization is derived based on empirical study of the high-impacts bugs, and they divide the bug in terms of both the process (e.g., surprise bug and blocker bug) and products (e.g., performance bug and security bug). 

In summary, we conclude that:

\begin{tcolorbox}[breakable,left=5pt,right=5pt,top=5pt,bottom=5pt] 
\textbf{\underline{Finding 6}}: Majority of the primary studies have been focusing on general bugs, while ten other studies work on specific bug types, showcasing an increasing research trend on dealing with specific bug types in recent studies.
\end{tcolorbox}

\subsection{RQ5: Which evaluation metrics and procedures are used in performance evaluation of bug report analysis?}
\label{sec:RQ5}

% \subsection{\textbf{Purpose}}
\subsubsection{Evaluation Metrics}
It is important to understand what metrics/procedures have been used to evaluate machine learning based bug report analysis, as their meaning underpins the conclusions drawn. In this study, we categorize the evaluation metrics used in each paper based on the descriptions provided. If a study explicitly mentions an evaluation metric that is specifically tailored to aspects relevant to bug report analysis processes and strategies, we classify it under \textit{Bug report related evaluation metrics}. Otherwise, we classify the study under \textit{General evaluation metrics}.

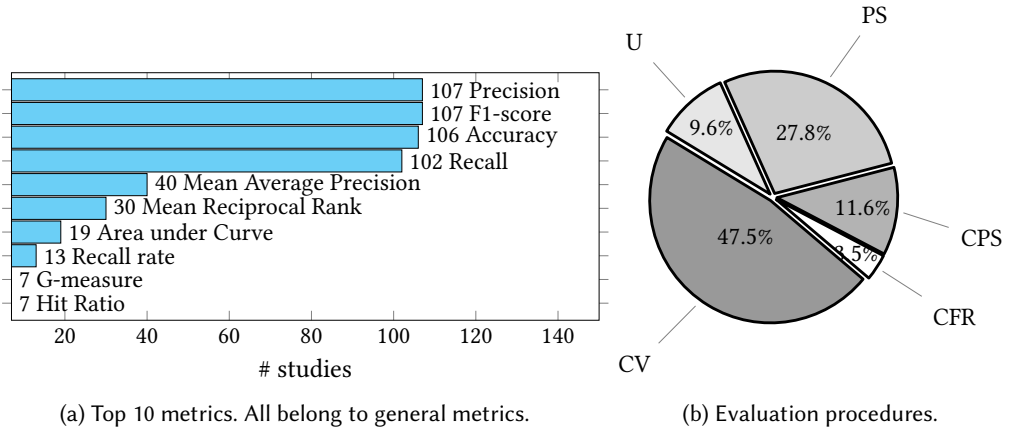
\begin{figure}[!t]
\centering
\begin{subfigure}[t]{0.6\columnwidth}
\newcommand\shift{1.5}
\begin{tikzpicture}[scale=1]
\begin{axis}[
    xbar=0.05cm,
    width=11cm,
    height=5.5cm,
    xmax=150,
    bar width=0.35cm,
 enlarge y limits=0.08,
 enlarge x limits=0,
    legend style={at={(1.05,0.7)},anchor=north,legend columns=1,font=\tiny},
    legend cell align={left},
   xlabel={\# studies},
     x label style={font=\Large},
     nodes near coords,
          point meta=explicit symbolic,
          every node near coord/.append style={anchor=west,font=\large},
       symbolic y coords={Hit Ratio,G-measure,Recall rate,AUC,MRR,MAP,Recall,Accuracy,F-measure,Precision},
      yticklabels={,,},
         y tick label style={font=\large},
          x tick label style={font=\large},
         ytick=data
    ]

\addplot [fill=cyan!50,draw=black] coordinates {(7,Hit Ratio)[7 Hit Ratio](7,G-measure)[7 G-measure](13,Recall rate)[13 Recall rate](19,AUC)[19 Area under Curve] (30,MRR)[30 Mean Reciprocal Rank] (40,MAP)[40 Mean Average Precision](102,Recall)[102 Recall](106,Accuracy)[106 Accuracy](107,F-measure)[107 F1-score](107,Precision)[107 Precision]};

 \end{axis}
 \end{tikzpicture}
 
    \subcaption{Top 10 metrics. All belong to general metrics.}
  \label{fig:RQ5}
    \end{subfigure}
~\hspace{-0.2cm}
\begin{subfigure}[t]{0.39\columnwidth}
\begin{tikzpicture}[scale=0.4]
\tikzstyle{every node}=[font=\scriptsize]
    \pie[
        /tikz/every pin/.style={align=center},
        text=pin,
        rotate=-40,
        explode=0.1,
        radius=3,
        color={white,black!30,black!20,black!10,black!40},
        ]
        {
        3.5/CFR,
        11.6/CPS,
        27.8/PS,
        9.6/U,
        47.5/CV
        }
    \end{tikzpicture}
   \subcaption{Evaluation procedures.}
 \label{fig:ep}
    \end{subfigure}
    \caption{Statistics of metrics and evaluation procedures.}
  \end{figure}

\label{sec:RQ5-1}
\paragraph{General Evaluation Metrics. }
In this section, we present the top 10 most used evaluation metrics, as shown in Figure~\ref{fig:RQ5}. These metrics are crucial for assessing the performance and effectiveness of bug report analysis tasks. In particular, Precision, F1-score, Accuracy and Recall are equally the most predominated metrics, which are widely used in most primary studies. %We categorize these metrics into two major groups: general evaluation metrics and bug report related evaluation metrics. 
%It is important to understand what metrics have been used to evaluate machine learning based bug report analysis, as they underpin the conclusions drawn.We first introduce the four most commonly used metrics.
\textbf{Precision} measures the accuracy of positive predictions, used in 107 studies~\cite{DBLP:journals/infsof/PanichellaCS21,DBLP:journals/tr/MaKYZZL23,DBLP:journals/jss/LiYSLW23,DBLP:journals/eswa/GuptaG21,DBLP:journals/infsof/SaborHH20}. It is the ratio of true positives to total positive predictions, helping to avoid false positives in bug report analysis. However, it neglects false negatives. The \textbf{F1-score}, the harmonic mean of precision and recall, is used in 107 studies~\cite{DBLP:conf/fedcsis/KumarKNMKP21,baarah2021sentiment,DBLP:conf/webi/ZhangZW21,DBLP:journals/access/KimY22d,DBLP:conf/dsa/ZhengCWFSC21}. It balances precision and recall, making it crucial for analyzing bug reports where both false positives and negatives are important, though it may not reflect specific operational priorities. \textbf{Accuracy} measures the overall correctness of predictions and is used in 106 studies~\cite{DBLP:journals/jss/AungWHS22,DBLP:journals/infsof/JahanshahiC22,DBLP:journals/kbs/MohsinSHJ22,DBLP:journals/eswa/MohsinS21,DBLP:journals/scp/SepahvandAJHB23}. It is the ratio of correct predictions to total instances but can be misleading in imbalanced datasets. \textbf{Recall}, used in 102 studies~\cite{DBLP:journals/infsof/NeysianiBA20,DBLP:journals/kbs/ZhuTWL22,DBLP:journals/infsof/WuZCZYM21,DBLP:journals/jss/ZhouLS20,DBLP:journals/compsec/SunOZLWZ23}, measures the system's ability to identify all relevant instances by the ratio of true positives to total actual positives. Its drawback is the potential increase in false positives, complicating bug triage.

 Some important but less frequently used metrics include: \textbf{MAP (Mean Average Precision)} evaluates the average precision at various recall levels, summarizing the precision-recall trade-off~\cite{DBLP:conf/sac/KimKL22a,DBLP:conf/qrs/WangZLYZ22,DBLP:conf/msr/RodriguesAFD20,DBLP:conf/scam/RazzaqBPCS21,DBLP:conf/qrs/ChenYYKZ22}. However, it might mask variations in performance across different bug report types. \textbf{MRR (Mean Reciprocal Rank)} calculates the average reciprocal rank of results for a sample of queries, focusing on the position of the first relevant result~\cite{DBLP:conf/scam/RazzaqBPCS21,DBLP:conf/qrs/ChenYYKZ22,DBLP:journals/access/LuoWC23,DBLP:conf/wcre/FlorezCTM21,DBLP:conf/iwpc/Zhang0YZZ20}. It may overlook the quality of all returned results. \textbf{AUC (Area Under the Curve)} measures system performance across different thresholds by calculating the area under the ROC curve~\cite{DBLP:journals/infsof/GomesTC21,DBLP:journals/infsof/ElmishaliK23,DBLP:journals/access/RochaC21,DBLP:journals/infsof/GeFQGQ22,DBLP:conf/qrs/SuHCQM22}. However, it focuses on ranking rather than actual predicted probabilities, which may not fully reflect the model's calibration.

\textbf{Recall rate} measures the proportion of true positives out of all actual positives, used in 13 studies~\cite{DBLP:conf/smc/AkilanSPM20,DBLP:conf/msr/RodriguesAFD20,DBLP:conf/icse/CooperBCMP21,DBLP:journals/tosem/ZhangHVIXTLJ23,DBLP:journals/jss/JiangSTSW23}. It ensures all relevant bug reports are identified but may lead to many false positives. The \textbf{G-measure}, the geometric mean of precision and recall, balances both metrics, giving equal weight to false positives and false negatives. It is used in seven studies~\cite{DBLP:journals/tse/PetersTYN19,DBLP:conf/dsa/ZhengCWFSC21,DBLP:journals/tse/WuZXL22,DBLP:journals/jss/TanXWZXL20,DBLP:journals/infsof/MaKYYLZ22,DBLP:journals/infsof/NeysianiBA20}, but may not align with specific operational priorities. The \textbf{Hit Ratio} measures the ratio of successful hits (retrieving at least one relevant bug report) to total queries, used in seven studies~\cite{DBLP:conf/qrs/WangZLYZ22,DBLP:conf/wcre/FlorezCTM21,DBLP:conf/icse/HaeringSM21,DBLP:conf/icsm/IsotaniWFNOS21,DBLP:journals/kbs/MohsinSHJ22}. It is useful for finding at least one relevant bug report but overlooks the completeness and relevance of all results.

\paragraph{Bug Report Related Evaluation Metrics. }
Bug report related evaluation metrics focus specifically on aspects relevant to bug report analysis processes and strategies. These metrics provide insights into the efficiency and effectiveness of bug assignment and prioritization techniques, which are essential for maintaining software quality and reliability.

\textbf{Overdue Rate} measures the percentage of bug reports that were not assigned to developers within the specified timeframe. This metric is essential for assessing the efficiency of bug assignment strategies~\cite{DBLP:journals/infsof/JahanshahiC22}. A high overdue rate indicates delays in bug triaging and assignment, potentially leading to prolonged resolution times and increased software instability. By monitoring and reducing the overdue rate, teams can ensure timely allocation of bug reports to developers, thereby enhancing the overall responsiveness of bug handling processes. \textbf{APFD (Average Percentage of Fault Detected)} calculates the average percentage of faults detected by a bug prioritization strategy. This metric considers the order in which faults are detected during the inspection process, reflecting the effectiveness of prioritization techniques in identifying critical bugs early~\cite{DBLP:journals/infsof/TongZ21}. A higher APFD score indicates that the prioritization strategy successfully detected faults sooner in the inspection process, leading to faster bug resolution and improved software quality. By optimizing bug prioritization strategies to maximize APFD scores, teams can prioritize resources effectively and focus on resolving the most impactful bugs promptly.

\begin{table}[t!]
  \caption{Summaries of the strengths, weaknesses, and best-suited usage scenarios for the top 10 metrics.}
% \vspace{-0.3cm}
\centering
\footnotesize
\begin{adjustbox}{width=\linewidth,center}
\begin{tabular}{p{2cm}p{4cm}p{3.5cm}p{3.5cm}}
\toprule

Category & Strength & Weakness & Best-Suited Scenario \\ \hline
Precision & Measures the accuracy of   positive predictions, helping to avoid false positives. & Neglects false negatives. & Scenarios where avoiding   false positives is crucial, such as bug report categorization. \\ \hline
F1-score & Balances precision and   recall, making it crucial for analyzing both false positives and negatives. & May not reflect specific   operational priorities. & Scenarios where both   false positives and negatives are important, such as Bug severity/priority prediction. \\ \hline
Accuracy & Measures the overall   correctness of predictions. & Can be misleading in   imbalanced datasets. & General performance   assessment in balanced datasets. \\ \hline
Recall & Measures the system's   ability to identify all relevant instances. & Potential increase in   false positives. & Scenarios where   identifying all relevant instances is important, such as bug assignment. \\ \hline
Mean Average Precision & Evaluates the average   precision at various recall levels, summarizing the precision-recall   trade-off. & Might mask variations in   performance across different bug report types. & Scenarios requiring a   summary of precision-recall trade-offs, such as bug report classification. \\ \hline
Mean Reciprocal Rank & Calculates the average   reciprocal rank of results, focusing on the position of the first relevant   result. & May overlook the quality   of all returned results. & Scenarios where the   position of the first relevant result is important, such as duplicate detection. \\ \hline
Area under curve & Measures system   performance across different thresholds by calculating the area under the ROC   curve. & Focuses on ranking rather   than actual predicted probabilities. & Scenarios requiring   performance assessment across different thresholds, such as bug report classification. \\ \hline
Recall rate & Measures the proportion   of true positives out of all actual positives. & May lead to many false   positives. & Scenarios where ensuring   all relevant bug reports are identified, such as duplicate detection. \\ \hline
G-measure & Balances precision and   recall, giving equal weight to false positives and false negatives. & May not align with   specific operational priorities. & Scenarios requiring a   balanced assessment of precision and recall, such as bug report classification. \\ \hline
Hit Ratio & Measures the ratio of   successful hits (retrieving at least one relevant bug report) to total   queries. & Overlooks the   completeness and relevance of all results. & Scenarios where finding   at least one relevant bug report is important, such as initial bug assignment.
\\ \bottomrule
\end{tabular}
\end{adjustbox}
\label{tb:pro_con_ml_metrics}
% \vspace{-0.3cm}
\end{table}

\paragraph{\textbf{Discussion}}
In fact, the result that F1-score, Recall, Precision and Accuracy are nearly equally the most predominated metrics is not surprising, as it is known that these metrics are effective for classification tasks in machine learning~\cite{DBLP:journals/sigkdd/FormanS10}. Accuracy, which measures the ratio of correct predictions over the total number of instances evaluated, is used quite often in many cases, but still significantly less common than the top three. This is not surprising either, as it has been shown that accuracy has less discriminating power than F1-score in most cases~\cite{joshi2002evaluating}
%Since F1-score is calculated based on Recall and Precision, it is therefore also reasonable that they are used in a similar amount of primary studies.
Meanwhile, despite their importance, bug report related evaluation metrics have not received significant attention in the bug report analysis literature. We found only two articles that utilized these unique metrics, indicating a lack of emphasis on evaluating bug assignment and prioritization strategies. Researchers and practitioners in this field should explore and employ these metrics more frequently to improve bug handling processes and enhance software quality.

The comprehensive review has revealed the unique strengths, weaknesses, and best-fit applications for each evaluation metric. These insights are succinctly captured in Table~\ref{tb:pro_con_ml_metrics}, serving as a practical guide for researchers and practitioners to determine the most suitable metric for evaluation in various contexts.

\subsubsection{Evaluation Procedures}
\label{sec:RQ5-2}
A related topic to the metrics is the evaluation procedures, which describe how the results measured by the metrics are validated using different data samples. This is non-trivial for providing reliable evaluation, given the stochastic nature of most machine learning algorithms.

In Figure~\ref{fig:ep}, we summarize the four different categories of evaluation procedures adopted in the primary studies, each of them are briefly explained as below:

\begin{itemize}
    \item \textbf{$k$-fold Cross-Validation (CV):} The training data is separated into $k$-folds. Each single fold is used for testing the model once, and the remaining $k-1$ folds are used as training data. Overall results of all folds are reported.
    \item \textbf{Chronologically Fold Rotation (CFR):} All training data is divided into $k$ equally-sized folds in chronological order, and three types are used in primary studies: 
    
    \begin{itemize}
        \item[---] The first $i$ folds form the training set and the $fold_{i+1}$ is used as a test set~\cite{hu2014effective,DBLP:conf/iwpc/ZhangCJLX17,zhang2016ksap,DBLP:journals/jss/ZhangCYLL16}.
        \item[---] Train a model on $fold_i$ and test it on $fold_{i+1}$~\cite{DBLP:conf/iwpc/TianWLG16}. 
        \item[---] Trained a model from $fold_i$ to $fold_{i+j-1}$ and test it on $fold_{i+j}$, for all $1 \le i \le k-j$~\cite{DBLP:conf/iwpc/TianWLG16}.
    \end{itemize}
    
    The overall results from the runs are reported.
    
    \item \textbf{Chronologically Percent Split (CPS):} Similar to the above, but instead of using folds, different percentages of split are used.
    \item \textbf{Percent Split (PS):} The training data is randomly split for training and testing to get the overall result, similar to multiple repeated hold-out.
    \item \textbf{Unspecified (U):} The evaluation procedure has not been clearly stated.
\end{itemize}

% \paragraph{\textbf{Discussion}}
%This is because, as demonstrated by \citeauthor{joshi2002evaluating}~\cite{joshi2002evaluating},  F1-score has better discriminating power than Accuracy in selecting and determining classifiers, which could be the reason of its wider adoption.

\begin{table}[t!]
  \caption{Summaries of the strengths, weaknesses, and best-suited usage scenarios for the evaluation procedures.}
% \vspace{-0.3cm}
\centering
\footnotesize
\begin{adjustbox}{width=\linewidth,center}
\begin{tabular}{p{2cm}p{4cm}p{3.5cm}p{3.5cm}}
\toprule

\textbf{Category} & \textbf{Strength} & \textbf{Weakness} & \textbf{Best-Suited Scenario} \\ \hline
CV & Provides a comprehensive   evaluation by using all data for both training and testing across different   folds. & Can be computationally   intensive, especially with large datasets. & General performance   assessment where reliable evaluation is crucial. \\ \hline
PS & Simple and quick to   implement, providing a straightforward evaluation. & May not provide a   reliable evaluation due to random splits and potential data imbalance. & Initial model evaluation   and scenarios with limited computational resources. \\ \hline
CPS & Maintains the temporal   order of data, providing a realistic evaluation for time-dependent data. & May not be suitable for   non-temporal data and can be less flexible. & Evaluating models on   time-series data or scenarios where chronological order is important. \\ \hline
U & Flexibility in evaluation   procedures, allowing for customized approaches. & Lack of standardization   can lead to inconsistent and unreliable results. & Scenarios where specific   evaluation procedures are not critical or well-defined. \\ \hline
CFR & Provides a realistic   evaluation by maintaining the chronological order of data and using different   fold rotations. & Can be complex to   implement and computationally intensive. & Evaluating models on   time-series data or scenarios requiring realistic temporal evaluation. 
\\ \bottomrule
\end{tabular}
\end{adjustbox}
\label{tb:pro_con_procedures}
% \vspace{-0.3cm}
\end{table}

\paragraph{\textbf{Discussion}}
Clearly, we note that cross-validation (normally 10 folds) ~\cite{DBLP:journals/sigkdd/FormanS10} is the evaluation procedure used in the majority of the primary studies (47.5\%), together with the evaluation metrics, particularly the Recall, Precision and F1-score. The other procedures form the minority with small margins between each other, including seven studies whose evaluation procedures have not been clearly stated. 

However, with the increasing application of deep learning models, cross-validation may become time-consuming. Deep learning models typically require large datasets to achieve optimal performance, and implementing cross-validation on such large datasets significantly increases computational pressure. Each fold in the cross-validation process necessitates retraining the model from scratch, which is computationally intensive and time-consuming, particularly for complex deep learning architectures. This constraint can limit the feasibility of using cross-validation as a standard evaluation method for deep learning models, prompting the need for alternative strategies that balance thorough model evaluation with computational efficiency.

Drawing from the review results, Table~\ref{tb:pro_con_procedures} encapsulates the strengths, weaknesses, and ideal scenarios for each evaluation procedure. This table acts as a valuable resource, enabling researchers and practitioners to identify the most appropriate evaluation method for various contexts and requirements.

Bringing together the above result from Section~\ref{sec:RQ5-1} and Section~\ref{sec:RQ5-2}, we can conclude that

\begin{tcolorbox}[breakable,left=5pt,right=5pt,top=5pt,bottom=5pt] 
\textbf{\underline{Finding 7}}: Precision, F1-score, accuracy, and recall are key metrics for evaluating machine learning models in bug report analysis tasks. Bug report related evaluation metrics have not received significant attention in the bug report analysis literature. The majority of studies prefer $k$-fold cross-validation for model evaluation.
\end{tcolorbox}

\subsection{RQ6: Which statistical tests and standardized effect size measurements are applied to mitigate bias?}
\label{sec:RQ6}

% \subsection{Purpose}

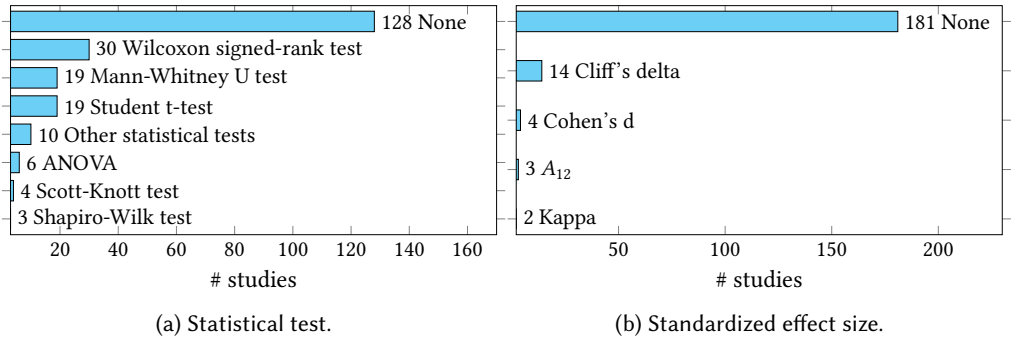
\begin{figure}[!t]
\centering
\begin{subfigure}[t]{0.5\columnwidth}
\newcommand\shift{1.5}
\begin{tikzpicture}[scale=1]
\begin{axis}[
    xbar=0.05cm,
    width=10cm,
    height=5.5cm,
    xmax=170,
    bar width=0.35cm,
 enlarge y limits=0.08,
 enlarge x limits=0,
    legend style={at={(1.05,0.7)},anchor=north,legend columns=1,font=\tiny},
    legend cell align={left},
   xlabel={\# studies},
     x label style={font=\Large},
     nodes near coords,
          point meta=explicit symbolic,
          every node near coord/.append style={anchor=west,font=\large},
      symbolic y coords={Shapiro-Wilk test,Scott-Knott test,ANOVA,Other statistical tests,Student t-test,Mann-Whitney U test,Wilcoxon signed-rank test,None},
      yticklabels={,,},
         y tick label style={font=\large},
          x tick label style={font=\large},
         ytick=data
    ]

\addplot [fill=cyan!50,draw=black] coordinates {(3,Shapiro-Wilk test)[3 Shapiro-Wilk test] (4,Scott-Knott test)[4 Scott-Knott test] (10,Other statistical tests)[10 Other statistical tests] (6,ANOVA)[6 ANOVA] (19,Student t-test)[19 Student t-test] (19,Mann-Whitney U test)[19 Mann-Whitney U test] (30,Wilcoxon signed-rank test)[30 Wilcoxon signed-rank test] (128,None)[128 None]};
 \end{axis}
 \end{tikzpicture}
 
    \subcaption{Statistical test.}
 \label{fig:RQ6}
    \end{subfigure}
~\hspace{-0.4cm}
\begin{subfigure}[t]{0.5\columnwidth}
\newcommand\shift{1.5}
\begin{tikzpicture}[scale=1]
\begin{axis}[
    xbar=0.05cm,
    width=10cm,
    height=5.5cm,
    xmax=230,
    bar width=0.35cm,
 enlarge y limits=0.08,
 enlarge x limits=0,
    legend style={at={(1.05,0.7)},anchor=north,legend columns=1,font=\tiny},
    legend cell align={left},
   xlabel={\# studies},
     x label style={font=\Large},
     nodes near coords,
          point meta=explicit symbolic,
          every node near coord/.append style={anchor=west,font=\large},
      symbolic y coords={kappa,a12,d,delta,None},
      yticklabels={,,},
         y tick label style={font=\large},
          x tick label style={font=\large},
         ytick=data
    ]

\addplot [fill=cyan!50,draw=black] coordinates {(2,kappa)[2 Kappa](3,a12)[3 $A_{12}$] (4,d)[4 Cohen's d] (14,delta)[14 Cliff's delta] (181,None)[181 None]};
 \end{axis}
 \end{tikzpicture}
 
   \subcaption{Standardized effect size.}
 \label{fig:es}
    \end{subfigure}
    \caption{Statistics of statistical test and standardized effect size.}
  \end{figure}

Statistical tests and standardized effect size measurements are essential to ensure the reliability of results, especially given that the stochastic nature of machine learning training pipelines. In Figure~\ref{fig:RQ6} and Figure~\ref{fig:es}, we summarize the statistical tests and effect size measurement methods used in the primary studies. As can be seen, although the results have shown that Wilcoxon signed-rank test is more commonly used than the others, surprisingly, a significant number of studies (128) have completely ignored the importance of statistical tests. As for the standardized effect size, most primary studies have not even mentioned such, constituting more than 90\% of the studies.

% \subsection{Results}

First, we examine the statistical tests employed in bug report analysis, which serve as crucial tools for evaluating the significance of findings and drawing reliable conclusions. These tests are categorized based on their applicability and assumptions. The following statistical tests are commonly utilized in bug report analysis:

\subsubsection{Statistical Test}
The \textbf{Wilcoxon Signed-Rank Test} is a non-parametric test comparing two related samples or repeated measurements from the same group to assess if their medians differ significantly. This test is useful for evaluating performance differences between two algorithms or models on the same dataset, cited in 30 studies~\cite{DBLP:conf/fedcsis/KumarKNMKP21,DBLP:conf/scam/RazzaqBPCS21,DBLP:journals/infsof/GomesTC21,DBLP:journals/jss/MiryeganehHH21,DBLP:journals/infsof/GeFQGQ22}. Its main disadvantage is the requirement for symmetrical data distribution around the median. The \textbf{Mann-Whitney U Test} compares two independent samples non-parametrically, appearing in 19 studies~\cite{DBLP:conf/wcre/FlorezCTM21,DBLP:journals/infsof/ElmishaliK23,DBLP:journals/tse/HuoTLLS21,DBLP:journals/infsof/PanichellaCS21,DBLP:journals/jss/KimGLKKBKT22}. It evaluates if the distributions of two groups are the same or if one group tends to have higher values. A limitation is its sensitivity to sample size differences. The \textbf{Student t-Test} determines if there's a significant difference between the means of two independent groups, used in 19 studies~\cite{DBLP:journals/infsof/NeysianiBA20,DBLP:journals/array/HirschH22,DBLP:conf/issre/WenYH16,DBLP:conf/iwpc/ZhangCJLX17,DBLP:journals/corr/XuanJHRZLW17}. It assumes normal data distribution, requiring a normality test before application, which can limit its use. \textbf{ANOVA (Analysis of Variance)} analyzes differences among group means in a sample, cited in six studies~\cite{DBLP:journals/access/UmerLS18,DBLP:journals/access/RamayUYZI19,DBLP:conf/msr/ZanjaniKB15,DBLP:journals/tr/UmerLI20,DBLP:conf/wcre/AggarwalRTHGS15,DBLP:conf/icse/WangCWW17}. It compares three or more groups simultaneously, but requires variance homogeneity among groups. The \textbf{Scott-Knott Test}, used post-ANOVA, identifies significant differences between group means by partitioning them into subgroups, used in four studies~\cite{DBLP:journals/tosem/AssiHGZ23,DBLP:conf/apsec/LongCC22,DBLP:journals/infsof/TongZ21,DBLP:journals/jss/ZhouLS20}. It handles data heterogeneity well but is sensitive to outliers. The \textbf{Shapiro-Wilk Test} assesses the normality of data samples, adopted in three studies~\cite{DBLP:journals/itiis/YangMLL19,DBLP:conf/sac/YangBLL17,DBLP:journals/corr/XuanJHRZLW17}. It validates assumptions for parametric tests like the t-test and ANOVA, but its sensitivity can lead to incorrect normality rejections with larger samples.

\subsubsection{Standard Effect Size Measurements}
Furthermore, we explain various effect size measurements commonly employed in bug report analysis. These measurements are pivotal for assessing the significance and magnitude of differences observed in bug-related metrics, aiding in the interpretation and comparison of results across different studies. We categorize these measurements into several widely used techniques: 

\textbf{Cliff’s delta} is a non-parametric measure used in 14 studies~\cite{DBLP:journals/iee/WuLM18,DBLP:journals/access/LiangSWY19,DBLP:conf/ease/ZhouNG15,DBLP:journals/infsof/XiaLSWY15,DBLP:journals/jcst/YangLXHS17} to quantify effect size by comparing the probabilities of observations between groups. It is robust against non-normal distributions, making it suitable for diverse bug report analyses. However, it may be less informative with very small sample sizes. \textbf{Cohen’s d}, used in four studies~\cite{DBLP:journals/tr/UmerLI20,DBLP:journals/infsof/TongZ21,DBLP:conf/qrs/SuHCQM22,DBLP:journals/tr/MaKYZZL23}, measures effect size as the difference between two means in standard deviation units. It helps gauge practical significance but assumes homogeneous, normally distributed data. $\mathbf{A_{12}}$ quantifies the strength of an effect, used in three studies, and helps evaluate the effectiveness of bug report analysis strategies but does not indicate effect direction. \textbf{Kappa}, found in two studies~\cite{DBLP:conf/wcre/AggarwalRTHGS15,DBLP:journals/ese/HindleAS16}, measures inter-rater agreement for categorical items, adjusting for chance agreement. It is useful for assessing consistency in bug report analysis but can be inaccurate with high prevalence or extreme distributions.

\begin{table}[t!]
  \caption{Summaries of the strengths, weaknesses, and best-suited usage scenarios for the most common statistical tests.}
% \vspace{-0.3cm}
\centering
\footnotesize
\begin{adjustbox}{width=\linewidth,center}
\begin{tabular}{p{2cm}p{4cm}p{3.5cm}p{3.5cm}}
\toprule

\textbf{Category} & \textbf{Strength} & \textbf{Weakness} & \textbf{Best-Suited Scenario} \\ \hline
Wilcoxon signed-rank test & Non-parametric test comparing two related samples or repeated measurements to assess median differences. & Requires symmetrical data distribution around the median. & Evaluating performance differences between two algorithms or models on the same dataset. \\ \hline
Mann-Whitney U test & Non-parametric test comparing two independent samples to evaluate if their distributions are the same. & Sensitive to sample size differences. & Comparing distributions of two independent groups. \\ \hline
Student t-test & Determines if there's a significant difference between the means of two independent groups. & Assumes normal data distribution, requiring a normality test before application. & Assessing mean differences between two groups with normally distributed data. \\ \hline
ANOVA & Analyzes differences among group means in a sample, comparing three or more groups simultaneously. & Requires variance homogeneity among groups. & Comparing means across multiple groups. \\ \hline
Scott-Knott test & Identifies significant differences between group means by partitioning them into subgroups. & Sensitive to outliers. & Post-ANOVA analysis to identify significant group differences. \\ \hline
Shapiro-Wilk test & Assesses the normality of data samples, validating assumptions for parametric tests. & Sensitivity can lead to incorrect normality rejections with larger samples. & Validating normality assumptions for parametric tests like the t-test and ANOVA. 
\\ \bottomrule
\end{tabular}
\end{adjustbox}
\label{tb:pro_con_ml_stat_test}
% \vspace{-0.3cm}
\end{table}

\paragraph{\textbf{Discussion}}

Our analysis reveals a varied landscape in the utilization of statistical tests and effect size measurements within the field of bug report analysis. Notably, while a range of statistical tests such as the Wilcoxon Signed-Rank Test, Mann-Whitney U Test, Student t-Test, ANOVA, Scott-Knott Test, and Shapiro-Wilk Test have been applied in numerous studies, a substantial proportion (59.8\%) of the research does not employ these robust methodologies. This omission may undermine the reliability and generalizability of the findings, especially given the stochastic nature of the underlying algorithms in machine learning applications.

Similarly, standardized effect size measurements of Cliff’s delta, Cohen’s d, $A_{12}$, and Kappa, though utilized in several studies, are conspicuously absent in the majority. This lack of standardized effect size reporting limits the ability of the research community to accurately assess the magnitude and practical significance of the findings, an essential aspect of empirical research.

The general neglect of statistical tests and effect size measurements suggests a need for more rigorous methodological standards in the domain of bug report analysis. Encouraging the adoption of a broader array of statistical tests and effect size measurements could enhance the interpretability and comparability of results across studies, fostering a more robust scientific dialogue and advancing the field.

\begin{table}[t!]
  \caption{Summaries of the strengths, weaknesses, and best-suited usage scenarios for the standardized effect size tests.}
% \vspace{-0.3cm}
\centering
\footnotesize
\begin{adjustbox}{width=\linewidth,center}
\begin{tabular}{p{2cm}p{4cm}p{3.5cm}p{3.5cm}}
\toprule

\textbf{Category} & \textbf{Strength} & \textbf{Weakness} & \textbf{Best-Suited Scenario} \\ \hline
Cliff's delta & Robust against non-normal   distributions, suitable for diverse bug report analyses. & May be less informative   with very small sample sizes. & Quantifying effect size   in non-normal distributions. \\ \hline
Cohen's d & Measures effect size as   the difference between two means in standard deviation units. & Assumes homogeneous,   normally distributed data. & Gauging practical   significance in normally distributed data. \\ \hline
A$_{12}$ & Quantifies the strength   of an effect, helping evaluate the effectiveness of strategies. & Does not indicate effect   direction. & Evaluating the   effectiveness of bug report analysis strategies. \\ \hline
Kappa & Measures inter-rater   agreement for categorical items, adjusting for chance agreement. & Can be inaccurate with   high prevalence or extreme distributions. & Assessing consistency in   bug report analysis.
\\ \bottomrule
\end{tabular}
\end{adjustbox}
\label{tb:pro_con_effect_size}
% \vspace{-0.3cm}
\end{table}

The comprehensive review has highlighted the unique benefits, drawbacks, and ideal applications for various model optimization methods and activation functions. These insights are clearly presented in Table~\ref{tb:pro_con_ml_stat_test} and Table~\ref{tb:pro_con_effect_size}, offering a practical guide for researchers and practitioners to choose the most suitable statistical test and effect size test in different contexts.

In summary, we therefore have:

\begin{tcolorbox}[breakable,left=5pt,right=5pt,top=5pt,bottom=5pt] 
\textbf{\underline{Finding 8}}: Wilcoxon signed-rank test and Cliff's delta are the most commonly used statistical test methods and effect size measurements. A significant proportion of studies have completely ignored the importance of statistical tests. Effect size has not been reported for most of the work.
\end{tcolorbox}

%\subsection{Lack of Statistical Test and Standardized Effect Size Measurement}

%\input{sec/smell}
%\input{sec/why}
\section{FUTURE RESEARCH OPPORTUNITIES}
\label{sec:future}

\subsection{Transformer-based Model Application for Bug Report Analysis}
As discussed in Section~\ref{sec:RQ1}, the landscape of machine learning in bug report analysis has witnessed significant advancements, particularly with the adoption of Transformer-based models, which offer promising enhancements over traditional methods like CNNs, LSTMs, and SVMs. These models excel in handling long-range dependencies and large datasets, which are common in bug report analysis. Despite existing research in areas such as bug report generation~\cite{DBLP:conf/noms/FukudaWHT22}, security bug report prediction~\cite{DBLP:conf/dsa/ZhengCWFSC21}, and bug severity prediction~\cite{DBLP:journals/access/WeiZR23}, many tasks remain unexplored. Future research should leverage the capabilities of Transformer architectures like BERT and GPT to tackle complex tasks such as bug triaging, duplicate report detection, and summarization of bug reports. Further exploration into the application of Transformers could significantly improve the accuracy and efficiency of parsing and understanding the textual context data within bug reports. Innovations could include the development of domain-specific models that are pre-trained on software development texts to better grasp the context and jargon specific to certain software projects. By effectively harnessing these models, researchers and practitioners could achieve greater precision in bug report analysis tasks, leading to faster and more efficient bug resolution processes. This would not only enhance the productivity of developers but also improve the overall software quality, making this an exciting avenue for future research.

\subsection{Automatic Bug Report Analysis Tools deployed in GitHub}
As analyzed in Section 4.2, our survey reveals a growing trend in the use of unstructured bug reports from platforms like GitHub. While structured bug reports benefit from predefined fields and standardized formats, unstructured reports are inherently more variable and complex, posing unique challenges for automated analysis.

Despite the above, we found only one study that design solution explicitly for structured bug reports: \citeauthor{xiao2018bug}~\cite{xiao2018bug} propose CNN Forest, which combines a convolutional neural network with an ensemble of random forests, achieving strong performance in semantic parsing and structural information extraction to enhance bug report localization.

Indeed, tools developed for analyzing structured bug reports can also be adapted to handle unstructured reports. For example, \citeauthor{DBLP:journals/tosem/ZhangHVIXTLJ23}~\cite{DBLP:journals/tosem/ZhangHVIXTLJ23} not only applied state-of-the-art duplicate bug report detection methods to structured bug reports from Bugzilla and Jira, but also to unstructured bug reports from GitHub. However, as \citeauthor{DBLP:journals/tosem/ZhangHVIXTLJ23}~\cite{DBLP:journals/tosem/ZhangHVIXTLJ23} observed, unstructured bug reports lack well-defined fields for maintaining duplicate references. As a result, it is difficult to extract duplicate issues following GitHub’s guidelines for marking duplicates. This implies that applying tools designed for structured reports to unstructured data introduces two key limitations:

\begin{itemize}
    \item \textbf{Inconsistent data formatting:} Structured tools rely on standardized fields, which are absent in unstructured reports. This inconsistency can lead to difficulties in accurately parsing and analyzing the data. For example, structured bug reports (e.g., from Jira) have standardized fields such as steps, expected outcomes, and environment, enabling accurate parsing. In contrast, unstructured reports (e.g., from GitHub) use free-form text without standardization without these fields, complicating data analysis and interpretation.
    \item \textbf{Incomplete information extraction:} Unstructured reports may lack critical details or present information in a non-standard manner, making it challenging for structured tools to extract necessary data effectively. For example, structured reports (e.g., from Bugzilla) require reporters to specify critical details such as OS version, severity, and reproduction steps explicitly, enabling efficient data extraction. Conversely, unstructured reports (e.g., from GitHub issues) often omit or vaguely describe essential details, hindering automated tools from effectively extracting key information.
    % \item \textbf{Scalability issues:} Tools optimized for structured data may not scale well when dealing with the variability and volume of unstructured reports, leading to performance bottlenecks \tao{give an example} [].
\end{itemize}

To overcome these limitations, future research should focus on developing advanced analysis tools tailored specifically for unstructured bug reports. Leveraging state-of-the-art natural language processing techniques, such as Transformer-based architectures and Large Language Models (e.g., BERT, GPT-4), can enhance the parsing and classification of the nuanced content found in GitHub issues. Additionally, conducting in-depth case studies and application-driven evaluations of such tools will provide concrete insights into their efficacy and practicality in real-world settings. Exploring hybrid approaches that combine rule-based and learning-based methods may offer balanced solutions to handle the variability inherent in unstructured bug reports.

\subsection{Bug Report Analysis with LLMs}
As emphasized in Section~\ref{sec:RQ3}, the incorporation of Large Language Models (LLMs) such as GPT-4 or LLAMA 3 into bug report analysis presents a cutting-edge frontier. These models' deep understanding of natural language makes them ideally suited for bug report analysis tasks. However, no primary studies have explored this area, indicating that this field has not kept up with recent trends. To bridge this gap, future research could explore the automation of these discriminative tasks using LLMs to enhance accuracy and process efficiency.

Moreover, LLMs could be pivotal in generative tasks within bug report analysis, such as summarizing complex reports or generating descriptive titles. Such applications could significantly reduce the workload on developers by automating routine analyses and allowing them to focus on more critical tasks. Exploring the use of LLMs to link bug reports with specific code commits or development tasks could also provide groundbreaking improvements in how bugs are traced and resolved.

Meanwhile, predictive tasks that estimate bug severity or prioritize issues based on historical data could benefit from the advanced capabilities of LLMs. LLMs can easily be used for the fields in bug reports, aiding in automated classification efficiently. The potential for LLMs to transform bug report analysis is vast, promising substantial advances in software maintenance and development efficiency.

\subsection{Bug Report Analysis for Specific Bugs}
As identified in Section~\ref{sec:RQ4}, only 10\% of primary studies have focused on analyzing specific types of bugs. This observation underscores a substantial opportunity for researchers to explore the investigation of various specific bug types more thoroughly.

For example, future research could prioritize the development of sophisticated machine learning algorithms that can identify and classify a wider range of specific bug types. By doing so, it will be possible to address less common but potentially critical specific bugs that may be overlooked by more generalized models. This focused approach could also facilitate the creation of specialized diagnostic tools that provide developers with precise insights into the nature of the bugs, enabling more targeted and effective interventions.

Moreover, there is significant potential in employing advanced analytics to understand the underlying patterns and causes of specific bugs. Researchers could utilize data mining techniques to extract valuable insights from historical specific bug data, helping to predict the likelihood of specific bug occurrences and their potential impacts on software systems. Such strategic research endeavors will not only fill the existing gap but will also elevate the standards and capabilities of specific bug report analysis frameworks.

\subsection{Dedicated Metrics for Bug Report Analysis}
As noted in Section~\ref{sec:RQ5}, aside from a few studies using dedicated metrics for bug report analysis like Overdue Rate~\cite{DBLP:journals/infsof/JahanshahiC22} and APFD~\cite{DBLP:journals/infsof/TongZ21}, there is a lack of research utilizing bug report-related metrics. Creating a dedicated metric for evaluating the performance of bug report analysis tools could lead to more detailed insights and improvements in software quality assurance. This type of metric would ideally reflect the unique aspects of bug reports, such as their urgency, complexity, and impact on the software project.

Hence, we believe that future research should focus on developing composite metrics that integrate various dimensions of bug report analysis, such as the accuracy of bug report generation, the similarity of duplicate bug reports, and the relevance of the recommendations provided. This would provide a more holistic view of a tool's performance and its practical value to developers. A possible opportunity is to consider multiple metrics simultaneously and their joint assessment~\cite{DBLP:journals/tse/LiCY22}.

Additionally, it would be beneficial to validate these metrics through extensive empirical studies, involving diverse software projects and development environments. This would ensure that the metrics are robust and universally applicable across different contexts and technologies.

\subsection{Explainable Deep Learning for Bug Report Analysis}
Integrating explainability into deep learning models used in bug report analysis is crucial for their acceptance and effectiveness in practical settings. Researchers should focus on developing methods that not only perform well but also provide transparent and understandable decision-making processes.

This could involve using techniques such as attention mechanisms that highlight the parts of the bug report most relevant to the model's decision, or developing hybrid models that combine deep learning with rule-based reasoning to provide clear explanations for their outputs.

Moreover, collaborations between machine learning researchers, software developers, and end-users are essential to ensure that the explanations provided by these models are meaningful and actionable in real-world software development scenarios. This interdisciplinary approach can help refine the models to better meet the practical needs of software maintenance and development.

\section{Threats to Validity}
\label{sec:threat}

Threats to construct validity can be raised by the research methodology applied. We have mitigated such threats by following the systematic review protocol proposed by Kitchenham et al.~\cite{DBLP:journals/infsof/KitchenhamBBTBL09}. 

Threats to internal validity may be introduced by having inappropriate classification and interpretation of the studies. We have limited this by conducting three iterations of paper reviews by the authors. Error checks and investigations were also conducted to correct any issues found during the search. 
 
Threats to external validity may restrict the generalizability of the revealed statistics and the justification of future research opportunities. We have mitigated such by conducting the survey wider and deeper: it covers 1,825 searched papers published between 2014 and 2023 from three repositories; while at the same time, extracting 204 prominent primary studies following the exclusion and inclusion procedure. This, together with the evidence from the best practice in software engineering and machine learning, have provided rich sources for us to discuss the results and the future research opportunities identified.
\section{Conclusion}
\label{sec:conclusion}

In this work, we perform a systematic literature review on bug report analysis with machine learning over the past ten years. The survey covers 1,825 papers from three repositories, based on which we extracted a more precise set of 585 candidate studies, and finally, 204 primary studies were selected for detailed analysis. Notably, we obtain the following key findings for our research questions:

\begin{itemize}
    \item \textbf{RQ1:} CNN, LSTM and $k$NN are the top three commonly used machine learning algorithms for bug report analysis. Word2vec and TF-IDF are the most widely used methods to represent features. Stop word removal is the most widely used preprocessing method, followed by tokenization and stemming. 
    \item \textbf{RQ2:} Eclipse and Mozilla Core are the most widely evaluated software projects.
    \item \textbf{RQ3:} Bug categorization task receives the most attention.
    \item \textbf{RQ4:} Majority of the studies focuses on general bug types without a specific focus.   
    \item \textbf{RQ5:} Precision, F1-score, accuracy, and recall are much more popular than the other metrics in evaluation. The majority of studies prefer $k$-fold cross-validation for model evaluation.
    \item \textbf{RQ6:} Statistical test and effect size are ignored by majority of the studies.
\end{itemize}

Based on the result analysis of the primary studies, we offer insights for each research question. Additionally, we propose six future research directions. Through the statistics and future research directions discussed in this work, we hope to provide insights and visionary impact that excite a more significant growth of research on specialized machine learning for bug report analysis, as well as its interaction to the other software engineering problems, such as configuration management~\cite{DBLP:conf/icse/ChenChen26,DBLP:conf/icse/XiangChen26,DBLP:conf/icse/LiangChen25,10832565,DBLP:journals/tosem/GongC25,DBLP:conf/sigsoft/0001L24,DBLP:journals/corr/abs-2112-07303}, general software optimization~\cite{DBLP:journals/tosem/ChenL23}, performance prediction~\cite{DBLP:journals/pacmse/Gong024,DBLP:journals/tse/GongCB25,DBLP:conf/sigsoft/Gong023}, testing~\cite{DBLP:conf/icse/MaChen25,DBLP:conf/esem/DuC24}, requirements handling~\cite{DBLP:journals/tosem/ChenL23a} and self-adaptation~\cite{DBLP:journals/tsc/ChenB17,DBLP:conf/icse/Ye0L25,DBLP:conf/wcre/Chen22,DBLP:journals/tse/ChenB17,DBLP:journals/tosem/ChenLBY18}.

\begin{acks}
This work was supported by an NSFC Grant (62372084) and a UKRI Grant (10054084).
\end{acks}

\bibliographystyle{ACM-Reference-Format}
\bibliography{main}


\begin{thebibliography}{211}

%%% the Web address.

\ifx \showCODEN    \undefined \def \showCODEN     #1{\unskip}     \fi
\ifx \showDOI      \undefined \def \showDOI       #1{#1}\fi
\ifx \showISBNx    \undefined \def \showISBNx     #1{\unskip}     \fi
\ifx \showISBNxiii \undefined \def \showISBNxiii  #1{\unskip}     \fi
\ifx \showISSN     \undefined \def \showISSN      #1{\unskip}     \fi
\ifx \showLCCN     \undefined \def \showLCCN      #1{\unskip}     \fi
\ifx \shownote     \undefined \def \shownote      #1{#1}          \fi
\ifx \showarticletitle \undefined \def \showarticletitle #1{#1}   \fi
\ifx \showURL      \undefined \def \showURL       {\relax}        \fi
% invisible to TeX
\providecommand\bibfield[2]{#2}
\providecommand\bibinfo[2]{#2}
\providecommand\natexlab[1]{#1}
\providecommand\showeprint[2][]{arXiv:#2}

\bibitem[Afric et~al\mbox{.}(2023)]%
        {DBLP:journals/access/AfricVSD23}
\bibfield{author}{\bibinfo{person}{Petar Afric}, \bibinfo{person}{Davor Vukadin}, \bibinfo{person}{Marin Silic}, {and} \bibinfo{person}{Goran Delac}.} \bibinfo{year}{2023}\natexlab{}.
\newblock \showarticletitle{Empirical Study: How Issue Classification Influences Software Defect Prediction}.
\newblock \bibinfo{journal}{\emph{{IEEE} Access}}  \bibinfo{volume}{11} (\bibinfo{year}{2023}), \bibinfo{pages}{11732--11748}.
\newblock


\bibitem[Aggarwal et~al\mbox{.}(2015)]%
        {DBLP:conf/wcre/AggarwalRTHGS15}
\bibfield{author}{\bibinfo{person}{Karan Aggarwal}, \bibinfo{person}{Tanner Rutgers}, \bibinfo{person}{Finbarr Timbers}, \bibinfo{person}{Abram Hindle}, \bibinfo{person}{Russell Greiner}, {and} \bibinfo{person}{Eleni Stroulia}.} \bibinfo{year}{2015}\natexlab{}.
\newblock \showarticletitle{Detecting duplicate bug reports with software engineering domain knowledge}. In \bibinfo{booktitle}{\emph{{SANER} 2015}}. \bibinfo{pages}{211--220}.
\newblock


\bibitem[Akilan et~al\mbox{.}(2020)]%
        {DBLP:conf/smc/AkilanSPM20}
\bibfield{author}{\bibinfo{person}{Thangarajah Akilan}, \bibinfo{person}{Dhruvit Shah}, \bibinfo{person}{Nishi Patel}, {and} \bibinfo{person}{Rinkal Mehta}.} \bibinfo{year}{2020}\natexlab{}.
\newblock \showarticletitle{Fast Detection of Duplicate Bug Reports using LDA-based Topic Modeling and Classification}. In \bibinfo{booktitle}{\emph{{IEEE} SMC 2020}}.
\newblock


\bibitem[Aljedaani et~al\mbox{.}(2022)]%
        {DBLP:conf/w4a/AljedaaniML0J22}
\bibfield{author}{\bibinfo{person}{Wajdi Aljedaani}, \bibinfo{person}{Mohamed~Wiem Mkaouer}, \bibinfo{person}{Stephanie Ludi}, \bibinfo{person}{Ali Ouni}, {and} \bibinfo{person}{Ilyes Jenhani}.} \bibinfo{year}{2022}\natexlab{}.
\newblock \showarticletitle{On the identification of accessibility bug reports in open source systems}. In \bibinfo{booktitle}{\emph{W4A'22: 19th Web for All Conference}}.
\newblock


\bibitem[Alkhazi et~al\mbox{.}(2020)]%
        {DBLP:journals/asc/AlkhaziDAAYM20}
\bibfield{author}{\bibinfo{person}{Bader Alkhazi}, \bibinfo{person}{Andrew DiStasi}, \bibinfo{person}{Wajdi Aljedaani}, \bibinfo{person}{Hussein Alrubaye}, \bibinfo{person}{Xin Ye}, {and} \bibinfo{person}{Mohamed~Wiem Mkaouer}.} \bibinfo{year}{2020}\natexlab{}.
\newblock \showarticletitle{Learning to rank developers for bug report assignment}.
\newblock \bibinfo{journal}{\emph{Applied Software Computing}}  \bibinfo{volume}{95} (\bibinfo{year}{2020}), \bibinfo{pages}{106667}.
\newblock


\bibitem[Anh and Luyen(2021)]%
        {DBLP:conf/apsec/AnhL21}
\bibfield{author}{\bibinfo{person}{Bui Thi~Mai Anh} {and} \bibinfo{person}{Nguyen~Viet Luyen}.} \bibinfo{year}{2021}\natexlab{}.
\newblock \showarticletitle{An Imbalanced Deep Learning Model for Bug Localization}. In \bibinfo{booktitle}{\emph{28th Asia-Pacific Software Engineering Conference Workshops, {APSEC} Workshops 2021, Taipei, Taiwan, December 6-9, 2021}}.
\newblock


\bibitem[Arcega et~al\mbox{.}(2018)]%
        {DBLP:conf/models/ArcegaFC18}
\bibfield{author}{\bibinfo{person}{Lorena Arcega}, \bibinfo{person}{Jaime Font}, {and} \bibinfo{person}{Carlos Cetina}.} \bibinfo{year}{2018}\natexlab{}.
\newblock \showarticletitle{Evolutionary Algorithm for Bug Localization in the Reconfigurations of Models at Runtime}. In \bibinfo{booktitle}{\emph{{ACM/IEEE} MODELS '18}}.
\newblock


\bibitem[Ardimento and Dinapoli(2017)]%
        {ardimento2017knowledge}
\bibfield{author}{\bibinfo{person}{Pasquale Ardimento} {and} \bibinfo{person}{Andrea Dinapoli}.} \bibinfo{year}{2017}\natexlab{}.
\newblock \showarticletitle{Knowledge extraction from on-line open source bug tracking systems to predict bug-fixing time}. In \bibinfo{booktitle}{\emph{Proceedings of the 7th international conference on web intelligence, mining and semantics}}. \bibinfo{pages}{1--9}.
\newblock


\bibitem[Assi et~al\mbox{.}(2023)]%
        {DBLP:journals/tosem/AssiHGZ23}
\bibfield{author}{\bibinfo{person}{Maram Assi}, \bibinfo{person}{Safwat Hassan}, \bibinfo{person}{Stefanos Georgiou}, {and} \bibinfo{person}{Ying Zou}.} \bibinfo{year}{2023}\natexlab{}.
\newblock \showarticletitle{Predicting the Change Impact of Resolving Defects by Leveraging the Topics of Issue Reports in Open Source Software Systems}.
\newblock \bibinfo{journal}{\emph{{ACM} Transactions on Software Engineering and Methodology}}  \bibinfo{volume}{32} (\bibinfo{year}{2023}), \bibinfo{pages}{141:1--141:34}.
\newblock


\bibitem[Aung et~al\mbox{.}(2022)]%
        {DBLP:journals/jss/AungWHS22}
\bibfield{author}{\bibinfo{person}{Thazin Win~Win Aung}, \bibinfo{person}{Yao Wan}, \bibinfo{person}{Huan Huo}, {and} \bibinfo{person}{Yulei Sui}.} \bibinfo{year}{2022}\natexlab{}.
\newblock \showarticletitle{Multi-triage: {A} multi-task learning framework for bug triage}.
\newblock \bibinfo{journal}{\emph{Journal of Systems and Software}}  \bibinfo{volume}{184} (\bibinfo{year}{2022}), \bibinfo{pages}{111133}.
\newblock


\bibitem[Baarah et~al\mbox{.}(2021)]%
        {baarah2021sentiment}
\bibfield{author}{\bibinfo{person}{Aladdin Baarah}, \bibinfo{person}{Ahmad Aloqaily}, \bibinfo{person}{Hala Zyod}, {and} \bibinfo{person}{Nasser Mustafa}.} \bibinfo{year}{2021}\natexlab{}.
\newblock \showarticletitle{Sentiment-Based Neural Network Approach for Predicting the Severity of Bug Reports}. In \bibinfo{booktitle}{\emph{2021 ICDS}}.
\newblock


\bibitem[Banerjee et~al\mbox{.}(2017)]%
        {banerjee2017automated}
\bibfield{author}{\bibinfo{person}{Sean Banerjee}, \bibinfo{person}{Zahid Syed}, \bibinfo{person}{Jordan Helmick}, \bibinfo{person}{Mark Culp}, \bibinfo{person}{Kenneth Ryan}, {and} \bibinfo{person}{Bojan Cukic}.} \bibinfo{year}{2017}\natexlab{}.
\newblock \showarticletitle{Automated triaging of very large bug repositories}.
\newblock \bibinfo{journal}{\emph{Information and software technology}}  \bibinfo{volume}{89} (\bibinfo{year}{2017}), \bibinfo{pages}{1--13}.
\newblock


\bibitem[Chaparro et~al\mbox{.}(2017)]%
        {chaparro2017detecting}
\bibfield{author}{\bibinfo{person}{Oscar Chaparro}, \bibinfo{person}{Jing Lu}, \bibinfo{person}{Fiorella Zampetti}, \bibinfo{person}{Laura Moreno}, \bibinfo{person}{Massimiliano Di~Penta}, \bibinfo{person}{Andrian Marcus}, \bibinfo{person}{Gabriele Bavota}, {and} \bibinfo{person}{Vincent Ng}.} \bibinfo{year}{2017}\natexlab{}.
\newblock \showarticletitle{Detecting missing information in bug descriptions}. In \bibinfo{booktitle}{\emph{Proceedings of the 2017 11th joint meeting on foundations of software engineering}}. \bibinfo{pages}{396--407}.
\newblock


\bibitem[Chen et~al\mbox{.}(2022)]%
        {DBLP:conf/qrs/ChenYYKZ22}
\bibfield{author}{\bibinfo{person}{Hao Chen}, \bibinfo{person}{Haiyang Yang}, \bibinfo{person}{Zilun Yan}, \bibinfo{person}{Li Kuang}, {and} \bibinfo{person}{Lingyan Zhang}.} \bibinfo{year}{2022}\natexlab{}.
\newblock \showarticletitle{{CGMBL:} Combining {GAN} and Method Name for Bug Localization}. In \bibinfo{booktitle}{\emph{22nd {IEEE} International Conference on Software Quality, Reliability and Security, {QRS} 2022, Guangzhou, China, December 5-9, 2022}}. \bibinfo{pages}{231--241}.
\newblock


\bibitem[Chen and Chen(2026)]%
        {DBLP:conf/icse/ChenChen26}
\bibfield{author}{\bibinfo{person}{Pengzhou Chen} {and} \bibinfo{person}{Tao Chen}.} \bibinfo{year}{2026}\natexlab{}.
\newblock \showarticletitle{PromiseTune: Unveiling Causally Promising and Explainable Configuration Tuning}. In \bibinfo{booktitle}{\emph{48th {IEEE/ACM} International Conference on Software Engineering (ICSE)}}. \bibinfo{publisher}{{ACM}}.
\newblock


\bibitem[Chen et~al\mbox{.}(2024)]%
        {DBLP:journals/corr/abs-2112-07303}
\bibfield{author}{\bibinfo{person}{Pengzhou Chen}, \bibinfo{person}{Tao Chen}, {and} \bibinfo{person}{Miqing Li}.} \bibinfo{year}{2024}\natexlab{}.
\newblock \showarticletitle{{MMO:} Meta Multi-Objectivization for Software Configuration Tuning}.
\newblock \bibinfo{journal}{\emph{{IEEE} Trans. Software Eng.}} \bibinfo{volume}{50}, \bibinfo{number}{6} (\bibinfo{year}{2024}), \bibinfo{pages}{1478--1504}.
\newblock
\urldef\tempurl%
\url{https://doi.org/10.1109/TSE.2024.3388910}
\showDOI{\tempurl}


\bibitem[Chen et~al\mbox{.}(2025)]%
        {10832565}
\bibfield{author}{\bibinfo{person}{Pengzhou Chen}, \bibinfo{person}{Jingzhi Gong}, {and} \bibinfo{person}{Tao Chen}.} \bibinfo{year}{2025}\natexlab{}.
\newblock \showarticletitle{Accuracy Can Lie: On the Impact of Surrogate Model in Configuration Tuning}.
\newblock \bibinfo{journal}{\emph{IEEE Transactions on Software Engineering}} \bibinfo{volume}{51}, \bibinfo{number}{2} (\bibinfo{year}{2025}), \bibinfo{pages}{548--580}.
\newblock
\urldef\tempurl%
\url{https://doi.org/10.1109/TSE.2025.3525955}
\showDOI{\tempurl}


\bibitem[Chen et~al\mbox{.}(2020)]%
        {DBLP:conf/kbse/ChenXYJCX20}
\bibfield{author}{\bibinfo{person}{Songqiang Chen}, \bibinfo{person}{Xiaoyuan Xie}, \bibinfo{person}{Bangguo Yin}, \bibinfo{person}{Yuanxiang Ji}, \bibinfo{person}{Lin Chen}, {and} \bibinfo{person}{Baowen Xu}.} \bibinfo{year}{2020}\natexlab{}.
\newblock \showarticletitle{Stay Professional and Efficient: Automatically Generate Titles for Your Bug Reports}. In \bibinfo{booktitle}{\emph{35th {IEEE/ACM} ASE}}.
\newblock


\bibitem[Chen(2022)]%
        {DBLP:conf/wcre/Chen22}
\bibfield{author}{\bibinfo{person}{Tao Chen}.} \bibinfo{year}{2022}\natexlab{}.
\newblock \showarticletitle{Lifelong Dynamic Optimization for Self-Adaptive Systems: Fact or Fiction?}. In \bibinfo{booktitle}{\emph{{IEEE} International Conference on Software Analysis, Evolution and Reengineering, {SANER} 2022, Honolulu, HI, USA, March 15-18, 2022}}. \bibinfo{publisher}{{IEEE}}, \bibinfo{pages}{78--89}.
\newblock
\urldef\tempurl%
\url{https://doi.org/10.1109/SANER53432.2022.00022}
\showDOI{\tempurl}


\bibitem[Chen and Bahsoon(2017a)]%
        {DBLP:journals/tse/ChenB17}
\bibfield{author}{\bibinfo{person}{Tao Chen} {and} \bibinfo{person}{Rami Bahsoon}.} \bibinfo{year}{2017}\natexlab{a}.
\newblock \showarticletitle{Self-Adaptive and Online QoS Modeling for Cloud-Based Software Services}.
\newblock \bibinfo{journal}{\emph{{IEEE} Trans. Software Eng.}} \bibinfo{volume}{43}, \bibinfo{number}{5} (\bibinfo{year}{2017}), \bibinfo{pages}{453--475}.
\newblock
\urldef\tempurl%
\url{https://doi.org/10.1109/TSE.2016.2608826}
\showDOI{\tempurl}


\bibitem[Chen and Bahsoon(2017b)]%
        {DBLP:journals/tsc/ChenB17}
\bibfield{author}{\bibinfo{person}{Tao Chen} {and} \bibinfo{person}{Rami Bahsoon}.} \bibinfo{year}{2017}\natexlab{b}.
\newblock \showarticletitle{Self-Adaptive Trade-off Decision Making for Autoscaling Cloud-Based Services}.
\newblock \bibinfo{journal}{\emph{{IEEE} Trans. Serv. Comput.}} \bibinfo{volume}{10}, \bibinfo{number}{4} (\bibinfo{year}{2017}), \bibinfo{pages}{618--632}.
\newblock
\urldef\tempurl%
\url{https://doi.org/10.1109/TSC.2015.2499770}
\showDOI{\tempurl}


\bibitem[Chen and Guestrin(2016)]%
        {DBLP:journals/corr/ChenG16}
\bibfield{author}{\bibinfo{person}{Tianqi Chen} {and} \bibinfo{person}{Carlos Guestrin}.} \bibinfo{year}{2016}\natexlab{}.
\newblock \showarticletitle{XGBoost: {A} Scalable Tree Boosting System}.
\newblock  (\bibinfo{year}{2016}), \bibinfo{pages}{785--794}.
\newblock


\bibitem[Chen et~al\mbox{.}(2018)]%
        {DBLP:journals/tosem/ChenLBY18}
\bibfield{author}{\bibinfo{person}{Tao Chen}, \bibinfo{person}{Ke Li}, \bibinfo{person}{Rami Bahsoon}, {and} \bibinfo{person}{Xin Yao}.} \bibinfo{year}{2018}\natexlab{}.
\newblock \showarticletitle{{FEMOSAA:} Feature-Guided and Knee-Driven Multi-Objective Optimization for Self-Adaptive Software}.
\newblock \bibinfo{journal}{\emph{{ACM} Trans. Softw. Eng. Methodol.}} \bibinfo{volume}{27}, \bibinfo{number}{2} (\bibinfo{year}{2018}), \bibinfo{pages}{5:1--5:50}.
\newblock
\urldef\tempurl%
\url{https://doi.org/10.1145/3204459}
\showDOI{\tempurl}


\bibitem[Chen and Li(2023a)]%
        {DBLP:journals/tosem/ChenL23a}
\bibfield{author}{\bibinfo{person}{Tao Chen} {and} \bibinfo{person}{Miqing Li}.} \bibinfo{year}{2023}\natexlab{a}.
\newblock \showarticletitle{Do Performance Aspirations Matter for Guiding Software Configuration Tuning? An Empirical Investigation under Dual Performance Objectives}.
\newblock \bibinfo{journal}{\emph{{ACM} Trans. Softw. Eng. Methodol.}} \bibinfo{volume}{32}, \bibinfo{number}{3} (\bibinfo{year}{2023}), \bibinfo{pages}{68:1--68:41}.
\newblock
\urldef\tempurl%
\url{https://doi.org/10.1145/3571853}
\showDOI{\tempurl}


\bibitem[Chen and Li(2023b)]%
        {DBLP:journals/tosem/ChenL23}
\bibfield{author}{\bibinfo{person}{Tao Chen} {and} \bibinfo{person}{Miqing Li}.} \bibinfo{year}{2023}\natexlab{b}.
\newblock \showarticletitle{The Weights Can Be Harmful: Pareto Search versus Weighted Search in Multi-objective Search-based Software Engineering}.
\newblock \bibinfo{journal}{\emph{{ACM} Trans. Softw. Eng. Methodol.}} \bibinfo{volume}{32}, \bibinfo{number}{1} (\bibinfo{year}{2023}), \bibinfo{pages}{5:1--5:40}.
\newblock
\urldef\tempurl%
\url{https://doi.org/10.1145/3514233}
\showDOI{\tempurl}


\bibitem[Chen and Li(2024)]%
        {DBLP:conf/sigsoft/0001L24}
\bibfield{author}{\bibinfo{person}{Tao Chen} {and} \bibinfo{person}{Miqing Li}.} \bibinfo{year}{2024}\natexlab{}.
\newblock \showarticletitle{Adapting Multi-objectivized Software Configuration Tuning}.
\newblock \bibinfo{journal}{\emph{Proc. {ACM} Softw. Eng.}} \bibinfo{volume}{1}, \bibinfo{number}{{FSE}} (\bibinfo{year}{2024}), \bibinfo{pages}{539--561}.
\newblock
\urldef\tempurl%
\url{https://doi.org/10.1145/3643751}
\showDOI{\tempurl}


\bibitem[Ciancarini et~al\mbox{.}(2017)]%
        {DBLP:conf/ijcnn/CiancariniPRS17}
\bibfield{author}{\bibinfo{person}{Paolo Ciancarini}, \bibinfo{person}{Francesco Poggi}, \bibinfo{person}{Davide Rossi}, {and} \bibinfo{person}{Alberto Sillitti}.} \bibinfo{year}{2017}\natexlab{}.
\newblock \showarticletitle{Analyzing and predicting concurrency bugs in open source systems}. In \bibinfo{booktitle}{\emph{2017 International Joint Conference on Neural Networks, {IJCNN} 2017 Anchorage, AK, USA, May 14-19, 2017}}. \bibinfo{pages}{721--728}.
\newblock


\bibitem[Cooper et~al\mbox{.}(2021)]%
        {DBLP:conf/icse/CooperBCMP21}
\bibfield{author}{\bibinfo{person}{Nathan Cooper}, \bibinfo{person}{Carlos Bernal{-}C{\'{a}}rdenas}, \bibinfo{person}{Oscar Chaparro}, \bibinfo{person}{Kevin Moran}, {and} \bibinfo{person}{Denys Poshyvanyk}.} \bibinfo{year}{2021}\natexlab{}.
\newblock \showarticletitle{It Takes Two to {TANGO:} Combining Visual and Textual Information for Detecting Duplicate Video-Based Bug Reports}. In \bibinfo{booktitle}{\emph{43rd {ICSE}}}.
\newblock


\bibitem[Dai et~al\mbox{.}(2023)]%
        {DBLP:journals/jss/DaiLXLWZ23}
\bibfield{author}{\bibinfo{person}{Jie Dai}, \bibinfo{person}{Qingshan Li}, \bibinfo{person}{Hui Xue}, \bibinfo{person}{Zhao Luo}, \bibinfo{person}{Yinglin Wang}, {and} \bibinfo{person}{Siyuan Zhan}.} \bibinfo{year}{2023}\natexlab{}.
\newblock \showarticletitle{Graph collaborative filtering-based bug triaging}.
\newblock \bibinfo{journal}{\emph{Journal of Systems and Software}}  \bibinfo{volume}{200} (\bibinfo{year}{2023}), \bibinfo{pages}{111667}.
\newblock


\bibitem[Deshmukh et~al\mbox{.}(2017)]%
        {deshmukh2017towards}
\bibfield{author}{\bibinfo{person}{Jayati Deshmukh}, \bibinfo{person}{Sanjay Podder}, \bibinfo{person}{Shubhashis Sengupta}, \bibinfo{person}{Neville Dubash}, {et~al\mbox{.}}} \bibinfo{year}{2017}\natexlab{}.
\newblock \showarticletitle{Towards accurate duplicate bug retrieval using deep learning techniques}. In \bibinfo{booktitle}{\emph{2017 IEEE International conference on software maintenance and evolution (ICSME)}}. \bibinfo{pages}{115--124}.
\newblock


\bibitem[Dipongkor et~al\mbox{.}(2023)]%
        {DBLP:journals/access/DipongkorIHYM23}
\bibfield{author}{\bibinfo{person}{Atish~Kumar Dipongkor}, \bibinfo{person}{Md.~Saiful Islam}, \bibinfo{person}{Ishtiaque Hussain}, \bibinfo{person}{Sira Yongchareon}, {and} \bibinfo{person}{Sajib Mistry}.} \bibinfo{year}{2023}\natexlab{}.
\newblock \showarticletitle{On Fusing Artificial and Convolutional Neural Network Features for Automatic Bug Assignments}.
\newblock \bibinfo{journal}{\emph{{IEEE} Access}}  \bibinfo{volume}{11} (\bibinfo{year}{2023}), \bibinfo{pages}{49493--49508}.
\newblock


\bibitem[Du and Chen(2024)]%
        {DBLP:conf/esem/DuC24}
\bibfield{author}{\bibinfo{person}{Chengwen Du} {and} \bibinfo{person}{Tao Chen}.} \bibinfo{year}{2024}\natexlab{}.
\newblock \showarticletitle{Contexts Matter: An Empirical Study on Contextual Influence in Fairness Testing for Deep Learning Systems}. In \bibinfo{booktitle}{\emph{Proceedings of the 18th {ACM/IEEE} International Symposium on Empirical Software Engineering and Measurement, {ESEM} 2024, Barcelona, Spain, October 24-25, 2024}}, \bibfield{editor}{\bibinfo{person}{Xavier Franch}, \bibinfo{person}{Maya Daneva}, \bibinfo{person}{Silverio Mart{\'{\i}}nez{-}Fern{\'{a}}ndez}, {and} \bibinfo{person}{Luigi Quaranta}} (Eds.). \bibinfo{publisher}{{ACM}}, \bibinfo{pages}{107--118}.
\newblock
\urldef\tempurl%
\url{https://doi.org/10.1145/3674805.3686673}
\showDOI{\tempurl}


\bibitem[Du et~al\mbox{.}(2022)]%
        {DBLP:journals/tr/DuZXZT22}
\bibfield{author}{\bibinfo{person}{Xiaoting Du}, \bibinfo{person}{Zheng Zheng}, \bibinfo{person}{Guanping Xiao}, \bibinfo{person}{Zenghui Zhou}, {and} \bibinfo{person}{Kishor~S. Trivedi}.} \bibinfo{year}{2022}\natexlab{}.
\newblock \showarticletitle{DeepSIM: Deep Semantic Information-Based Automatic Mandelbug Classification}.
\newblock \bibinfo{journal}{\emph{{IEEE} Transactions on Reliability}}  \bibinfo{volume}{71} (\bibinfo{year}{2022}), \bibinfo{pages}{1540--1554}.
\newblock


\bibitem[Elmishali and Kalech(2023)]%
        {DBLP:journals/infsof/ElmishaliK23}
\bibfield{author}{\bibinfo{person}{Amir Elmishali} {and} \bibinfo{person}{Meir Kalech}.} \bibinfo{year}{2023}\natexlab{}.
\newblock \showarticletitle{Issues-Driven features for software fault prediction}.
\newblock \bibinfo{journal}{\emph{Information and Software Technology}}  \bibinfo{volume}{155} (\bibinfo{year}{2023}), \bibinfo{pages}{107102}.
\newblock


\bibitem[Fan et~al\mbox{.}(2017)]%
        {DBLP:conf/esem/FanYYWW17}
\bibfield{author}{\bibinfo{person}{Qiang Fan}, \bibinfo{person}{Yue Yu}, \bibinfo{person}{Gang Yin}, \bibinfo{person}{Tao Wang}, {and} \bibinfo{person}{Huaimin Wang}.} \bibinfo{year}{2017}\natexlab{}.
\newblock \showarticletitle{Where Is the Road for Issue Reports Classification Based on Text Mining?}. In \bibinfo{booktitle}{\emph{{ESEM} 2017}}.
\newblock


\bibitem[Fan et~al\mbox{.}(2020)]%
        {DBLP:journals/tse/FanXLH20}
\bibfield{author}{\bibinfo{person}{Yuanrui Fan}, \bibinfo{person}{Xin Xia}, \bibinfo{person}{David Lo}, {and} \bibinfo{person}{Ahmed~E. Hassan}.} \bibinfo{year}{2020}\natexlab{}.
\newblock \showarticletitle{Chaff from the Wheat: Characterizing and Determining Valid Bug Reports}.
\newblock \bibinfo{journal}{\emph{{IEEE} Transactions on Software Engineering}}  \bibinfo{volume}{46} (\bibinfo{year}{2020}), \bibinfo{pages}{495--525}.
\newblock


\bibitem[Fang et~al\mbox{.}(2023)]%
        {DBLP:conf/icse/FangZTJXS23}
\bibfield{author}{\bibinfo{person}{Sen Fang}, \bibinfo{person}{Tao Zhang}, \bibinfo{person}{Youshuai Tan}, \bibinfo{person}{He Jiang}, \bibinfo{person}{Xin Xia}, {and} \bibinfo{person}{Xiaobing Sun}.} \bibinfo{year}{2023}\natexlab{}.
\newblock \showarticletitle{RepresentThemAll: {A} Universal Learning Representation of Bug Reports}. In \bibinfo{booktitle}{\emph{45th {ICSE}}}.
\newblock


\bibitem[Feng et~al\mbox{.}(2020)]%
        {DBLP:conf/emnlp/FengGTDFGS0LJZ20}
\bibfield{author}{\bibinfo{person}{Zhangyin Feng}, \bibinfo{person}{Daya Guo}, \bibinfo{person}{Duyu Tang}, \bibinfo{person}{Nan Duan}, \bibinfo{person}{Xiaocheng Feng}, \bibinfo{person}{Ming Gong}, \bibinfo{person}{Linjun Shou}, \bibinfo{person}{Bing Qin}, \bibinfo{person}{Ting Liu}, \bibinfo{person}{Daxin Jiang}, {and} \bibinfo{person}{Ming Zhou}.} \bibinfo{year}{2020}\natexlab{}.
\newblock \showarticletitle{CodeBERT: {A} Pre-Trained Model for Programming and Natural Languages}. In \bibinfo{booktitle}{\emph{Findings of the Association for Computational Linguistics: {EMNLP} 2020, Online Event, 16-20 November 2020}} \emph{(\bibinfo{series}{Findings of {ACL}}, Vol.~\bibinfo{volume}{{EMNLP} 2020})}, \bibfield{editor}{\bibinfo{person}{Trevor Cohn}, \bibinfo{person}{Yulan He}, {and} \bibinfo{person}{Yang Liu}} (Eds.). \bibinfo{publisher}{Association for Computational Linguistics}, \bibinfo{pages}{1536--1547}.
\newblock


\bibitem[Florea et~al\mbox{.}(2017)]%
        {DBLP:conf/icaisc/FloreaAA17}
\bibfield{author}{\bibinfo{person}{Adrian{-}Catalin Florea}, \bibinfo{person}{John Anvik}, {and} \bibinfo{person}{Razvan Andonie}.} \bibinfo{year}{2017}\natexlab{}.
\newblock \showarticletitle{Spark-Based Cluster Implementation of a Bug Report Assignment Recommender System}. In \bibinfo{booktitle}{\emph{ICAISC 2017}}, Vol.~\bibinfo{volume}{10246}. \bibinfo{pages}{31--42}.
\newblock


\bibitem[Florez et~al\mbox{.}(2021)]%
        {DBLP:conf/wcre/FlorezCTM21}
\bibfield{author}{\bibinfo{person}{Juan~Manuel Florez}, \bibinfo{person}{Oscar Chaparro}, \bibinfo{person}{Christoph Treude}, {and} \bibinfo{person}{Andrian Marcus}.} \bibinfo{year}{2021}\natexlab{}.
\newblock \showarticletitle{Combining Query Reduction and Expansion for Text-Retrieval-Based Bug Localization}. In \bibinfo{booktitle}{\emph{{SANER} 2021}}. \bibinfo{pages}{166--176}.
\newblock


\bibitem[Forman and Scholz(2010)]%
        {DBLP:journals/sigkdd/FormanS10}
\bibfield{author}{\bibinfo{person}{George Forman} {and} \bibinfo{person}{Martin Scholz}.} \bibinfo{year}{2010}\natexlab{}.
\newblock \showarticletitle{Apples-to-apples in cross-validation studies: pitfalls in classifier performance measurement}.
\newblock \bibinfo{journal}{\emph{{SIGKDD} Explorations}}  \bibinfo{volume}{12} (\bibinfo{year}{2010}), \bibinfo{pages}{49--57}.
\newblock


\bibitem[Freund and Schapire(1995)]%
        {freund1995desicion}
\bibfield{author}{\bibinfo{person}{Yoav Freund} {and} \bibinfo{person}{Robert~E Schapire}.} \bibinfo{year}{1995}\natexlab{}.
\newblock \showarticletitle{A desicion-theoretic generalization of on-line learning and an application to boosting}. In \bibinfo{booktitle}{\emph{European conference on computational learning theory}}. \bibinfo{pages}{23--37}.
\newblock


\bibitem[Friedman(2001)]%
        {friedman2001greedy}
\bibfield{author}{\bibinfo{person}{Jerome~H Friedman}.} \bibinfo{year}{2001}\natexlab{}.
\newblock \showarticletitle{Greedy function approximation: a gradient boosting machine}.
\newblock \bibinfo{journal}{\emph{Annals of statistics}} (\bibinfo{year}{2001}), \bibinfo{pages}{1189--1232}.
\newblock


\bibitem[Fukuda et~al\mbox{.}(2022)]%
        {DBLP:conf/noms/FukudaWHT22}
\bibfield{author}{\bibinfo{person}{Nobukazu Fukuda}, \bibinfo{person}{Chao Wu}, \bibinfo{person}{Shingo Horiuchi}, {and} \bibinfo{person}{Kenichi Tayama}.} \bibinfo{year}{2022}\natexlab{}.
\newblock \showarticletitle{Fault Report Generation for {ICT} Systems by Jointly Learning Time-series and Text Data}. In \bibinfo{booktitle}{\emph{2022 {IEEE/IFIP} Network Operations and Management Symposium, {NOMS} 2022, Budapest, Hungary, April 25-29, 2022}}. \bibinfo{pages}{1--9}.
\newblock


\bibitem[Gao et~al\mbox{.}(2018)]%
        {DBLP:conf/bdccf/GaoL0GG18}
\bibfield{author}{\bibinfo{person}{Guofeng Gao}, \bibinfo{person}{Hui Li}, \bibinfo{person}{Rong Chen}, \bibinfo{person}{Xin Ge}, {and} \bibinfo{person}{Shikai Guo}.} \bibinfo{year}{2018}\natexlab{}.
\newblock \showarticletitle{Identification of High Priority Bug Reports via Integration Method}. In \bibinfo{booktitle}{\emph{Big Data - 6th {CCF} Conference, Big Data 2018, Xi'an, China, October 11-13, 2018, Proceedings}}, Vol.~\bibinfo{volume}{945}. \bibinfo{pages}{336--349}.
\newblock


\bibitem[Garcia et~al\mbox{.}(2018)]%
        {DBLP:journals/jss/GarciaSN18}
\bibfield{author}{\bibinfo{person}{Harold~Valdivia Garcia}, \bibinfo{person}{Emad Shihab}, {and} \bibinfo{person}{Meiyappan Nagappan}.} \bibinfo{year}{2018}\natexlab{}.
\newblock \showarticletitle{Characterizing and predicting blocking bugs in open source projects}.
\newblock \bibinfo{journal}{\emph{Journal of Systems and Software}}  \bibinfo{volume}{143} (\bibinfo{year}{2018}), \bibinfo{pages}{44--58}.
\newblock


\bibitem[Gawron et~al\mbox{.}(2017)]%
        {DBLP:conf/crisis/Gawron0M17}
\bibfield{author}{\bibinfo{person}{Marian Gawron}, \bibinfo{person}{Feng Cheng}, {and} \bibinfo{person}{Christoph Meinel}.} \bibinfo{year}{2017}\natexlab{}.
\newblock \showarticletitle{Automatic Vulnerability Classification Using Machine Learning}. In \bibinfo{booktitle}{\emph{Risks and Security of Internet and Systems - 12th International Conference CRiSIS 2017, Dinard, France, September 19-21, 2017, Revised Selected Papers}}, Vol.~\bibinfo{volume}{10694}. \bibinfo{pages}{3--17}.
\newblock


\bibitem[Ge et~al\mbox{.}(2022)]%
        {DBLP:journals/infsof/GeFQGQ22}
\bibfield{author}{\bibinfo{person}{Xiuting Ge}, \bibinfo{person}{Chunrong Fang}, \bibinfo{person}{Meiyuan Qian}, \bibinfo{person}{Yu Ge}, {and} \bibinfo{person}{Mingshuang Qing}.} \bibinfo{year}{2022}\natexlab{}.
\newblock \showarticletitle{Locality-based security bug report identification via active learning}.
\newblock \bibinfo{journal}{\emph{Information and Software Technology}}  \bibinfo{volume}{147} (\bibinfo{year}{2022}), \bibinfo{pages}{106899}.
\newblock


\bibitem[Gomes et~al\mbox{.}(2019)]%
        {DBLP:journals/infsof/GomesTC19}
\bibfield{author}{\bibinfo{person}{Luiz Alberto~Ferreira Gomes}, \bibinfo{person}{Ricardo da Silva~Torres}, {and} \bibinfo{person}{Mario~L{\'{u}}cio C{\^{o}}rtes}.} \bibinfo{year}{2019}\natexlab{}.
\newblock \showarticletitle{Bug report severity level prediction in open source software: {A survey and research opportunities}}.
\newblock \bibinfo{journal}{\emph{Information and Software Technology}}  \bibinfo{volume}{115} (\bibinfo{year}{2019}), \bibinfo{pages}{58--78}.
\newblock


\bibitem[Gomes et~al\mbox{.}(2021)]%
        {DBLP:journals/infsof/GomesTC21}
\bibfield{author}{\bibinfo{person}{Luiz Alberto~Ferreira Gomes}, \bibinfo{person}{Ricardo da Silva~Torres}, {and} \bibinfo{person}{Mario~L{\'{u}}cio C{\^{o}}rtes}.} \bibinfo{year}{2021}\natexlab{}.
\newblock \showarticletitle{On the prediction of long-lived bugs: An analysis and comparative study using {FLOSS} projects}.
\newblock \bibinfo{journal}{\emph{Information and Software Technology}}  \bibinfo{volume}{132} (\bibinfo{year}{2021}), \bibinfo{pages}{106508}.
\newblock


\bibitem[Gong and Chen(2023)]%
        {DBLP:conf/sigsoft/Gong023}
\bibfield{author}{\bibinfo{person}{Jingzhi Gong} {and} \bibinfo{person}{Tao Chen}.} \bibinfo{year}{2023}\natexlab{}.
\newblock \showarticletitle{Predicting Software Performance with Divide-and-Learn}. In \bibinfo{booktitle}{\emph{Proceedings of the 31st {ACM} Joint European Software Engineering Conference and Symposium on the Foundations of Software Engineering, {ESEC/FSE} 2023, San Francisco, CA, USA, December 3-9, 2023}}. \bibinfo{pages}{858--870}.
\newblock
\urldef\tempurl%
\url{https://doi.org/10.1145/3611643.3616334}
\showDOI{\tempurl}


\bibitem[Gong and Chen(2024)]%
        {DBLP:journals/pacmse/Gong024}
\bibfield{author}{\bibinfo{person}{Jingzhi Gong} {and} \bibinfo{person}{Tao Chen}.} \bibinfo{year}{2024}\natexlab{}.
\newblock \showarticletitle{Predicting Configuration Performance in Multiple Environments with Sequential Meta-Learning}.
\newblock \bibinfo{journal}{\emph{Proceedings of {ACM} Software Engineering}} \bibinfo{volume}{1}, \bibinfo{number}{{FSE}} (\bibinfo{year}{2024}), \bibinfo{pages}{359--382}.
\newblock
\urldef\tempurl%
\url{https://doi.org/10.1145/3643743}
\showDOI{\tempurl}


\bibitem[Gong and Chen(2025)]%
        {DBLP:journals/tosem/GongC25}
\bibfield{author}{\bibinfo{person}{Jingzhi Gong} {and} \bibinfo{person}{Tao Chen}.} \bibinfo{year}{2025}\natexlab{}.
\newblock \showarticletitle{Deep Configuration Performance Learning: {A} Systematic Survey and Taxonomy}.
\newblock \bibinfo{journal}{\emph{{ACM} Trans. Softw. Eng. Methodol.}} \bibinfo{volume}{34}, \bibinfo{number}{1} (\bibinfo{year}{2025}), \bibinfo{pages}{25:1--25:62}.
\newblock
\urldef\tempurl%
\url{https://doi.org/10.1145/3702986}
\showDOI{\tempurl}


\bibitem[Gong et~al\mbox{.}(2025)]%
        {DBLP:journals/tse/GongCB25}
\bibfield{author}{\bibinfo{person}{Jingzhi Gong}, \bibinfo{person}{Tao Chen}, {and} \bibinfo{person}{Rami Bahsoon}.} \bibinfo{year}{2025}\natexlab{}.
\newblock \showarticletitle{Dividable Configuration Performance Learning}.
\newblock \bibinfo{journal}{\emph{{IEEE} Trans. Software Eng.}} \bibinfo{volume}{51}, \bibinfo{number}{1} (\bibinfo{year}{2025}), \bibinfo{pages}{106--134}.
\newblock
\urldef\tempurl%
\url{https://doi.org/10.1109/TSE.2024.3491945}
\showDOI{\tempurl}


\bibitem[Guo et~al\mbox{.}(2018)]%
        {DBLP:journals/access/Guo0W0L18}
\bibfield{author}{\bibinfo{person}{Shikai Guo}, \bibinfo{person}{Rong Chen}, \bibinfo{person}{Miaomiao Wei}, \bibinfo{person}{Hui Li}, {and} \bibinfo{person}{Yaqing Liu}.} \bibinfo{year}{2018}\natexlab{}.
\newblock \showarticletitle{Ensemble Data Reduction Techniques and Multi-RSMOTE via Fuzzy Integral for Bug Report Classification}.
\newblock \bibinfo{journal}{\emph{{IEEE} Access}} (\bibinfo{year}{2018}).
\newblock


\bibitem[Gupta et~al\mbox{.}(2022)]%
        {DBLP:journals/jksucis/GuptaIF22}
\bibfield{author}{\bibinfo{person}{Chetna Gupta}, \bibinfo{person}{Pedro R.~M. In{\'{a}}cio}, {and} \bibinfo{person}{M{\'{a}}rio~M. Freire}.} \bibinfo{year}{2022}\natexlab{}.
\newblock \showarticletitle{Improving software maintenance with improved bug triaging}.
\newblock \bibinfo{journal}{\emph{Journal of King Saud University-Computer and Information Sciences}}  \bibinfo{volume}{34} (\bibinfo{year}{2022}), \bibinfo{pages}{8757--8764}.
\newblock


\bibitem[Gupta and Gupta(2021)]%
        {DBLP:journals/eswa/GuptaG21}
\bibfield{author}{\bibinfo{person}{Som Gupta} {and} \bibinfo{person}{Sanjai~Kumar Gupta}.} \bibinfo{year}{2021}\natexlab{}.
\newblock \showarticletitle{An approach to generate the bug report summaries using two-level feature extraction}.
\newblock \bibinfo{journal}{\emph{Expert Systems with Applications}}  \bibinfo{volume}{176} (\bibinfo{year}{2021}), \bibinfo{pages}{114816}.
\newblock


\bibitem[Haering et~al\mbox{.}(2021)]%
        {DBLP:conf/icse/HaeringSM21}
\bibfield{author}{\bibinfo{person}{Marlo Haering}, \bibinfo{person}{Christoph Stanik}, {and} \bibinfo{person}{Walid Maalej}.} \bibinfo{year}{2021}\natexlab{}.
\newblock \showarticletitle{Automatically Matching Bug Reports With Related App Reviews}. In \bibinfo{booktitle}{\emph{43rd {IEEE/ACM} International Conference on Software Engineering {ICSE} 2021, Madrid, Spain, 22-30 May 2021}}. \bibinfo{pages}{970--981}.
\newblock


\bibitem[He et~al\mbox{.}(2020a)]%
        {DBLP:conf/issre/HeXF0YL20}
\bibfield{author}{\bibinfo{person}{Jianjun He}, \bibinfo{person}{Ling Xu}, \bibinfo{person}{Yuanrui Fan}, \bibinfo{person}{Zhou Xu}, \bibinfo{person}{Meng Yan}, {and} \bibinfo{person}{Yan Lei}.} \bibinfo{year}{2020}\natexlab{a}.
\newblock \showarticletitle{Deep Learning Based Valid Bug Reports Determination and Explanation}. In \bibinfo{booktitle}{\emph{31st {ISSRE} 2020, Coimbra, Portugal, October 12-15, 2020}}. \bibinfo{pages}{184--194}.
\newblock


\bibitem[He et~al\mbox{.}(2020b)]%
        {he2020duplicate}
\bibfield{author}{\bibinfo{person}{Jianjun He}, \bibinfo{person}{Ling Xu}, \bibinfo{person}{Meng Yan}, \bibinfo{person}{Xin Xia}, {and} \bibinfo{person}{Yan Lei}.} \bibinfo{year}{2020}\natexlab{b}.
\newblock \showarticletitle{Duplicate bug report detection using dual-channel convolutional neural networks}. In \bibinfo{booktitle}{\emph{Proceedings of the 28th International Conference on Program Comprehension}}. \bibinfo{pages}{117--127}.
\newblock


\bibitem[Hindle et~al\mbox{.}(2016)]%
        {DBLP:journals/ese/HindleAS16}
\bibfield{author}{\bibinfo{person}{Abram Hindle}, \bibinfo{person}{Anahita Alipour}, {and} \bibinfo{person}{Eleni Stroulia}.} \bibinfo{year}{2016}\natexlab{}.
\newblock \showarticletitle{A contextual approach towards more accurate duplicate bug report detection and ranking}.
\newblock \bibinfo{journal}{\emph{Empirical Software Engineering}}  \bibinfo{volume}{21} (\bibinfo{year}{2016}), \bibinfo{pages}{368--410}.
\newblock


\bibitem[Hirakawa et~al\mbox{.}(2020)]%
        {DBLP:conf/iccel/HirakawaTN20}
\bibfield{author}{\bibinfo{person}{Rin Hirakawa}, \bibinfo{person}{Keitaro Tominaga}, {and} \bibinfo{person}{Yoshihisa Nakatoh}.} \bibinfo{year}{2020}\natexlab{}.
\newblock \showarticletitle{Study on Automatic Defect Report Classification System with Self Attention Visualization}. In \bibinfo{booktitle}{\emph{2020 {IEEE} International Conference on Consumer Electronics (ICCE), Las Vegas, NV, USA, January 4-6, 2020}}. \bibinfo{publisher}{{IEEE}}, \bibinfo{pages}{1--2}.
\newblock


\bibitem[Hirsch and Hofer(2022a)]%
        {DBLP:journals/array/HirschH22}
\bibfield{author}{\bibinfo{person}{Thomas Hirsch} {and} \bibinfo{person}{Birgit Hofer}.} \bibinfo{year}{2022}\natexlab{a}.
\newblock \showarticletitle{Using textual bug reports to predict the fault category of software bugs}.
\newblock \bibinfo{journal}{\emph{Array}}  \bibinfo{volume}{15} (\bibinfo{year}{2022}), \bibinfo{pages}{100189}.
\newblock


\bibitem[Hirsch and Hofer(2022b)]%
        {hirsch2022using}
\bibfield{author}{\bibinfo{person}{Thomas Hirsch} {and} \bibinfo{person}{Birgit Hofer}.} \bibinfo{year}{2022}\natexlab{b}.
\newblock \showarticletitle{Using textual bug reports to predict the fault category of software bugs}.
\newblock \bibinfo{journal}{\emph{Array}}  \bibinfo{volume}{15} (\bibinfo{year}{2022}), \bibinfo{pages}{100189}.
\newblock


\bibitem[Hu et~al\mbox{.}(2014)]%
        {hu2014effective}
\bibfield{author}{\bibinfo{person}{Hao Hu}, \bibinfo{person}{Hongyu Zhang}, \bibinfo{person}{Jifeng Xuan}, {and} \bibinfo{person}{Weigang Sun}.} \bibinfo{year}{2014}\natexlab{}.
\newblock \showarticletitle{Effective bug triage based on historical bug-fix information}. In \bibinfo{booktitle}{\emph{2014 IEEE 25th International Symposium on Software Reliability Engineering}}. \bibinfo{pages}{122--132}.
\newblock


\bibitem[Huang et~al\mbox{.}(2022)]%
        {DBLP:conf/internetware/HuangSFYYZ22}
\bibfield{author}{\bibinfo{person}{Zijie Huang}, \bibinfo{person}{Zhiqing Shao}, \bibinfo{person}{Guisheng Fan}, \bibinfo{person}{Huiqun Yu}, \bibinfo{person}{Kang Yang}, {and} \bibinfo{person}{Ziyi Zhou}.} \bibinfo{year}{2022}\natexlab{}.
\newblock \showarticletitle{Bug Report Priority Prediction Using Developer-Oriented Socio-Technical Features}. In \bibinfo{booktitle}{\emph{Internetware 2022, Hohhot China, June 11 - 12, 2022}}. \bibinfo{pages}{202--211}.
\newblock


\bibitem[Huo et~al\mbox{.}(2021)]%
        {DBLP:journals/tse/HuoTLLS21}
\bibfield{author}{\bibinfo{person}{Xuan Huo}, \bibinfo{person}{Ferdian Thung}, \bibinfo{person}{Ming Li}, \bibinfo{person}{David Lo}, {and} \bibinfo{person}{Shu{-}Ting Shi}.} \bibinfo{year}{2021}\natexlab{}.
\newblock \showarticletitle{Deep Transfer Bug Localization}.
\newblock \bibinfo{journal}{\emph{{IEEE} Transactions on Software Engineering}}  \bibinfo{volume}{47} (\bibinfo{year}{2021}), \bibinfo{pages}{1368--1380}.
\newblock


\bibitem[Huo et~al\mbox{.}(2019)]%
        {huo2019deep}
\bibfield{author}{\bibinfo{person}{Xuan Huo}, \bibinfo{person}{Ferdian Thung}, \bibinfo{person}{Ming Li}, \bibinfo{person}{David Lo}, {and} \bibinfo{person}{Shu-Ting Shi}.} \bibinfo{year}{2019}\natexlab{}.
\newblock \showarticletitle{Deep transfer bug localization}.
\newblock \bibinfo{journal}{\emph{IEEE Transactions on Software Engineering}} (\bibinfo{year}{2019}).
\newblock


\bibitem[Isotani et~al\mbox{.}(2021)]%
        {DBLP:conf/icsm/IsotaniWFNOS21}
\bibfield{author}{\bibinfo{person}{Haruna Isotani}, \bibinfo{person}{Hironori Washizaki}, \bibinfo{person}{Yoshiaki Fukazawa}, \bibinfo{person}{Tsutomu Nomoto}, \bibinfo{person}{Saori Ouji}, {and} \bibinfo{person}{Shinobu Saito}.} \bibinfo{year}{2021}\natexlab{}.
\newblock \showarticletitle{Duplicate Bug Report Detection by Using Sentence Embedding and Fine-tuning}. In \bibinfo{booktitle}{\emph{{IEEE} International Conference on Software Maintenance and Evolution, {ICSME} 2021, Luxembourg, September 27 - October 1, 2021}}. \bibinfo{publisher}{{IEEE}}, \bibinfo{pages}{535--544}.
\newblock


\bibitem[Jahanshahi and Cevik(2022)]%
        {DBLP:journals/infsof/JahanshahiC22}
\bibfield{author}{\bibinfo{person}{Hadi Jahanshahi} {and} \bibinfo{person}{Mucahit Cevik}.} \bibinfo{year}{2022}\natexlab{}.
\newblock \showarticletitle{{S-DABT:} Schedule and Dependency-aware Bug Triage in open-source bug tracking systems}.
\newblock \bibinfo{journal}{\emph{Information and Software Technology}}  \bibinfo{volume}{151} (\bibinfo{year}{2022}), \bibinfo{pages}{107025}.
\newblock


\bibitem[Jahanshahi et~al\mbox{.}(2021)]%
        {jahanshahi2021dabt}
\bibfield{author}{\bibinfo{person}{Hadi Jahanshahi}, \bibinfo{person}{Kritika Chhabra}, \bibinfo{person}{Mucahit Cevik}, {and} \bibinfo{person}{Ay{\th}e Ba{\th}ar}.} \bibinfo{year}{2021}\natexlab{}.
\newblock \showarticletitle{DABT: A dependency-aware bug triaging method}. In \bibinfo{booktitle}{\emph{Proceedings of the 25th International Conference on Evaluation and Assessment in Software Engineering}}. \bibinfo{pages}{221--230}.
\newblock


\bibitem[Jain et~al\mbox{.}(2019a)]%
        {DBLP:journals/access/JainKSNT19}
\bibfield{author}{\bibinfo{person}{Deepak~Kumar Jain}, \bibinfo{person}{Akshi Kumar}, \bibinfo{person}{Saurabh~Raj Sangwan}, \bibinfo{person}{Gia~Nhu Nguyen}, {and} \bibinfo{person}{Prayag Tiwari}.} \bibinfo{year}{2019}\natexlab{a}.
\newblock \showarticletitle{A Particle Swarm Optimized Learning Model of Fault Classification in Web-Apps}.
\newblock \bibinfo{journal}{\emph{{IEEE} Access}}  \bibinfo{volume}{7} (\bibinfo{year}{2019}), \bibinfo{pages}{18480--18489}.
\newblock


\bibitem[Jain et~al\mbox{.}(2019b)]%
        {jain2019particle}
\bibfield{author}{\bibinfo{person}{Deepak~Kumar Jain}, \bibinfo{person}{Akshi Kumar}, \bibinfo{person}{Saurabh~Raj Sangwan}, \bibinfo{person}{Gia~Nhu Nguyen}, {and} \bibinfo{person}{Prayag Tiwari}.} \bibinfo{year}{2019}\natexlab{b}.
\newblock \showarticletitle{A particle swarm optimized learning model of fault classification in Web-Apps}.
\newblock \bibinfo{journal}{\emph{IEEE Access}}  \bibinfo{volume}{7} (\bibinfo{year}{2019}), \bibinfo{pages}{18480--18489}.
\newblock


\bibitem[Jayarajah et~al\mbox{.}(2016)]%
        {jayarajah2016duplicate}
\bibfield{author}{\bibinfo{person}{Kasthuri Jayarajah}, \bibinfo{person}{Meera Radhakrishnan}, {and} \bibinfo{person}{Camellia Zakaria}.} \bibinfo{year}{2016}\natexlab{}.
\newblock \showarticletitle{Duplicate issue detection for the Android open source project}. In \bibinfo{booktitle}{\emph{Proceedings of the 5th International Workshop on Software Mining}}. \bibinfo{pages}{24--31}.
\newblock


\bibitem[Ji et~al\mbox{.}(2018)]%
        {DBLP:conf/compsac/JiPCM18}
\bibfield{author}{\bibinfo{person}{Tao Ji}, \bibinfo{person}{Jinkun Pan}, \bibinfo{person}{Liqian Chen}, {and} \bibinfo{person}{Xiaoguang Mao}.} \bibinfo{year}{2018}\natexlab{}.
\newblock \showarticletitle{Identifying Supplementary Bug-fix Commits}. In \bibinfo{booktitle}{\emph{2018 {IEEE} 42nd Annual Computer Software and Applications Conference {COMPSAC} 2018, Tokyo, Japan, 23-27 July 2018, Volume 1}}. \bibinfo{pages}{184--193}.
\newblock


\bibitem[Jiang et~al\mbox{.}(2020)]%
        {jiang2020ltrwes}
\bibfield{author}{\bibinfo{person}{Yuan Jiang}, \bibinfo{person}{Pengcheng Lu}, \bibinfo{person}{Xiaohong Su}, {and} \bibinfo{person}{Tiantian Wang}.} \bibinfo{year}{2020}\natexlab{}.
\newblock \showarticletitle{LTRWES: A new framework for security bug report detection}.
\newblock \bibinfo{journal}{\emph{Information and Software Technology}}  \bibinfo{volume}{124} (\bibinfo{year}{2020}), \bibinfo{pages}{106314}.
\newblock


\bibitem[Jiang et~al\mbox{.}(2023)]%
        {DBLP:journals/jss/JiangSTSW23}
\bibfield{author}{\bibinfo{person}{Yuan Jiang}, \bibinfo{person}{Xiaohong Su}, \bibinfo{person}{Christoph Treude}, \bibinfo{person}{Chao Shang}, {and} \bibinfo{person}{Tiantian Wang}.} \bibinfo{year}{2023}\natexlab{}.
\newblock \showarticletitle{Does Deep Learning improve the performance of duplicate bug report detection? An empirical study}.
\newblock \bibinfo{journal}{\emph{Journal of Systems and Software}}  \bibinfo{volume}{198} (\bibinfo{year}{2023}), \bibinfo{pages}{111607}.
\newblock


\bibitem[Jindal and Kaur(2020)]%
        {DBLP:journals/access/JindalK20}
\bibfield{author}{\bibinfo{person}{Shubhra~Goyal Jindal} {and} \bibinfo{person}{Arvinder Kaur}.} \bibinfo{year}{2020}\natexlab{}.
\newblock \showarticletitle{Automatic Keyword and Sentence-Based Text Summarization for Software Bug Reports}.
\newblock \bibinfo{journal}{\emph{{IEEE} Access}}  \bibinfo{volume}{8} (\bibinfo{year}{2020}), \bibinfo{pages}{65352--65370}.
\newblock


\bibitem[Jonsson et~al\mbox{.}(2016a)]%
        {DBLP:journals/ese/JonssonBBSER16}
\bibfield{author}{\bibinfo{person}{Leif Jonsson}, \bibinfo{person}{Markus Borg}, \bibinfo{person}{David Broman}, \bibinfo{person}{Kristian Sandahl}, \bibinfo{person}{Sigrid Eldh}, {and} \bibinfo{person}{Per Runeson}.} \bibinfo{year}{2016}\natexlab{a}.
\newblock \showarticletitle{Automated bug assignment: Ensemble-based machine learning in large scale industrial contexts}.
\newblock \bibinfo{journal}{\emph{Empirical Software Engineering}}  \bibinfo{volume}{21} (\bibinfo{year}{2016}), \bibinfo{pages}{1533--1578}.
\newblock


\bibitem[Jonsson et~al\mbox{.}(2016b)]%
        {jonsson2016automated}
\bibfield{author}{\bibinfo{person}{Leif Jonsson}, \bibinfo{person}{Markus Borg}, \bibinfo{person}{David Broman}, \bibinfo{person}{Kristian Sandahl}, \bibinfo{person}{Sigrid Eldh}, {and} \bibinfo{person}{Per Runeson}.} \bibinfo{year}{2016}\natexlab{b}.
\newblock \showarticletitle{Automated bug assignment: Ensemble-based machine learning in large scale industrial contexts}.
\newblock \bibinfo{journal}{\emph{Empirical Software Engineering}}  \bibinfo{volume}{21} (\bibinfo{year}{2016}), \bibinfo{pages}{1533--1578}.
\newblock


\bibitem[Jonsson et~al\mbox{.}({[n.\,d.]})]%
        {DBLP:conf/qrs/JonssonBMSVE16}
\bibfield{author}{\bibinfo{person}{Leif Jonsson}, \bibinfo{person}{David Broman}, \bibinfo{person}{M{\aa}ns Magnusson}, \bibinfo{person}{Kristian Sandahl}, \bibinfo{person}{Mattias Villani}, {and} \bibinfo{person}{Sigrid Eldh}.} \bibinfo{year}{[n.\,d.]}\natexlab{}.
\newblock \showarticletitle{Automatic Localization of Bugs to Faulty Components in Large Scale Software Systems Using Bayesian Classification}. In \bibinfo{booktitle}{\emph{{QRS} 2016}}. \bibinfo{pages}{423--430}.
\newblock


\bibitem[Joshi(2002)]%
        {joshi2002evaluating}
\bibfield{author}{\bibinfo{person}{Mahesh~V Joshi}.} \bibinfo{year}{2002}\natexlab{}.
\newblock \showarticletitle{On evaluating performance of classifiers for rare classes}. In \bibinfo{booktitle}{\emph{2002 IEEE International Conference on Data Mining, 2002. Proceedings}}. \bibinfo{pages}{641--644}.
\newblock


\bibitem[Kallis et~al\mbox{.}(2021)]%
        {DBLP:journals/scp/KallisSCP21}
\bibfield{author}{\bibinfo{person}{Rafael Kallis}, \bibinfo{person}{Andrea~Di Sorbo}, \bibinfo{person}{Gerardo Canfora}, {and} \bibinfo{person}{Sebastiano Panichella}.} \bibinfo{year}{2021}\natexlab{}.
\newblock \showarticletitle{Predicting issue types on GitHub}.
\newblock \bibinfo{journal}{\emph{Science of Computer Programming}}  \bibinfo{volume}{205} (\bibinfo{year}{2021}), \bibinfo{pages}{102598}.
\newblock


\bibitem[Kashiwa et~al\mbox{.}(2014)]%
        {DBLP:conf/icsm/KashiwaYKO14}
\bibfield{author}{\bibinfo{person}{Yutaro Kashiwa}, \bibinfo{person}{Hayato Yoshiyuki}, \bibinfo{person}{Yusuke Kukita}, {and} \bibinfo{person}{Masao Ohira}.} \bibinfo{year}{2014}\natexlab{}.
\newblock \showarticletitle{A Pilot Study of Diversity in High Impact Bugs}. In \bibinfo{booktitle}{\emph{30th {IEEE} ICSME, BC, Canada, September 29 - October 3, 2014}}. \bibinfo{pages}{536--540}.
\newblock


\bibitem[Kaur and Jindal(2019)]%
        {DBLP:journals/saem/KaurJ19}
\bibfield{author}{\bibinfo{person}{Arvinder Kaur} {and} \bibinfo{person}{Shubhra~Goyal Jindal}.} \bibinfo{year}{2019}\natexlab{}.
\newblock \showarticletitle{Text analytics based severity prediction of software bugs for apache projects}.
\newblock \bibinfo{journal}{\emph{International Journal of System Assurance Engineering and Management}}  \bibinfo{volume}{10} (\bibinfo{year}{2019}), \bibinfo{pages}{765--782}.
\newblock


\bibitem[Khatiwada et~al\mbox{.}(2020)]%
        {DBLP:conf/iwpc/KhatiwadaTM20}
\bibfield{author}{\bibinfo{person}{Saket Khatiwada}, \bibinfo{person}{Miroslav Tushev}, {and} \bibinfo{person}{Anas Mahmoud}.} \bibinfo{year}{2020}\natexlab{}.
\newblock \showarticletitle{On Combining {IR} Methods to Improve Bug Localization}. In \bibinfo{booktitle}{\emph{{ICPC} '20: 28th International Conference on Program Comprehension Seoul, Republic of Korea, July 13-15, 2020}}. \bibinfo{pages}{252--262}.
\newblock


\bibitem[Kim and Yang(2022a)]%
        {DBLP:journals/access/KimY22d}
\bibfield{author}{\bibinfo{person}{Jungyeon Kim} {and} \bibinfo{person}{Geunseok Yang}.} \bibinfo{year}{2022}\natexlab{a}.
\newblock \showarticletitle{Bug Severity Prediction Algorithm Using Topic-Based Feature Selection and {CNN-LSTM} Algorithm}.
\newblock \bibinfo{journal}{\emph{{IEEE} Access}}  \bibinfo{volume}{10} (\bibinfo{year}{2022}), \bibinfo{pages}{94643--94651}.
\newblock


\bibitem[Kim et~al\mbox{.}(2022a)]%
        {DBLP:journals/jss/KimGLKKBKT22}
\bibfield{author}{\bibinfo{person}{Kisub Kim}, \bibinfo{person}{Sankalp Ghatpande}, \bibinfo{person}{Kui Liu}, \bibinfo{person}{Anil Koyuncu}, \bibinfo{person}{Dongsun Kim}, \bibinfo{person}{Tegawend{\'{e}}~F. Bissyand{\'{e}}}, \bibinfo{person}{Jacques Klein}, {and} \bibinfo{person}{Yves~Le Traon}.} \bibinfo{year}{2022}\natexlab{a}.
\newblock \showarticletitle{DigBug - Pre/post-processing operator selection for accurate bug localization}.
\newblock \bibinfo{journal}{\emph{Journal of Systems and Software}}  \bibinfo{volume}{189} (\bibinfo{year}{2022}), \bibinfo{pages}{111300}.
\newblock


\bibitem[Kim et~al\mbox{.}(2021)]%
        {kim2021novel}
\bibfield{author}{\bibinfo{person}{Misoo Kim}, \bibinfo{person}{Youngkyoung Kim}, {and} \bibinfo{person}{Eunseok Lee}.} \bibinfo{year}{2021}\natexlab{}.
\newblock \showarticletitle{A novel automatic query expansion with word embedding for ir-based bug localization}. In \bibinfo{booktitle}{\emph{2021 IEEE 32nd International Symposium on Software Reliability Engineering (ISSRE)}}. IEEE, \bibinfo{pages}{276--287}.
\newblock


\bibitem[Kim and Yang(2022b)]%
        {DBLP:journals/access/KimY22f}
\bibfield{author}{\bibinfo{person}{Taemin Kim} {and} \bibinfo{person}{Geunseok Yang}.} \bibinfo{year}{2022}\natexlab{b}.
\newblock \showarticletitle{Predicting Duplicate in Bug Report Using Topic-Based Duplicate Learning With Fine Tuning-Based {BERT} Algorithm}.
\newblock \bibinfo{journal}{\emph{{IEEE} Access}}  \bibinfo{volume}{10} (\bibinfo{year}{2022}), \bibinfo{pages}{129666--129675}.
\newblock


\bibitem[Kim et~al\mbox{.}(2022b)]%
        {DBLP:conf/sac/KimKL22a}
\bibfield{author}{\bibinfo{person}{Youngkyoung Kim}, \bibinfo{person}{Misoo Kim}, {and} \bibinfo{person}{Eunseok Lee}.} \bibinfo{year}{2022}\natexlab{b}.
\newblock \showarticletitle{Feature assortment for deep learning-based bug localization with a program graph}. In \bibinfo{booktitle}{\emph{{SAC} '22: The 37th {ACM/SIGAPP} Symposium on Applied Computing, Virtual Event, April 25 - 29, 2022}}. \bibinfo{pages}{1536--1544}.
\newblock


\bibitem[Kitchenham et~al\mbox{.}(2009)]%
        {DBLP:journals/infsof/KitchenhamBBTBL09}
\bibfield{author}{\bibinfo{person}{Barbara~A. Kitchenham}, \bibinfo{person}{Pearl Brereton}, \bibinfo{person}{David Budgen}, \bibinfo{person}{Mark Turner}, \bibinfo{person}{John Bailey}, {and} \bibinfo{person}{Stephen~G. Linkman}.} \bibinfo{year}{2009}\natexlab{}.
\newblock \showarticletitle{Systematic literature reviews in software engineering - {A} systematic literature review}.
\newblock \bibinfo{journal}{\emph{Information {\&} Software Technology}}  \bibinfo{volume}{51} (\bibinfo{year}{2009}), \bibinfo{pages}{7--15}.
\newblock


\bibitem[Kochhar et~al\mbox{.}(2014a)]%
        {DBLP:conf/iceccs/KochharTL14}
\bibfield{author}{\bibinfo{person}{Pavneet~Singh Kochhar}, \bibinfo{person}{Ferdian Thung}, {and} \bibinfo{person}{David Lo}.} \bibinfo{year}{2014}\natexlab{a}.
\newblock \showarticletitle{Automatic Fine-Grained Issue Report Reclassification}. In \bibinfo{booktitle}{\emph{2014 19th International Conference on Engineering of Complex Computer Systems, Tianjin, China, August 4-7, 2014}}. \bibinfo{pages}{126--135}.
\newblock


\bibitem[Kochhar et~al\mbox{.}(2014b)]%
        {kochhar2014automatic}
\bibfield{author}{\bibinfo{person}{Pavneet~Singh Kochhar}, \bibinfo{person}{Ferdian Thung}, {and} \bibinfo{person}{David Lo}.} \bibinfo{year}{2014}\natexlab{b}.
\newblock \showarticletitle{Automatic fine-grained issue report reclassification}. In \bibinfo{booktitle}{\emph{2014 19th International Conference on Engineering of Complex Computer Systems}}. \bibinfo{pages}{126--135}.
\newblock


\bibitem[Kukkar et~al\mbox{.}(2020)]%
        {kukkar2020duplicate}
\bibfield{author}{\bibinfo{person}{Ashima Kukkar}, \bibinfo{person}{Rajni Mohana}, \bibinfo{person}{Yugal Kumar}, \bibinfo{person}{Anand Nayyar}, \bibinfo{person}{Muhammad Bilal}, {and} \bibinfo{person}{Kyung-Sup Kwak}.} \bibinfo{year}{2020}\natexlab{}.
\newblock \showarticletitle{Duplicate bug report detection and classification system based on deep learning technique}.
\newblock \bibinfo{journal}{\emph{IEEE Access}}  \bibinfo{volume}{8} (\bibinfo{year}{2020}), \bibinfo{pages}{200749--200763}.
\newblock


\bibitem[Kumar et~al\mbox{.}(2021)]%
        {DBLP:conf/fedcsis/KumarKNMKP21}
\bibfield{author}{\bibinfo{person}{Lov Kumar}, \bibinfo{person}{Mukesh Kumar}, \bibinfo{person}{Lalita Bhanu~Murthy Neti}, \bibinfo{person}{Sanjay Misra}, \bibinfo{person}{Vipul Kocher}, {and} \bibinfo{person}{Srinivas Padmanabhuni}.} \bibinfo{year}{2021}\natexlab{}.
\newblock \showarticletitle{An Empirical Study on Application of Word Embedding Techniques for Prediction of Software Defect Severity Level}. In \bibinfo{booktitle}{\emph{Proceedings of the 16th Conference on Computer Science and Intelligence Systems, Online, September 2-5, 2021}}, Vol.~\bibinfo{volume}{25}. \bibinfo{pages}{477--484}.
\newblock


\bibitem[Lam et~al\mbox{.}(2017)]%
        {DBLP:conf/iwpc/LamNNN17}
\bibfield{author}{\bibinfo{person}{An~Ngoc Lam}, \bibinfo{person}{Anh~Tuan Nguyen}, \bibinfo{person}{Hoan~Anh Nguyen}, {and} \bibinfo{person}{Tien~N. Nguyen}.} \bibinfo{year}{2017}\natexlab{}.
\newblock \showarticletitle{Bug localization with combination of deep learning and information retrieval}. In \bibinfo{booktitle}{\emph{Proceedings of the 25th International Conference on Program Comprehension {ICPC} 2017, Buenos Aires, Argentina, May 22-23, 2017}}. \bibinfo{pages}{218--229}.
\newblock


\bibitem[Le et~al\mbox{.}(2014)]%
        {le2014predicting}
\bibfield{author}{\bibinfo{person}{Tien-Duy~B Le}, \bibinfo{person}{Ferdian Thung}, {and} \bibinfo{person}{David Lo}.} \bibinfo{year}{2014}\natexlab{}.
\newblock \showarticletitle{Predicting effectiveness of ir-based bug localization techniques}. In \bibinfo{booktitle}{\emph{2014 IEEE 25th International Symposium on Software Reliability Engineering}}. \bibinfo{pages}{335--345}.
\newblock


\bibitem[Le et~al\mbox{.}(2017)]%
        {le2017will}
\bibfield{author}{\bibinfo{person}{Tien-Duy~B Le}, \bibinfo{person}{Ferdian Thung}, {and} \bibinfo{person}{David Lo}.} \bibinfo{year}{2017}\natexlab{}.
\newblock \showarticletitle{Will this localization tool be effective for this bug? Mitigating the impact of unreliability of information retrieval based bug localization tools}.
\newblock \bibinfo{journal}{\emph{Empirical Software Engineering}}  \bibinfo{volume}{22} (\bibinfo{year}{2017}), \bibinfo{pages}{2237--2279}.
\newblock


\bibitem[Lewis(1998)]%
        {DBLP:conf/ecml/Lewis98}
\bibfield{author}{\bibinfo{person}{David~D. Lewis}.} \bibinfo{year}{1998}\natexlab{}.
\newblock \showarticletitle{Naive (Bayes) at Forty: The Independence Assumption in Information Retrieval}. In \bibinfo{booktitle}{\emph{Machine Learning: ECML-98, 10th European Conference on Machine Learning Chemnitz, Germany, April 21-23, 1998, Proceedings}}, Vol.~\bibinfo{volume}{1398}. \bibinfo{pages}{4--15}.
\newblock


\bibitem[Li et~al\mbox{.}(2023)]%
        {DBLP:journals/jss/LiYSLW23}
\bibfield{author}{\bibinfo{person}{Hongyan Li}, \bibinfo{person}{Meng Yan}, \bibinfo{person}{Weifeng Sun}, \bibinfo{person}{Xiao Liu}, {and} \bibinfo{person}{Yunsong Wu}.} \bibinfo{year}{2023}\natexlab{}.
\newblock \showarticletitle{A first look at bug report templates on GitHub}.
\newblock \bibinfo{journal}{\emph{Journal of Systems and Software}}  \bibinfo{volume}{202} (\bibinfo{year}{2023}), \bibinfo{pages}{111709}.
\newblock


\bibitem[Li et~al\mbox{.}(2022b)]%
        {DBLP:journals/tse/LiCY22}
\bibfield{author}{\bibinfo{person}{Miqing Li}, \bibinfo{person}{Tao Chen}, {and} \bibinfo{person}{Xin Yao}.} \bibinfo{year}{2022}\natexlab{b}.
\newblock \showarticletitle{How to Evaluate Solutions in Pareto-Based Search-Based Software Engineering: {A} Critical Review and Methodological Guidance}.
\newblock \bibinfo{journal}{\emph{{IEEE} Trans. Software Eng.}} \bibinfo{volume}{48}, \bibinfo{number}{5} (\bibinfo{year}{2022}), \bibinfo{pages}{1771--1799}.
\newblock
\urldef\tempurl%
\url{https://doi.org/10.1109/TSE.2020.3036108}
\showDOI{\tempurl}


\bibitem[Li et~al\mbox{.}(2018)]%
        {DBLP:conf/iwpc/LiJLRL18}
\bibfield{author}{\bibinfo{person}{Xiaochen Li}, \bibinfo{person}{He Jiang}, \bibinfo{person}{Dong Liu}, \bibinfo{person}{Zhilei Ren}, {and} \bibinfo{person}{Ge Li}.} \bibinfo{year}{2018}\natexlab{}.
\newblock \showarticletitle{Unsupervised deep bug report summarization}. In \bibinfo{booktitle}{\emph{Proceedings of the 26th Conference on Program Comprehension, {ICPC} 2018, Gothenburg, Sweden, May 27-28, 2018}}. \bibinfo{pages}{144--155}.
\newblock


\bibitem[Li et~al\mbox{.}(2019)]%
        {DBLP:conf/issta/LiLZZ19}
\bibfield{author}{\bibinfo{person}{Xia Li}, \bibinfo{person}{Wei Li}, \bibinfo{person}{Yuqun Zhang}, {and} \bibinfo{person}{Lingming Zhang}.} \bibinfo{year}{2019}\natexlab{}.
\newblock \showarticletitle{DeepFL: integrating multiple fault diagnosis dimensions for deep fault localization}. In \bibinfo{booktitle}{\emph{Proceedings of the 28th {ACM} {SIGSOFT} International Symposium on Software Testing and Analysis, {ISSTA} 2019, Beijing, China, July 15-19, 2019}}. \bibinfo{pages}{169--180}.
\newblock


\bibitem[Li et~al\mbox{.}(2022a)]%
        {DBLP:conf/esem/LiCHWWWW22}
\bibfield{author}{\bibinfo{person}{Yingling Li}, \bibinfo{person}{Xing Che}, \bibinfo{person}{Yuekai Huang}, \bibinfo{person}{Junjie Wang}, \bibinfo{person}{Song Wang}, \bibinfo{person}{Yawen Wang}, {and} \bibinfo{person}{Qing Wang}.} \bibinfo{year}{2022}\natexlab{a}.
\newblock \showarticletitle{A Tale of Two Tasks: Automated Issue Priority Prediction with Deep Multi-task Learning}. In \bibinfo{booktitle}{\emph{Proceedings of the 16th ACM/IEEE international symposium on empirical software engineering and measurement}}. \bibinfo{pages}{1--11}.
\newblock


\bibitem[Li et~al\mbox{.}(2022c)]%
        {li2022deeplabel}
\bibfield{author}{\bibinfo{person}{Zhong Li}, \bibinfo{person}{Minxue Pan}, \bibinfo{person}{Yu Pei}, \bibinfo{person}{Tian Zhang}, \bibinfo{person}{Linzhang Wang}, {and} \bibinfo{person}{Xuandong Li}.} \bibinfo{year}{2022}\natexlab{c}.
\newblock \showarticletitle{Deeplabel: Automated issue classification for issue tracking systems}. In \bibinfo{booktitle}{\emph{Proceedings of the 13th Asia-Pacific Symposium on Internetware}}. \bibinfo{pages}{231--241}.
\newblock


\bibitem[Liang et~al\mbox{.}(2025)]%
        {DBLP:conf/icse/LiangChen25}
\bibfield{author}{\bibinfo{person}{Hongyuan Liang}, \bibinfo{person}{Yue Huang}, {and} \bibinfo{person}{Tao Chen}.} \bibinfo{year}{2025}\natexlab{}.
\newblock \showarticletitle{The Same Only Different: On Information Modality for Configuration Performance Analysis}. In \bibinfo{booktitle}{\emph{47th {IEEE/ACM} International Conference on Software Engineering, {ICSE} 2025, Ottawa, ON, Canada, April 26 - May 6, 2025}}. \bibinfo{publisher}{{IEEE}}, \bibinfo{pages}{2522--2534}.
\newblock
\urldef\tempurl%
\url{https://doi.org/10.1109/ICSE55347.2025.00212}
\showDOI{\tempurl}


\bibitem[Liang et~al\mbox{.}(2019)]%
        {DBLP:journals/access/LiangSWY19}
\bibfield{author}{\bibinfo{person}{Hongliang Liang}, \bibinfo{person}{Lu Sun}, \bibinfo{person}{Meilin Wang}, {and} \bibinfo{person}{Yuxing Yang}.} \bibinfo{year}{2019}\natexlab{}.
\newblock \showarticletitle{Deep Learning With Customized Abstract Syntax Tree for Bug Localization}.
\newblock \bibinfo{journal}{\emph{{IEEE} Access}}  \bibinfo{volume}{7} (\bibinfo{year}{2019}), \bibinfo{pages}{116309--116320}.
\newblock


\bibitem[Lin et~al\mbox{.}(2016a)]%
        {DBLP:journals/jss/LinYLC16}
\bibfield{author}{\bibinfo{person}{Meng{-}Jie Lin}, \bibinfo{person}{Cheng{-}Zen Yang}, \bibinfo{person}{Chao{-}Yuan Lee}, {and} \bibinfo{person}{Chun{-}Chang Chen}.} \bibinfo{year}{2016}\natexlab{a}.
\newblock \showarticletitle{Enhancements for duplication detection in bug reports with manifold correlation features}.
\newblock \bibinfo{journal}{\emph{Journal of Systems and Software}}  \bibinfo{volume}{121} (\bibinfo{year}{2016}), \bibinfo{pages}{223--233}.
\newblock


\bibitem[Lin et~al\mbox{.}(2016b)]%
        {lin2016enhancements}
\bibfield{author}{\bibinfo{person}{Meng-Jie Lin}, \bibinfo{person}{Cheng-Zen Yang}, \bibinfo{person}{Chao-Yuan Lee}, {and} \bibinfo{person}{Chun-Chang Chen}.} \bibinfo{year}{2016}\natexlab{b}.
\newblock \showarticletitle{Enhancements for duplication detection in bug reports with manifold correlation features}.
\newblock \bibinfo{journal}{\emph{Journal of Systems and Software}}  \bibinfo{volume}{121} (\bibinfo{year}{2016}), \bibinfo{pages}{223--233}.
\newblock


\bibitem[Long and Chen(2022)]%
        {DBLP:conf/ease/LongC22}
\bibfield{author}{\bibinfo{person}{Guoming Long} {and} \bibinfo{person}{Tao Chen}.} \bibinfo{year}{2022}\natexlab{}.
\newblock \showarticletitle{On Reporting Performance and Accuracy Bugs for Deep Learning Frameworks: An Exploratory Study from GitHub}. In \bibinfo{booktitle}{\emph{{EASE} 2022: The International Conference on Evaluation and Assessment in Software Engineering 2022, Gothenburg, Sweden, June 13 - 15, 2022}}, \bibfield{editor}{\bibinfo{person}{Miroslaw Staron}, \bibinfo{person}{Christian Berger}, \bibinfo{person}{Jocelyn Simmonds}, {and} \bibinfo{person}{Rafael Prikladnicki}} (Eds.). \bibinfo{publisher}{{ACM}}, \bibinfo{pages}{90--99}.
\newblock
\urldef\tempurl%
\url{https://doi.org/10.1145/3530019.3530029}
\showDOI{\tempurl}


\bibitem[Long et~al\mbox{.}(2022)]%
        {DBLP:conf/apsec/LongCC22}
\bibfield{author}{\bibinfo{person}{Guoming Long}, \bibinfo{person}{Tao Chen}, {and} \bibinfo{person}{Georgina Cosma}.} \bibinfo{year}{2022}\natexlab{}.
\newblock \showarticletitle{Multifaceted Hierarchical Report Identification for Non-Functional Bugs in Deep Learning Frameworks}. In \bibinfo{booktitle}{\emph{29th Asia-Pacific Software Engineering Conference, {APSEC} 2022, Virtual Event, Japan, December 6-9, 2022}}. \bibinfo{pages}{289--298}.
\newblock


\bibitem[Luo et~al\mbox{.}(2023)]%
        {DBLP:journals/access/LuoWC23}
\bibfield{author}{\bibinfo{person}{Zhengmao Luo}, \bibinfo{person}{Wenyao Wang}, {and} \bibinfo{person}{Cen Caichun}.} \bibinfo{year}{2023}\natexlab{}.
\newblock \showarticletitle{Improving Bug Localization With Effective Contrastive Learning Representation}.
\newblock \bibinfo{journal}{\emph{{IEEE} Access}}  \bibinfo{volume}{11} (\bibinfo{year}{2023}), \bibinfo{pages}{32523--32533}.
\newblock


\bibitem[Ma et~al\mbox{.}(2022)]%
        {DBLP:journals/infsof/MaKYYLZ22}
\bibfield{author}{\bibinfo{person}{Xiaoxue Ma}, \bibinfo{person}{Jacky Keung}, \bibinfo{person}{Zhen Yang}, \bibinfo{person}{Xiao Yu}, \bibinfo{person}{Yishu Li}, {and} \bibinfo{person}{Hao Zhang}.} \bibinfo{year}{2022}\natexlab{}.
\newblock \showarticletitle{{CASMS:} Combining clustering with attention semantic model for identifying security bug reports}.
\newblock \bibinfo{journal}{\emph{Information and Software Technology}}  \bibinfo{volume}{147} (\bibinfo{year}{2022}), \bibinfo{pages}{106906}.
\newblock


\bibitem[Ma et~al\mbox{.}(2023)]%
        {DBLP:journals/tr/MaKYZZL23}
\bibfield{author}{\bibinfo{person}{Xiaoxue Ma}, \bibinfo{person}{Jacky~Wai Keung}, \bibinfo{person}{Xiao Yu}, \bibinfo{person}{Huiqi Zou}, \bibinfo{person}{Jingyu Zhang}, {and} \bibinfo{person}{Yishu Li}.} \bibinfo{year}{2023}\natexlab{}.
\newblock \showarticletitle{AttSum: {A} Deep Attention-Based Summarization Model for Bug Report Title Generation}.
\newblock \bibinfo{journal}{\emph{{IEEE} Transactions on Reliability}}  \bibinfo{volume}{72} (\bibinfo{year}{2023}), \bibinfo{pages}{1663--1677}.
\newblock


\bibitem[Ma et~al\mbox{.}(2025)]%
        {DBLP:conf/icse/MaChen25}
\bibfield{author}{\bibinfo{person}{Youpeng Ma}, \bibinfo{person}{Tao Chen}, {and} \bibinfo{person}{Ke Li}.} \bibinfo{year}{2025}\natexlab{}.
\newblock \showarticletitle{Faster Configuration Performance Bug Testing with Neural Dual-Level Prioritization}. In \bibinfo{booktitle}{\emph{47th {IEEE/ACM} International Conference on Software Engineering, {ICSE} 2025, Ottawa, ON, Canada, April 26 - May 6, 2025}}. \bibinfo{publisher}{{IEEE}}, \bibinfo{pages}{988--1000}.
\newblock
\urldef\tempurl%
\url{https://doi.org/10.1109/ICSE55347.2025.00201}
\showDOI{\tempurl}


\bibitem[Mani et~al\mbox{.}(2019)]%
        {DBLP:conf/comad/ManiSA19}
\bibfield{author}{\bibinfo{person}{Senthil Mani}, \bibinfo{person}{Anush Sankaran}, {and} \bibinfo{person}{Rahul Aralikatte}.} \bibinfo{year}{2019}\natexlab{}.
\newblock \showarticletitle{DeepTriage: Exploring the Effectiveness of Deep Learning for Bug Triaging}. In \bibinfo{booktitle}{\emph{Proceedings of the {ACM} India Joint International Conference on Data Science and Management of Data, {COMAD/CODS} 2019, Kolkata, India January 3-5, 2019}}. \bibinfo{pages}{171--179}.
\newblock


\bibitem[Messaoud et~al\mbox{.}(2023)]%
        {DBLP:journals/tr/MessaoudMJMG23}
\bibfield{author}{\bibinfo{person}{Montassar~Ben Messaoud}, \bibinfo{person}{Asma Miladi}, \bibinfo{person}{Ilyes Jenhani}, \bibinfo{person}{Mohamed~Wiem Mkaouer}, {and} \bibinfo{person}{Lobna Ghadhab}.} \bibinfo{year}{2023}\natexlab{}.
\newblock \showarticletitle{Duplicate Bug Report Detection Using an Attention-Based Neural Language Model}.
\newblock \bibinfo{journal}{\emph{{IEEE} Transactions on Reliability}}  \bibinfo{volume}{72} (\bibinfo{year}{2023}), \bibinfo{pages}{846--858}.
\newblock


\bibitem[Mikolov et~al\mbox{.}(2013)]%
        {DBLP:journals/corr/abs-1301-3781}
\bibfield{author}{\bibinfo{person}{Tom{\'{a}}s Mikolov}, \bibinfo{person}{Kai Chen}, \bibinfo{person}{Greg Corrado}, {and} \bibinfo{person}{Jeffrey Dean}.} \bibinfo{year}{2013}\natexlab{}.
\newblock \showarticletitle{Efficient Estimation of Word Representations in Vector Space}. In \bibinfo{booktitle}{\emph{1st International Conference on Learning Representations, {ICLR} 2013 Scottsdale, Arizona, USA, May 2-4, 2013, Workshop Track Proceedings}}.
\newblock


\bibitem[Miryeganeh et~al\mbox{.}(2021)]%
        {DBLP:journals/jss/MiryeganehHH21}
\bibfield{author}{\bibinfo{person}{Nima Miryeganeh}, \bibinfo{person}{Sepehr Hashtroudi}, {and} \bibinfo{person}{Hadi Hemmati}.} \bibinfo{year}{2021}\natexlab{}.
\newblock \showarticletitle{GloBug: Using global data in Fault Localization}.
\newblock \bibinfo{journal}{\emph{Journal of Systems and Software}}  \bibinfo{volume}{177} (\bibinfo{year}{2021}), \bibinfo{pages}{110961}.
\newblock


\bibitem[Mohsen et~al\mbox{.}(2023)]%
        {DBLP:journals/access/MohsenHWMM23}
\bibfield{author}{\bibinfo{person}{Amr~Mansour Mohsen}, \bibinfo{person}{Hesham~A. Hassan}, \bibinfo{person}{Khaled Wassif}, \bibinfo{person}{Ramadan Moawad}, {and} \bibinfo{person}{Soha Makady}.} \bibinfo{year}{2023}\natexlab{}.
\newblock \showarticletitle{Enhancing Bug Localization Using Phase-Based Approach}.
\newblock \bibinfo{journal}{\emph{{IEEE} Access}}  \bibinfo{volume}{11} (\bibinfo{year}{2023}), \bibinfo{pages}{35901--35913}.
\newblock


\bibitem[Mohsin and Shi(2021)]%
        {DBLP:journals/eswa/MohsinS21}
\bibfield{author}{\bibinfo{person}{Hufsa Mohsin} {and} \bibinfo{person}{Chongyang Shi}.} \bibinfo{year}{2021}\natexlab{}.
\newblock \showarticletitle{{SPBC:} {A} self-paced learning model for bug classification from historical repositories of open-source software}.
\newblock \bibinfo{journal}{\emph{Expert Systems with Applications}}  \bibinfo{volume}{167} (\bibinfo{year}{2021}), \bibinfo{pages}{113808}.
\newblock


\bibitem[Mohsin et~al\mbox{.}(2022)]%
        {DBLP:journals/kbs/MohsinSHJ22}
\bibfield{author}{\bibinfo{person}{Hufsa Mohsin}, \bibinfo{person}{Chongyang Shi}, \bibinfo{person}{Shufeng Hao}, {and} \bibinfo{person}{He Jiang}.} \bibinfo{year}{2022}\natexlab{}.
\newblock \showarticletitle{{SPAN:} {A} self-paced association augmentation and node embedding-based model for software bug classification and assignment}.
\newblock \bibinfo{journal}{\emph{Knowledge-Based Systems}}  \bibinfo{volume}{236} (\bibinfo{year}{2022}), \bibinfo{pages}{107711}.
\newblock


\bibitem[Neysiani et~al\mbox{.}(2020)]%
        {DBLP:journals/infsof/NeysianiBA20}
\bibfield{author}{\bibinfo{person}{Behzad~Soleimani Neysiani}, \bibinfo{person}{Seyed~Morteza Babamir}, {and} \bibinfo{person}{Masayoshi Aritsugi}.} \bibinfo{year}{2020}\natexlab{}.
\newblock \showarticletitle{Efficient feature extraction model for validation performance improvement of duplicate bug report detection in software bug triage systems}.
\newblock \bibinfo{journal}{\emph{Information and Software Technology}}  \bibinfo{volume}{126} (\bibinfo{year}{2020}), \bibinfo{pages}{106344}.
\newblock


\bibitem[Ohira et~al\mbox{.}(2015)]%
        {inproceedings}
\bibfield{author}{\bibinfo{person}{Masao Ohira}, \bibinfo{person}{Yutaro Kashiwa}, \bibinfo{person}{Yosuke Yamatani}, \bibinfo{person}{Hayato Yoshiyuki}, \bibinfo{person}{Yoshiya Maeda}, \bibinfo{person}{Nachai Limsettho}, \bibinfo{person}{Keisuke Fujino}, \bibinfo{person}{Hideaki Hata}, \bibinfo{person}{Akinori Ihara}, {and} \bibinfo{person}{Kenichi Matsumoto}.} \bibinfo{year}{2015}\natexlab{}.
\newblock \showarticletitle{A Dataset of High Impact Bugs: Manually-Classified Issue Reports}. \bibinfo{pages}{518--521}.
\newblock


\bibitem[Pandey et~al\mbox{.}(2017)]%
        {pandey2017automated}
\bibfield{author}{\bibinfo{person}{Nitish Pandey}, \bibinfo{person}{Debarshi~Kumar Sanyal}, \bibinfo{person}{Abir Hudait}, {and} \bibinfo{person}{Amitava Sen}.} \bibinfo{year}{2017}\natexlab{}.
\newblock \showarticletitle{Automated classification of software issue reports using machine learning techniques: an empirical study}.
\newblock \bibinfo{journal}{\emph{Innovations in Systems and Software Engineering}}  \bibinfo{volume}{13} (\bibinfo{year}{2017}), \bibinfo{pages}{279--297}.
\newblock


\bibitem[Panichella et~al\mbox{.}(2021)]%
        {DBLP:journals/infsof/PanichellaCS21}
\bibfield{author}{\bibinfo{person}{Sebastiano Panichella}, \bibinfo{person}{Gerardo Canfora}, {and} \bibinfo{person}{Andrea~Di Sorbo}.} \bibinfo{year}{2021}\natexlab{}.
\newblock \showarticletitle{\emph{Won't We Fix this Issue?}" Qualitative characterization and automated identification of wontfix issues on GitHub}.
\newblock \bibinfo{journal}{\emph{Information and Software Technology}}  \bibinfo{volume}{139} (\bibinfo{year}{2021}), \bibinfo{pages}{106665}.
\newblock


\bibitem[Park et~al\mbox{.}(2019)]%
        {DBLP:conf/ccs/ParkKS19}
\bibfield{author}{\bibinfo{person}{Sunnyeo Park}, \bibinfo{person}{Dohyeok Kim}, {and} \bibinfo{person}{Sooel Son}.} \bibinfo{year}{2019}\natexlab{}.
\newblock \showarticletitle{An Empirical Study of Prioritizing JavaScript Engine Crashes via Machine Learning}. In \bibinfo{booktitle}{\emph{Proceedings of the 2019 {ACM} Asia Conference on Computer and Communications Security, AsiaCCS 2019, Auckland, New Zealand, July 09-12, 2019}}. \bibinfo{pages}{646--657}.
\newblock


\bibitem[Peters et~al\mbox{.}(2017)]%
        {peters2017text}
\bibfield{author}{\bibinfo{person}{Fayola Peters}, \bibinfo{person}{Thein Tun}, \bibinfo{person}{Yijun Yu}, {and} \bibinfo{person}{Bashar Nuseibeh}.} \bibinfo{year}{2017}\natexlab{}.
\newblock \showarticletitle{Text filtering and ranking for security bug report prediction}.
\newblock \bibinfo{journal}{\emph{IEEE Transactions on Software Engineering}} (\bibinfo{year}{2017}).
\newblock


\bibitem[Peters et~al\mbox{.}(2019)]%
        {DBLP:journals/tse/PetersTYN19}
\bibfield{author}{\bibinfo{person}{Fayola Peters}, \bibinfo{person}{Thein~Than Tun}, \bibinfo{person}{Yijun Yu}, {and} \bibinfo{person}{Bashar Nuseibeh}.} \bibinfo{year}{2019}\natexlab{}.
\newblock \showarticletitle{Text Filtering and Ranking for Security Bug Report Prediction}.
\newblock \bibinfo{journal}{\emph{{IEEE} Transactions on Software Engineering}}  \bibinfo{volume}{45} (\bibinfo{year}{2019}), \bibinfo{pages}{615--631}.
\newblock


\bibitem[Polisetty et~al\mbox{.}(2019)]%
        {polisetty2019usefulness}
\bibfield{author}{\bibinfo{person}{Sravya Polisetty}, \bibinfo{person}{Andriy Miranskyy}, {and} \bibinfo{person}{Ay{\c{s}}e Ba{\c{s}}ar}.} \bibinfo{year}{2019}\natexlab{}.
\newblock \showarticletitle{On usefulness of the deep-learning-based bug localization models to practitioners}. In \bibinfo{booktitle}{\emph{Proceedings of the Fifteenth International Conference on Predictive Models and Data Analytics in Software Engineering}}. \bibinfo{pages}{16--25}.
\newblock


\bibitem[Ramay et~al\mbox{.}(2019a)]%
        {DBLP:journals/access/RamayUYZI19}
\bibfield{author}{\bibinfo{person}{Waheed~Yousuf Ramay}, \bibinfo{person}{Qasim Umer}, \bibinfo{person}{Xu{-}Cheng Yin}, \bibinfo{person}{Chao Zhu}, {and} \bibinfo{person}{Inam Illahi}.} \bibinfo{year}{2019}\natexlab{a}.
\newblock \showarticletitle{Deep Neural Network-Based Severity Prediction of Bug Reports}.
\newblock \bibinfo{journal}{\emph{{IEEE} Access}}  \bibinfo{volume}{7} (\bibinfo{year}{2019}), \bibinfo{pages}{46846--46857}.
\newblock


\bibitem[Ramay et~al\mbox{.}(2019b)]%
        {ramay2019deep}
\bibfield{author}{\bibinfo{person}{Waheed~Yousuf Ramay}, \bibinfo{person}{Qasim Umer}, \bibinfo{person}{Xu~Cheng Yin}, \bibinfo{person}{Chao Zhu}, {and} \bibinfo{person}{Inam Illahi}.} \bibinfo{year}{2019}\natexlab{b}.
\newblock \showarticletitle{Deep Neural Network-Based Severity Prediction of Bug Reports}.
\newblock \bibinfo{journal}{\emph{IEEE Access}}  \bibinfo{volume}{7} (\bibinfo{year}{2019}), \bibinfo{pages}{46846--46857}.
\newblock


\bibitem[Rao et~al\mbox{.}(2015)]%
        {DBLP:journals/sigsoft/RaoMK15}
\bibfield{author}{\bibinfo{person}{Shivani Rao}, \bibinfo{person}{Henry Medeiros}, {and} \bibinfo{person}{Avinash Kak}.} \bibinfo{year}{2015}\natexlab{}.
\newblock \showarticletitle{Comparing incremental latent semantic analysis algorithms for efficient retrieval from software libraries for bug localization}.
\newblock \bibinfo{journal}{\emph{ACM SIGSOFT Software Engineering Notes}} \bibinfo{volume}{40}, \bibinfo{number}{1} (\bibinfo{year}{2015}), \bibinfo{pages}{1--8}.
\newblock


\bibitem[Razzaq et~al\mbox{.}(2021)]%
        {DBLP:conf/scam/RazzaqBPCS21}
\bibfield{author}{\bibinfo{person}{Abdul Razzaq}, \bibinfo{person}{Jim Buckley}, \bibinfo{person}{James~Vincent Patten}, \bibinfo{person}{Muslim Chochlov}, {and} \bibinfo{person}{Ashish~Rajendra Sai}.} \bibinfo{year}{2021}\natexlab{}.
\newblock \showarticletitle{BoostNSift: {A} Query Boosting and Code Sifting Technique for Method Level Bug Localization}. In \bibinfo{booktitle}{\emph{21st {IEEE} International Working Conference on Source Code Analysis and Manipulation, {SCAM} 2021, Luxembourg, September 27-28, 2021}}. \bibinfo{pages}{81--91}.
\newblock


\bibitem[Rish et~al\mbox{.}(2001)]%
        {rish2001empirical}
\bibfield{author}{\bibinfo{person}{Irina Rish} {et~al\mbox{.}}} \bibinfo{year}{2001}\natexlab{}.
\newblock \showarticletitle{An empirical study of the naive Bayes classifier}. In \bibinfo{booktitle}{\emph{IJCAI 2001 workshop on empirical methods in artificial intelligence}}, Vol.~\bibinfo{volume}{3}. \bibinfo{pages}{41--46}.
\newblock


\bibitem[Rocha and da~Costa~Carvalho(2021)]%
        {DBLP:journals/access/RochaC21}
\bibfield{author}{\bibinfo{person}{Thiago~Marques Rocha} {and} \bibinfo{person}{Andr{\'{e}}~Luiz da Costa~Carvalho}.} \bibinfo{year}{2021}\natexlab{}.
\newblock \showarticletitle{SiameseQAT: {A} Semantic Context-Based Duplicate Bug Report Detection Using Replicated Cluster Information}.
\newblock \bibinfo{journal}{\emph{{IEEE} Access}}  \bibinfo{volume}{9} (\bibinfo{year}{2021}), \bibinfo{pages}{44610--44630}.
\newblock


\bibitem[Rodrigues et~al\mbox{.}(2020)]%
        {DBLP:conf/msr/RodriguesAFD20}
\bibfield{author}{\bibinfo{person}{Irving~Muller Rodrigues}, \bibinfo{person}{Daniel Aloise}, \bibinfo{person}{Eraldo~Rezende Fernandes}, {and} \bibinfo{person}{Michel~R. Dagenais}.} \bibinfo{year}{2020}\natexlab{}.
\newblock \showarticletitle{A Soft Alignment Model for Bug Deduplication}. In \bibinfo{booktitle}{\emph{{MSR} '20: 17th International Conference on Mining Software Repositories Seoul, Republic of Korea, 29-30 June, 2020}}. \bibinfo{pages}{43--53}.
\newblock


\bibitem[Roy and Rossi(2014)]%
        {roy2014towards}
\bibfield{author}{\bibinfo{person}{Nivir Kanti~Singha Roy} {and} \bibinfo{person}{Bruno Rossi}.} \bibinfo{year}{2014}\natexlab{}.
\newblock \showarticletitle{Towards an improvement of bug severity classification}. In \bibinfo{booktitle}{\emph{2014 40th EUROMICRO Conference on Software Engineering and Advanced Applications}}. \bibinfo{pages}{269--276}.
\newblock


\bibitem[Sabor et~al\mbox{.}(2016)]%
        {DBLP:conf/cascon/SaborHH16}
\bibfield{author}{\bibinfo{person}{Korosh~Koochekian Sabor}, \bibinfo{person}{Mohammad Hamdaqa}, {and} \bibinfo{person}{Abdelwahab Hamou{-}Lhadj}.} \bibinfo{year}{2016}\natexlab{}.
\newblock \showarticletitle{Automatic prediction of the severity of bugs using stack traces}. In \bibinfo{booktitle}{\emph{Proceedings of the 26th Annual International Conference on Computer Science and Software Engineering, {CASCON} 2016, Toronto, Ontario, Canada, October 31 - November 2, 2016}}, \bibfield{editor}{\bibinfo{person}{Marcellus Mindel}, \bibinfo{person}{Blake Jones}, \bibinfo{person}{Hausi~A. M{\"{u}}ller}, {and} \bibinfo{person}{Vio Onut}} (Eds.). \bibinfo{publisher}{{IBM} / {ACM}}, \bibinfo{pages}{96--105}.
\newblock
\urldef\tempurl%
\url{http://dl.acm.org/citation.cfm?id=3049887}
\showURL{%
\tempurl}


\bibitem[Sabor et~al\mbox{.}(2020)]%
        {DBLP:journals/infsof/SaborHH20}
\bibfield{author}{\bibinfo{person}{Korosh~Koochekian Sabor}, \bibinfo{person}{Mohammad Hamdaqa}, {and} \bibinfo{person}{Abdelwahab Hamou{-}Lhadj}.} \bibinfo{year}{2020}\natexlab{}.
\newblock \showarticletitle{Automatic prediction of the severity of bugs using stack traces and categorical features}.
\newblock \bibinfo{journal}{\emph{Information and Software Technology}}  \bibinfo{volume}{123} (\bibinfo{year}{2020}), \bibinfo{pages}{106205}.
\newblock


\bibitem[Sepahvand et~al\mbox{.}(2023)]%
        {DBLP:journals/scp/SepahvandAJHB23}
\bibfield{author}{\bibinfo{person}{Reza Sepahvand}, \bibinfo{person}{Reza Akbari}, \bibinfo{person}{Behnaz Jamasb}, \bibinfo{person}{Sattar Hashemi}, {and} \bibinfo{person}{Omid Boushehrian}.} \bibinfo{year}{2023}\natexlab{}.
\newblock \showarticletitle{Using word embedding and convolution neural network for bug triaging by considering design flaws}.
\newblock \bibinfo{journal}{\emph{Science of Computer Programming}}  \bibinfo{volume}{228} (\bibinfo{year}{2023}), \bibinfo{pages}{102945}.
\newblock


\bibitem[Sharma et~al\mbox{.}(2015)]%
        {sharma2015novel}
\bibfield{author}{\bibinfo{person}{Gitika Sharma}, \bibinfo{person}{Sumit Sharma}, {and} \bibinfo{person}{Shruti Gujral}.} \bibinfo{year}{2015}\natexlab{}.
\newblock \showarticletitle{A novel way of assessing software bug severity using dictionary of critical terms}.
\newblock \bibinfo{journal}{\emph{Procedia Computer Science}}  \bibinfo{volume}{70} (\bibinfo{year}{2015}), \bibinfo{pages}{632--639}.
\newblock


\bibitem[Sharma et~al\mbox{.}(2014)]%
        {DBLP:conf/iccsa/SharmaKSS14}
\bibfield{author}{\bibinfo{person}{Meera Sharma}, \bibinfo{person}{Madhu Kumari}, \bibinfo{person}{Rajeev~Kumar Singh}, {and} \bibinfo{person}{V.~B. Singh}.} \bibinfo{year}{2014}\natexlab{}.
\newblock \showarticletitle{Multiattribute Based Machine Learning Models for Severity Prediction in Cross Project Context}. In \bibinfo{booktitle}{\emph{{ICCSA} 2014}}.
\newblock


\bibitem[Shi et~al\mbox{.}(2018a)]%
        {DBLP:journals/asc/ShiKBZ18}
\bibfield{author}{\bibinfo{person}{Zhendong Shi}, \bibinfo{person}{Jacky Keung}, \bibinfo{person}{Kwabena~Ebo Bennin}, {and} \bibinfo{person}{Xingjun Zhang}.} \bibinfo{year}{2018}\natexlab{a}.
\newblock \showarticletitle{Comparing learning to rank techniques in hybrid bug localization}.
\newblock \bibinfo{journal}{\emph{Applied Soft Computing}}  \bibinfo{volume}{62} (\bibinfo{year}{2018}), \bibinfo{pages}{636--648}.
\newblock


\bibitem[Shi et~al\mbox{.}(2018b)]%
        {shi2018comparing}
\bibfield{author}{\bibinfo{person}{Zhendong Shi}, \bibinfo{person}{Jacky Keung}, \bibinfo{person}{Kwabena~Ebo Bennin}, {and} \bibinfo{person}{Xingjun Zhang}.} \bibinfo{year}{2018}\natexlab{b}.
\newblock \showarticletitle{Comparing learning to rank techniques in hybrid bug localization}.
\newblock \bibinfo{journal}{\emph{Applied Soft Computing}}  \bibinfo{volume}{62} (\bibinfo{year}{2018}), \bibinfo{pages}{636--648}.
\newblock


\bibitem[Singh et~al\mbox{.}(2017)]%
        {DBLP:journals/jikm/SinghMS17}
\bibfield{author}{\bibinfo{person}{V.~B. Singh}, \bibinfo{person}{Sanjay Misra}, {and} \bibinfo{person}{Meera Sharma}.} \bibinfo{year}{2017}\natexlab{}.
\newblock \showarticletitle{Bug Severity Assessment in Cross Project Context and Identifying Training Candidates}.
\newblock \bibinfo{journal}{\emph{Journal of Information \& Knowledge Management}}  \bibinfo{volume}{16} (\bibinfo{year}{2017}), \bibinfo{pages}{1750005:1--1750005:30}.
\newblock


\bibitem[Strate and Laplante(2013)]%
        {DBLP:journals/tr/StrateL13}
\bibfield{author}{\bibinfo{person}{Jonathan~D. Strate} {and} \bibinfo{person}{Phillip~A. Laplante}.} \bibinfo{year}{2013}\natexlab{}.
\newblock \showarticletitle{A Literature Review of Research in Software Defect Reporting}.
\newblock \bibinfo{journal}{\emph{{IEEE} Transactions on Reliabilityility}}  \bibinfo{volume}{62} (\bibinfo{year}{2013}), \bibinfo{pages}{444--454}.
\newblock


\bibitem[Su et~al\mbox{.}(2022)]%
        {DBLP:conf/qrs/SuHCQM22}
\bibfield{author}{\bibinfo{person}{Yu Su}, \bibinfo{person}{Xinping Hu}, \bibinfo{person}{Xiang Chen}, \bibinfo{person}{Yubin Qu}, {and} \bibinfo{person}{Qianshuang Meng}.} \bibinfo{year}{2022}\natexlab{}.
\newblock \showarticletitle{{CIL-BSP:} Bug Report Severity Prediction based on Class Imbalanced Learning}. In \bibinfo{booktitle}{\emph{22nd {IEEE} International Conference on Software Quality, Reliability and Security, {QRS} 2022 - Companion, Guangzhou, China, December 5-9 2022}}. \bibinfo{pages}{298--306}.
\newblock


\bibitem[Sun et~al\mbox{.}(2023)]%
        {DBLP:journals/compsec/SunOZLWZ23}
\bibfield{author}{\bibinfo{person}{Hongyu Sun}, \bibinfo{person}{Guoliang Ou}, \bibinfo{person}{Ziqiu Zheng}, \bibinfo{person}{Lei Liao}, \bibinfo{person}{He Wang}, {and} \bibinfo{person}{Yuqing Zhang}.} \bibinfo{year}{2023}\natexlab{}.
\newblock \showarticletitle{Inconsistent measurement and incorrect detection of software names in security vulnerability reports}.
\newblock \bibinfo{journal}{\emph{Computers \& Security}}  \bibinfo{volume}{135} (\bibinfo{year}{2023}), \bibinfo{pages}{103477}.
\newblock


\bibitem[Tan et~al\mbox{.}(2020)]%
        {DBLP:journals/jss/TanXWZXL20}
\bibfield{author}{\bibinfo{person}{Youshuai Tan}, \bibinfo{person}{Sijie Xu}, \bibinfo{person}{Zhaowei Wang}, \bibinfo{person}{Tao Zhang}, \bibinfo{person}{Zhou Xu}, {and} \bibinfo{person}{Xiapu Luo}.} \bibinfo{year}{2020}\natexlab{}.
\newblock \showarticletitle{Bug severity prediction using question-and-answer pairs from Stack Overflow}.
\newblock \bibinfo{journal}{\emph{Journal of Systems and Software}}  \bibinfo{volume}{165} (\bibinfo{year}{2020}), \bibinfo{pages}{110567}.
\newblock


\bibitem[Tarar et~al\mbox{.}(2019)]%
        {DBLP:conf/icetc/TararAB19}
\bibfield{author}{\bibinfo{person}{M.~Irtaza~Nawaz Tarar}, \bibinfo{person}{Mubashir Ali}, {and} \bibinfo{person}{Wasi~Haider Butt}.} \bibinfo{year}{2019}\natexlab{}.
\newblock \showarticletitle{Bug Report Summarization: {A} systematic Literature Review}. In \bibinfo{booktitle}{\emph{{ICETC} 2019, 11th International Conference on Education Technology and Computers, Amsterdam, The Netherlands, October 28-31, 2019}}. \bibinfo{pages}{257--261}.
\newblock


\bibitem[Thung et~al\mbox{.}(2015)]%
        {DBLP:conf/iwpc/ThungLL15}
\bibfield{author}{\bibinfo{person}{Ferdian Thung}, \bibinfo{person}{Xuan{-}Bach~Dinh Le}, {and} \bibinfo{person}{David Lo}.} \bibinfo{year}{2015}\natexlab{}.
\newblock \showarticletitle{Active semi-supervised defect categorization}. In \bibinfo{booktitle}{\emph{Proceedings of the 2015 {IEEE} 23rd International Conference on Program Comprehension, {ICPC} 2015, May 16-24, 2015}}. \bibinfo{pages}{60--70}.
\newblock


\bibitem[Tian et~al\mbox{.}(2015)]%
        {DBLP:journals/ese/TianLXS15}
\bibfield{author}{\bibinfo{person}{Yuan Tian}, \bibinfo{person}{David Lo}, \bibinfo{person}{Xin Xia}, {and} \bibinfo{person}{Chengnian Sun}.} \bibinfo{year}{2015}\natexlab{}.
\newblock \showarticletitle{Automated prediction of bug report priority using multi-factor analysis}.
\newblock \bibinfo{journal}{\emph{Empirical Software Engineering}}  \bibinfo{volume}{20} (\bibinfo{year}{2015}), \bibinfo{pages}{1354--1383}.
\newblock


\bibitem[Tian et~al\mbox{.}(2016)]%
        {DBLP:conf/iwpc/TianWLG16}
\bibfield{author}{\bibinfo{person}{Yuan Tian}, \bibinfo{person}{Dinusha Wijedasa}, \bibinfo{person}{David Lo}, {and} \bibinfo{person}{Claire {Le Goues}}.} \bibinfo{year}{2016}\natexlab{}.
\newblock \showarticletitle{Learning to rank for bug report assignee recommendation}. In \bibinfo{booktitle}{\emph{24th {IEEE} International Conference on Program Comprehension, {ICPC} 2016, Austin, TX, USA, May 16-17, 2016}}. \bibinfo{pages}{1--10}.
\newblock


\bibitem[Tong and Zhang(2021)]%
        {DBLP:journals/infsof/TongZ21}
\bibfield{author}{\bibinfo{person}{Yao Tong} {and} \bibinfo{person}{Xiaofang Zhang}.} \bibinfo{year}{2021}\natexlab{}.
\newblock \showarticletitle{Crowdsourced test report prioritization considering bug severity}.
\newblock \bibinfo{journal}{\emph{Information and Software Technology}}  \bibinfo{volume}{139} (\bibinfo{year}{2021}), \bibinfo{pages}{106668}.
\newblock


\bibitem[Tsuruda et~al\mbox{.}(2015)]%
        {DBLP:conf/apsec/TsurudaMA15}
\bibfield{author}{\bibinfo{person}{Akihiro Tsuruda}, \bibinfo{person}{Yuki Manabe}, {and} \bibinfo{person}{Masayoshi Aritsugi}.} \bibinfo{year}{2015}\natexlab{}.
\newblock \showarticletitle{Can We Detect Bug Report Duplication with Unfinished Bug Reports?}. In \bibinfo{booktitle}{\emph{2015 Asia-Pacific Software Engineering Conference, {APSEC} 2015, December 1-4, 2015}}. \bibinfo{pages}{151--158}.
\newblock


\bibitem[Uddin et~al\mbox{.}(2017)]%
        {DBLP:journals/air/UddinGDNS17}
\bibfield{author}{\bibinfo{person}{Jamal Uddin}, \bibinfo{person}{Rozaida Ghazali}, \bibinfo{person}{Mustafa~Mat Deris}, \bibinfo{person}{Rashid Naseem}, {and} \bibinfo{person}{Habib Shah}.} \bibinfo{year}{2017}\natexlab{}.
\newblock \showarticletitle{A survey on bug prioritization}.
\newblock \bibinfo{journal}{\emph{Artificial Intelligence Review}}  \bibinfo{volume}{47} (\bibinfo{year}{2017}), \bibinfo{pages}{145--180}.
\newblock


\bibitem[Umer et~al\mbox{.}(2019)]%
        {umer2019cnn}
\bibfield{author}{\bibinfo{person}{Qasim Umer}, \bibinfo{person}{Hui Liu}, {and} \bibinfo{person}{Inam Illahi}.} \bibinfo{year}{2019}\natexlab{}.
\newblock \showarticletitle{CNN-Based Automatic Prioritization of Bug Reports}.
\newblock \bibinfo{journal}{\emph{IEEE Transactions on Reliability}} (\bibinfo{year}{2019}).
\newblock


\bibitem[Umer et~al\mbox{.}(2020)]%
        {DBLP:journals/tr/UmerLI20}
\bibfield{author}{\bibinfo{person}{Qasim Umer}, \bibinfo{person}{Hui Liu}, {and} \bibinfo{person}{Inam Illahi}.} \bibinfo{year}{2020}\natexlab{}.
\newblock \showarticletitle{CNN-Based Automatic Prioritization of Bug Reports}.
\newblock \bibinfo{journal}{\emph{{IEEE} Transactions on Reliability}}  \bibinfo{volume}{69} (\bibinfo{year}{2020}), \bibinfo{pages}{1341--1354}.
\newblock


\bibitem[Umer et~al\mbox{.}(2018)]%
        {DBLP:journals/access/UmerLS18}
\bibfield{author}{\bibinfo{person}{Qasim Umer}, \bibinfo{person}{Hui Liu}, {and} \bibinfo{person}{Yasir Sultan}.} \bibinfo{year}{2018}\natexlab{}.
\newblock \showarticletitle{Emotion Based Automated Priority Prediction for Bug Reports}.
\newblock \bibinfo{journal}{\emph{{IEEE} Access}}  \bibinfo{volume}{6} (\bibinfo{year}{2018}), \bibinfo{pages}{35743--35752}.
\newblock


\bibitem[Wang and Lee(2022)]%
        {DBLP:journals/access/WangL22d}
\bibfield{author}{\bibinfo{person}{Dae{-}Sung Wang} {and} \bibinfo{person}{Chan{-}Gun Lee}.} \bibinfo{year}{2022}\natexlab{}.
\newblock \showarticletitle{Automatic Component Prediction for Issue Reports Using Fine-Tuned Pretrained Language Models}.
\newblock \bibinfo{journal}{\emph{{IEEE} Access}}  \bibinfo{volume}{10} (\bibinfo{year}{2022}), \bibinfo{pages}{131456--131468}.
\newblock


\bibitem[Wang and Li(2021)]%
        {DBLP:conf/apsec/WangL21}
\bibfield{author}{\bibinfo{person}{Hongbing Wang} {and} \bibinfo{person}{Qi Li}.} \bibinfo{year}{2021}\natexlab{}.
\newblock \showarticletitle{Effective Bug Triage Based on a Hybrid Neural Network}. In \bibinfo{booktitle}{\emph{28th Asia-Pacific Software Engineering Conference, {APSEC} 2021, Taipei Taiwan, December 6-9, 2021}}. \bibinfo{pages}{82--91}.
\newblock


\bibitem[Wang et~al\mbox{.}(2016)]%
        {DBLP:conf/esem/WangCWW16}
\bibfield{author}{\bibinfo{person}{Junjie Wang}, \bibinfo{person}{Qiang Cui}, \bibinfo{person}{Qing Wang}, {and} \bibinfo{person}{Song Wang}.} \bibinfo{year}{2016}\natexlab{}.
\newblock \showarticletitle{Towards Effectively Test Report Classification to Assist Crowdsourced Testing}. In \bibinfo{booktitle}{\emph{{ESEM} 2016}}.
\newblock


\bibitem[Wang et~al\mbox{.}(2017)]%
        {DBLP:conf/icse/WangCWW17}
\bibfield{author}{\bibinfo{person}{Junjie Wang}, \bibinfo{person}{Qiang Cui}, \bibinfo{person}{Song Wang}, {and} \bibinfo{person}{Qing Wang}.} \bibinfo{year}{2017}\natexlab{}.
\newblock \showarticletitle{Domain Adaptation for Test Report Classification in Crowdsourced Testing}. In \bibinfo{booktitle}{\emph{{ICSE-SEIP} 2017}}.
\newblock


\bibitem[Wang et~al\mbox{.}(2014)]%
        {DBLP:conf/esem/WangZW14}
\bibfield{author}{\bibinfo{person}{Song Wang}, \bibinfo{person}{Wen Zhang}, {and} \bibinfo{person}{Qing Wang}.} \bibinfo{year}{2014}\natexlab{}.
\newblock \showarticletitle{FixerCache: unsupervised caching active developers for diverse bug triage}. In \bibinfo{booktitle}{\emph{{ESEM} '14}}.
\newblock


\bibitem[Wang et~al\mbox{.}(2023)]%
        {DBLP:journals/infsof/WangWH23}
\bibfield{author}{\bibinfo{person}{Wenyao Wang}, \bibinfo{person}{Chen{-}Hao Wu}, {and} \bibinfo{person}{Jie He}.} \bibinfo{year}{2023}\natexlab{}.
\newblock \showarticletitle{CLeBPI: Contrastive Learning for Bug Priority Inference}.
\newblock \bibinfo{journal}{\emph{Information and Software Technology}}  \bibinfo{volume}{164} (\bibinfo{year}{2023}), \bibinfo{pages}{107302}.
\newblock


\bibitem[Wang et~al\mbox{.}(2022)]%
        {DBLP:conf/qrs/WangZLYZ22}
\bibfield{author}{\bibinfo{person}{Xiaojuan Wang}, \bibinfo{person}{Wenyu Zhang}, \bibinfo{person}{Shanyan Lai}, \bibinfo{person}{Chunyang Ye}, {and} \bibinfo{person}{Hui Zhou}.} \bibinfo{year}{2022}\natexlab{}.
\newblock \showarticletitle{The Use of Pretrained Model for Matching App Reviews and Bug Reports}. In \bibinfo{booktitle}{\emph{22nd {IEEE} International Conference on Software Quality, Reliability and Security, {QRS} 2022, Guangzhou, China, December 5-9, 2022}}. \bibinfo{pages}{242--251}.
\newblock


\bibitem[Wei et~al\mbox{.}(2023)]%
        {DBLP:journals/access/WeiZR23}
\bibfield{author}{\bibinfo{person}{Ye Wei}, \bibinfo{person}{Chunfu Zhang}, {and} \bibinfo{person}{Teng Ren}.} \bibinfo{year}{2023}\natexlab{}.
\newblock \showarticletitle{Improving Bug Severity Prediction With Domain-Specific Representation Learning}.
\newblock \bibinfo{journal}{\emph{{IEEE} Access}}  \bibinfo{volume}{11} (\bibinfo{year}{2023}), \bibinfo{pages}{62829--62839}.
\newblock


\bibitem[Wen et~al\mbox{.}(2016)]%
        {DBLP:conf/issre/WenYH16}
\bibfield{author}{\bibinfo{person}{Wei Wen}, \bibinfo{person}{Tingting Yu}, {and} \bibinfo{person}{Jane~Huffman Hayes}.} \bibinfo{year}{2016}\natexlab{}.
\newblock \showarticletitle{CoLUA: Automatically Predicting Configuration Bug Reports and Extracting Configuration Options}. In \bibinfo{booktitle}{\emph{27th {IEEE} International Symposium on Software Reliability Engineering {ISSRE} 2016, Ottawa, ON, Canada, October 23-27, 2016}}. \bibinfo{pages}{150--161}.
\newblock


\bibitem[Wu et~al\mbox{.}(2018)]%
        {DBLP:journals/iee/WuLM18}
\bibfield{author}{\bibinfo{person}{Hongrun Wu}, \bibinfo{person}{Haiyang Liu}, {and} \bibinfo{person}{Yutao Ma}.} \bibinfo{year}{2018}\natexlab{}.
\newblock \showarticletitle{Empirical study on developer factors affecting tossing path length of bug reports}.
\newblock \bibinfo{journal}{\emph{{IET} Software}}  \bibinfo{volume}{12} (\bibinfo{year}{2018}), \bibinfo{pages}{258--270}.
\newblock


\bibitem[Wu et~al\mbox{.}(2022a)]%
        {DBLP:journals/kbs/WuMX0H22}
\bibfield{author}{\bibinfo{person}{Hongrun Wu}, \bibinfo{person}{Yutao Ma}, \bibinfo{person}{Zhenglong Xiang}, \bibinfo{person}{Chen Yang}, {and} \bibinfo{person}{Keqing He}.} \bibinfo{year}{2022}\natexlab{a}.
\newblock \showarticletitle{A spatial-temporal graph neural network framework for automated software bug triaging}.
\newblock \bibinfo{journal}{\emph{Knowledge-Based Systems}}  \bibinfo{volume}{241} (\bibinfo{year}{2022}), \bibinfo{pages}{108308}.
\newblock


\bibitem[Wu et~al\mbox{.}(2020)]%
        {DBLP:journals/jss/WuZCWM20}
\bibfield{author}{\bibinfo{person}{Xiaoxue Wu}, \bibinfo{person}{Wei Zheng}, \bibinfo{person}{Xiang Chen}, \bibinfo{person}{Fang Wang}, {and} \bibinfo{person}{Dejun Mu}.} \bibinfo{year}{2020}\natexlab{}.
\newblock \showarticletitle{CVE-assisted large-scale security bug report dataset construction method}.
\newblock \bibinfo{journal}{\emph{Journal of Systems and Software}}  \bibinfo{volume}{160} (\bibinfo{year}{2020}).
\newblock


\bibitem[Wu et~al\mbox{.}(2021)]%
        {DBLP:journals/infsof/WuZCZYM21}
\bibfield{author}{\bibinfo{person}{Xiaoxue Wu}, \bibinfo{person}{Wei Zheng}, \bibinfo{person}{Xiang Chen}, \bibinfo{person}{Yu Zhao}, \bibinfo{person}{Tingting Yu}, {and} \bibinfo{person}{Dejun Mu}.} \bibinfo{year}{2021}\natexlab{}.
\newblock \showarticletitle{Improving high-impact bug report prediction with combination of interactive machine learning and active learning}.
\newblock \bibinfo{journal}{\emph{Information and Software Technology}}  \bibinfo{volume}{133} (\bibinfo{year}{2021}), \bibinfo{pages}{106530}.
\newblock


\bibitem[Wu et~al\mbox{.}(2022b)]%
        {DBLP:journals/tse/WuZXL22}
\bibfield{author}{\bibinfo{person}{Xiaoxue Wu}, \bibinfo{person}{Wei Zheng}, \bibinfo{person}{Xin Xia}, {and} \bibinfo{person}{David Lo}.} \bibinfo{year}{2022}\natexlab{b}.
\newblock \showarticletitle{Data Quality Matters: {A} Case Study on Data Label Correctness for Security Bug Report Prediction}.
\newblock \bibinfo{journal}{\emph{{IEEE} Transactions on Software Engineering}}  \bibinfo{volume}{48} (\bibinfo{year}{2022}), \bibinfo{pages}{2541--2556}.
\newblock


\bibitem[Xi et~al\mbox{.}(2018)]%
        {DBLP:conf/internetware/Xi0XX018}
\bibfield{author}{\bibinfo{person}{Shengqu Xi}, \bibinfo{person}{Yuan Yao}, \bibinfo{person}{Xusheng Xiao}, \bibinfo{person}{Feng Xu}, {and} \bibinfo{person}{Jian Lu}.} \bibinfo{year}{2018}\natexlab{}.
\newblock \showarticletitle{An Effective Approach for Routing the Bug Reports to the Right Fixers}. In \bibinfo{booktitle}{\emph{Proceedings of the Tenth Asia-Pacific Symposium on Internetware, Internetware 2018, Beijing, China, September 16-16, 2018}}. \bibinfo{pages}{11:1--11:10}.
\newblock


\bibitem[Xia et~al\mbox{.}(2014)]%
        {DBLP:conf/compsac/XiaLQWZ14}
\bibfield{author}{\bibinfo{person}{Xin Xia}, \bibinfo{person}{David Lo}, \bibinfo{person}{Weiwei Qiu}, \bibinfo{person}{Xingen Wang}, {and} \bibinfo{person}{Bo Zhou}.} \bibinfo{year}{2014}\natexlab{}.
\newblock \showarticletitle{Automated Configuration Bug Report Prediction Using Text Mining}. In \bibinfo{booktitle}{\emph{{IEEE} 38th Annual Computer Software and Applications Conference {COMPSAC} 2014, Vasteras, Sweden, July 21-25, 2014}}. \bibinfo{pages}{107--116}.
\newblock


\bibitem[Xia et~al\mbox{.}(2016)]%
        {DBLP:journals/tr/XiaLSW16}
\bibfield{author}{\bibinfo{person}{Xin Xia}, \bibinfo{person}{David Lo}, \bibinfo{person}{Emad Shihab}, {and} \bibinfo{person}{Xinyu Wang}.} \bibinfo{year}{2016}\natexlab{}.
\newblock \showarticletitle{Automated Bug Report Field Reassignment and Refinement Prediction}.
\newblock \bibinfo{journal}{\emph{{IEEE} Transactions on Reliability}}  \bibinfo{volume}{65} (\bibinfo{year}{2016}), \bibinfo{pages}{1094--1113}.
\newblock


\bibitem[Xia et~al\mbox{.}(2015a)]%
        {DBLP:journals/infsof/XiaLSWY15}
\bibfield{author}{\bibinfo{person}{Xin Xia}, \bibinfo{person}{David Lo}, \bibinfo{person}{Emad Shihab}, \bibinfo{person}{Xinyu Wang}, {and} \bibinfo{person}{Xiaohu Yang}.} \bibinfo{year}{2015}\natexlab{a}.
\newblock \showarticletitle{ELBlocker: Predicting blocking bugs with ensemble imbalance learning}.
\newblock \bibinfo{journal}{\emph{Information and Software Technology}}  \bibinfo{volume}{61} (\bibinfo{year}{2015}), \bibinfo{pages}{93--106}.
\newblock


\bibitem[Xia et~al\mbox{.}(2015b)]%
        {DBLP:journals/ase/XiaLSWZ15}
\bibfield{author}{\bibinfo{person}{Xin Xia}, \bibinfo{person}{David Lo}, \bibinfo{person}{Emad Shihab}, \bibinfo{person}{Xinyu Wang}, {and} \bibinfo{person}{Bo Zhou}.} \bibinfo{year}{2015}\natexlab{b}.
\newblock \showarticletitle{Automatic, high accuracy prediction of reopened bugs}.
\newblock \bibinfo{journal}{\emph{Automated Software Engineering}}  \bibinfo{volume}{22} (\bibinfo{year}{2015}), \bibinfo{pages}{75--109}.
\newblock


\bibitem[Xia et~al\mbox{.}(2015c)]%
        {DBLP:journals/smr/XiaLWZ15}
\bibfield{author}{\bibinfo{person}{Xin Xia}, \bibinfo{person}{David Lo}, \bibinfo{person}{Xinyu Wang}, {and} \bibinfo{person}{Bo Zhou}.} \bibinfo{year}{2015}\natexlab{c}.
\newblock \showarticletitle{Dual analysis for recommending developers to resolve bugs}.
\newblock \bibinfo{journal}{\emph{Journal of Software: Evolution and Process}}  \bibinfo{volume}{27} (\bibinfo{year}{2015}), \bibinfo{pages}{195--220}.
\newblock


\bibitem[Xiang et~al\mbox{.}(2026)]%
        {DBLP:conf/icse/XiangChen26}
\bibfield{author}{\bibinfo{person}{Zezhen Xiang}, \bibinfo{person}{Jingzhi Gong}, {and} \bibinfo{person}{Tao Chen}.} \bibinfo{year}{2026}\natexlab{}.
\newblock \showarticletitle{Dually Hierarchical Drift Adaptation for Online Configuration Performance Learning}. In \bibinfo{booktitle}{\emph{48th {IEEE/ACM} International Conference on Software Engineering (ICSE)}}. \bibinfo{publisher}{{ACM}}.
\newblock


\bibitem[Xiao et~al\mbox{.}(2017)]%
        {xiao2017improving}
\bibfield{author}{\bibinfo{person}{Yan Xiao}, \bibinfo{person}{Jacky Keung}, \bibinfo{person}{Qing Mi}, {and} \bibinfo{person}{Kwabena~E Bennin}.} \bibinfo{year}{2017}\natexlab{}.
\newblock \showarticletitle{Improving bug localization with an enhanced convolutional neural network}. In \bibinfo{booktitle}{\emph{2017 24th Asia-Pacific Software Engineering Conference (APSEC)}}. \bibinfo{pages}{338--347}.
\newblock


\bibitem[Xiao et~al\mbox{.}(2018)]%
        {xiao2018bug}
\bibfield{author}{\bibinfo{person}{Yan Xiao}, \bibinfo{person}{Jacky Keung}, \bibinfo{person}{Qing Mi}, {and} \bibinfo{person}{Kwabena~E Bennin}.} \bibinfo{year}{2018}\natexlab{}.
\newblock \showarticletitle{Bug localization with semantic and structural features using convolutional neural network and cascade forest}. In \bibinfo{booktitle}{\emph{Proceedings of the 22nd International Conference on Evaluation and Assessment in Software Engineering 2018}}. \bibinfo{pages}{101--111}.
\newblock


\bibitem[Xu et~al\mbox{.}(2015)]%
        {DBLP:conf/apsec/XuLXSL15}
\bibfield{author}{\bibinfo{person}{Bowen Xu}, \bibinfo{person}{David Lo}, \bibinfo{person}{Xin Xia}, \bibinfo{person}{Ashish Sureka}, {and} \bibinfo{person}{Shanping Li}.} \bibinfo{year}{2015}\natexlab{}.
\newblock \showarticletitle{EFSPredictor: Predicting Configuration Bugs with Ensemble Feature Selection}. In \bibinfo{booktitle}{\emph{2015 Asia-Pacific Software Engineering Conference, {APSEC} 2015, December 1-4, 2015}}. \bibinfo{pages}{206--213}.
\newblock


\bibitem[Xuan et~al\mbox{.}(2014)]%
        {DBLP:journals/corr/XuanJHRZLW17}
\bibfield{author}{\bibinfo{person}{Jifeng Xuan}, \bibinfo{person}{He Jiang}, \bibinfo{person}{Yan Hu}, \bibinfo{person}{Zhilei Ren}, \bibinfo{person}{Weiqin Zou}, \bibinfo{person}{Zhongxuan Luo}, {and} \bibinfo{person}{Xindong Wu}.} \bibinfo{year}{2014}\natexlab{}.
\newblock \showarticletitle{Towards effective bug triage with software data reduction techniques}.
\newblock \bibinfo{journal}{\emph{IEEE transactions on knowledge and data engineering}} \bibinfo{volume}{27}, \bibinfo{number}{1} (\bibinfo{year}{2014}), \bibinfo{pages}{264--280}.
\newblock


\bibitem[Yang et~al\mbox{.}(2017a)]%
        {DBLP:conf/sac/YangBLL17}
\bibfield{author}{\bibinfo{person}{Geunseok Yang}, \bibinfo{person}{Seungsuk Baek}, \bibinfo{person}{Jung{-}Won Lee}, {and} \bibinfo{person}{Byungjeong Lee}.} \bibinfo{year}{2017}\natexlab{a}.
\newblock \showarticletitle{Analyzing emotion words to predict severity of software bugs: a case study of open source projects}. In \bibinfo{booktitle}{\emph{Proceedings of the Symposium on Applied Computing, {SAC} 2017, Marrakech Morocco, April 3-7, 2017}}. \bibinfo{pages}{1280--1287}.
\newblock


\bibitem[Yang et~al\mbox{.}(2019)]%
        {DBLP:journals/itiis/YangMLL19}
\bibfield{author}{\bibinfo{person}{Geunseok Yang}, \bibinfo{person}{Kyeongsic Min}, \bibinfo{person}{Jung{-}Won Lee}, {and} \bibinfo{person}{Byungjeong Lee}.} \bibinfo{year}{2019}\natexlab{}.
\newblock \showarticletitle{Applying Topic Modeling and Similarity for Predicting Bug Severity in Cross Projects}.
\newblock \bibinfo{journal}{\emph{{KSII} Trans. Internet Inf. Syst}}  \bibinfo{volume}{13} (\bibinfo{year}{2019}), \bibinfo{pages}{1583--1598}.
\newblock


\bibitem[Yang et~al\mbox{.}(2014)]%
        {DBLP:conf/compsac/YangZL14}
\bibfield{author}{\bibinfo{person}{Geunseok Yang}, \bibinfo{person}{Tao Zhang}, {and} \bibinfo{person}{Byungjeong Lee}.} \bibinfo{year}{2014}\natexlab{}.
\newblock \showarticletitle{Towards Semi-automatic Bug Triage and Severity Prediction Based on Topic Model and Multi-feature of Bug Reports}. In \bibinfo{booktitle}{\emph{{IEEE} 38th Annual Computer Software and Applications Conference {COMPSAC} 2014, Vasteras, Sweden, July 21-25, 2014}}. \bibinfo{pages}{97--106}.
\newblock


\bibitem[Yang et~al\mbox{.}(2017b)]%
        {DBLP:journals/jcst/YangLXHS17}
\bibfield{author}{\bibinfo{person}{Xinli Yang}, \bibinfo{person}{David Lo}, \bibinfo{person}{Xin Xia}, \bibinfo{person}{Qiao Huang}, {and} \bibinfo{person}{Jian{-}Ling Sun}.} \bibinfo{year}{2017}\natexlab{b}.
\newblock \showarticletitle{High-Impact Bug Report Identification with Imbalanced Learning Strategies}.
\newblock \bibinfo{journal}{\emph{Journal of Computer Science and Technology}}  \bibinfo{volume}{32} (\bibinfo{year}{2017}), \bibinfo{pages}{181--198}.
\newblock


\bibitem[Yang et~al\mbox{.}(2017c)]%
        {yang2017high}
\bibfield{author}{\bibinfo{person}{Xin-Li Yang}, \bibinfo{person}{David Lo}, \bibinfo{person}{Xin Xia}, \bibinfo{person}{Qiao Huang}, {and} \bibinfo{person}{Jian-Ling Sun}.} \bibinfo{year}{2017}\natexlab{c}.
\newblock \showarticletitle{High-impact bug report identification with imbalanced learning strategies}.
\newblock \bibinfo{journal}{\emph{Journal of Computer Science and Technology}}  \bibinfo{volume}{32} (\bibinfo{year}{2017}), \bibinfo{pages}{181--198}.
\newblock


\bibitem[Ye et~al\mbox{.}(2018)]%
        {DBLP:conf/icmla/0003FWBL18}
\bibfield{author}{\bibinfo{person}{Xin Ye}, \bibinfo{person}{Fan Fang}, \bibinfo{person}{John Wu}, \bibinfo{person}{Razvan~C. Bunescu}, {and} \bibinfo{person}{Chang Liu}.} \bibinfo{year}{2018}\natexlab{}.
\newblock \showarticletitle{Bug Report Classification Using {LSTM} Architecture for More Accurate Software Defect Locating}. In \bibinfo{booktitle}{\emph{17th {IEEE} International Conference on Machine Learning and Applications {ICMLA} 2018, Orlando, FL, USA, December 17-20, 2018}}. \bibinfo{pages}{1438--1445}.
\newblock


\bibitem[Ye et~al\mbox{.}(2025)]%
        {DBLP:conf/icse/Ye0L25}
\bibfield{author}{\bibinfo{person}{Yulong Ye}, \bibinfo{person}{Tao Chen}, {and} \bibinfo{person}{Miqing Li}.} \bibinfo{year}{2025}\natexlab{}.
\newblock \showarticletitle{Distilled Lifelong Self-Adaptation for Configurable Systems}. In \bibinfo{booktitle}{\emph{47th {IEEE/ACM} International Conference on Software Engineering, {ICSE} 2025, Ottawa, ON, Canada, April 26 - May 6, 2025}}. \bibinfo{publisher}{{IEEE}}, \bibinfo{pages}{1333--1345}.
\newblock
\urldef\tempurl%
\url{https://doi.org/10.1109/ICSE55347.2025.00094}
\showDOI{\tempurl}


\bibitem[Zaidi et~al\mbox{.}(2020)]%
        {DBLP:journals/access/ZaidiALWL20}
\bibfield{author}{\bibinfo{person}{Syed Farhan~Alam Zaidi}, \bibinfo{person}{Faraz~Malik Awan}, \bibinfo{person}{Minsoo Lee}, \bibinfo{person}{Honguk Woo}, {and} \bibinfo{person}{Chan{-}Gun Lee}.} \bibinfo{year}{2020}\natexlab{}.
\newblock \showarticletitle{Applying Convolutional Neural Networks With Different Word Representation Techniques to Recommend Bug Fixers}.
\newblock \bibinfo{journal}{\emph{{IEEE} Access}}  \bibinfo{volume}{8} (\bibinfo{year}{2020}), \bibinfo{pages}{213729--213747}.
\newblock


\bibitem[Zanjani et~al\mbox{.}(2015)]%
        {DBLP:conf/msr/ZanjaniKB15}
\bibfield{author}{\bibinfo{person}{Motahareh~Bahrami Zanjani}, \bibinfo{person}{Huzefa~H. Kagdi}, {and} \bibinfo{person}{Christian Bird}.} \bibinfo{year}{2015}\natexlab{}.
\newblock \showarticletitle{Using Developer-Interaction Trails to Triage Change Requests}. In \bibinfo{booktitle}{\emph{12th {IEEE/ACM} Working Conference on Mining Software Repositories {MSR} 2015, Florence, Italy, May 16-17, 2015}}. \bibinfo{pages}{88--98}.
\newblock


\bibitem[Zhang et~al\mbox{.}(2015)]%
        {Zhang2015}
\bibfield{author}{\bibinfo{person}{Jie Zhang}, \bibinfo{person}{XiaoYin Wang}, \bibinfo{person}{Dan Hao}, \bibinfo{person}{Bing Xie}, \bibinfo{person}{Lu Zhang}, {and} \bibinfo{person}{Hong Mei}.} \bibinfo{year}{2015}\natexlab{}.
\newblock \showarticletitle{A survey on bug-report analysis}.
\newblock \bibinfo{journal}{\emph{Science China Information Sciences}}  \bibinfo{volume}{58} (\bibinfo{year}{2015}), \bibinfo{pages}{1--24}.
\newblock


\bibitem[Zhang et~al\mbox{.}(2020)]%
        {DBLP:conf/iwpc/Zhang0YZZ20}
\bibfield{author}{\bibinfo{person}{Jinglei Zhang}, \bibinfo{person}{Rui Xie}, \bibinfo{person}{Wei Ye}, \bibinfo{person}{Yuhan Zhang}, {and} \bibinfo{person}{Shikun Zhang}.} \bibinfo{year}{2020}\natexlab{}.
\newblock \showarticletitle{Exploiting Code Knowledge Graph for Bug Localization via Bi-directional Attention}. In \bibinfo{booktitle}{\emph{{ICPC} '20: 28th International Conference on Program Comprehension Seoul, Republic of Korea, July 13-15, 2020}}. \bibinfo{pages}{219--229}.
\newblock


\bibitem[Zhang and Ma(2008)]%
        {zhang2008improved}
\bibfield{author}{\bibinfo{person}{R Zhang} {and} \bibinfo{person}{J Ma}.} \bibinfo{year}{2008}\natexlab{}.
\newblock \showarticletitle{An improved SVM method P-SVM for classification of remotely sensed data}.
\newblock \bibinfo{journal}{\emph{International Journal of Remote Sensing}}  \bibinfo{volume}{29} (\bibinfo{year}{2008}), \bibinfo{pages}{6029--6036}.
\newblock


\bibitem[Zhang et~al\mbox{.}(2017)]%
        {DBLP:conf/iwpc/ZhangCJLX17}
\bibfield{author}{\bibinfo{person}{Tao Zhang}, \bibinfo{person}{Jiachi Chen}, \bibinfo{person}{He Jiang}, \bibinfo{person}{Xiapu Luo}, {and} \bibinfo{person}{Xin Xia}.} \bibinfo{year}{2017}\natexlab{}.
\newblock \showarticletitle{Bug report enrichment with application of automated fixer recommendation}. In \bibinfo{booktitle}{\emph{Proceedings of the 25th International Conference on Program Comprehension {ICPC} 2017, Buenos Aires, Argentina, May 22-23, 2017}}. \bibinfo{pages}{230--240}.
\newblock


\bibitem[Zhang et~al\mbox{.}(2019)]%
        {DBLP:journals/software/ZhangCLL19}
\bibfield{author}{\bibinfo{person}{Tao Zhang}, \bibinfo{person}{Jiachi Chen}, \bibinfo{person}{Xiapu Luo}, {and} \bibinfo{person}{Tao Li}.} \bibinfo{year}{2019}\natexlab{}.
\newblock \showarticletitle{Bug Reports for Desktop Software and Mobile Apps in GitHub: What's the Difference?}
\newblock \bibinfo{journal}{\emph{{IEEE} Softw.}} \bibinfo{volume}{36}, \bibinfo{number}{1} (\bibinfo{year}{2019}), \bibinfo{pages}{63--71}.
\newblock
\urldef\tempurl%
\url{https://doi.org/10.1109/MS.2017.377142400}
\showDOI{\tempurl}


\bibitem[Zhang et~al\mbox{.}(2016a)]%
        {DBLP:journals/jss/ZhangCYLL16}
\bibfield{author}{\bibinfo{person}{Tao Zhang}, \bibinfo{person}{Jiachi Chen}, \bibinfo{person}{Geunseok Yang}, \bibinfo{person}{Byungjeong Lee}, {and} \bibinfo{person}{Xiapu Luo}.} \bibinfo{year}{2016}\natexlab{a}.
\newblock \showarticletitle{Towards more accurate severity prediction and fixer recommendation of software bugs}.
\newblock \bibinfo{journal}{\emph{Journal of Systems and Software}}  \bibinfo{volume}{117} (\bibinfo{year}{2016}), \bibinfo{pages}{166--184}.
\newblock


\bibitem[Zhang et~al\mbox{.}(2023)]%
        {DBLP:journals/tosem/ZhangHVIXTLJ23}
\bibfield{author}{\bibinfo{person}{Ting Zhang}, \bibinfo{person}{DongGyun Han}, \bibinfo{person}{Venkatesh Vinayakarao}, \bibinfo{person}{Ivana~Clairine Irsan}, \bibinfo{person}{Bowen Xu}, \bibinfo{person}{Ferdian Thung}, \bibinfo{person}{David Lo}, {and} \bibinfo{person}{Lingxiao Jiang}.} \bibinfo{year}{2023}\natexlab{}.
\newblock \showarticletitle{Duplicate Bug Report Detection: How Far Are We?}
\newblock \bibinfo{journal}{\emph{{IEEE} Transactions on Software Engineering and Methodology}}  \bibinfo{volume}{32} (\bibinfo{year}{2023}), \bibinfo{pages}{97:1--97:32}.
\newblock


\bibitem[Zhang et~al\mbox{.}(2016b)]%
        {zhang2016ksap}
\bibfield{author}{\bibinfo{person}{Wen Zhang}, \bibinfo{person}{Song Wang}, {and} \bibinfo{person}{Qing Wang}.} \bibinfo{year}{2016}\natexlab{b}.
\newblock \showarticletitle{KSAP: An approach to bug report assignment using KNN search and heterogeneous proximity}.
\newblock \bibinfo{journal}{\emph{Information and Software Technology}}  \bibinfo{volume}{70} (\bibinfo{year}{2016}), \bibinfo{pages}{68--84}.
\newblock


\bibitem[Zhang et~al\mbox{.}(2021)]%
        {DBLP:conf/webi/ZhangZW21}
\bibfield{author}{\bibinfo{person}{Wen Zhang}, \bibinfo{person}{Jiangpeng Zhao}, {and} \bibinfo{person}{Song Wang}.} \bibinfo{year}{2021}\natexlab{}.
\newblock \showarticletitle{Sustriage: sustainable bug triage with multi-modal ensemble learning}. In \bibinfo{booktitle}{\emph{IEEE/WIC/ACM International Conference on Web Intelligence and Intelligent Agent Technology}}. \bibinfo{pages}{441--448}.
\newblock


\bibitem[Zheng et~al\mbox{.}(2021a)]%
        {DBLP:conf/dsa/ZhengCWFSC21}
\bibfield{author}{\bibinfo{person}{Wei Zheng}, \bibinfo{person}{Zheng Chen}, \bibinfo{person}{Xiaoxue Wu}, \bibinfo{person}{Weiqiang Fu}, \bibinfo{person}{Bowen Sun}, {and} \bibinfo{person}{Jingyuan Cheng}.} \bibinfo{year}{2021}\natexlab{a}.
\newblock \showarticletitle{A Domain Knowledge-Guided Lightweight Approach for Security Bug Reports Prediction}. In \bibinfo{booktitle}{\emph{{DSA} 2021, August 5-6, 2021}}. \bibinfo{pages}{359--368}.
\newblock


\bibitem[Zheng et~al\mbox{.}(2021b)]%
        {DBLP:conf/issre/ZhengZTCCWS21}
\bibfield{author}{\bibinfo{person}{Wei Zheng}, \bibinfo{person}{Manqing Zhang}, \bibinfo{person}{Hui Tang}, \bibinfo{person}{Yuanfang Cai}, \bibinfo{person}{Xiang Chen}, \bibinfo{person}{Xiaoxue Wu}, {and} \bibinfo{person}{Abubakar Omari~Abdallah Semasaba}.} \bibinfo{year}{2021}\natexlab{b}.
\newblock \showarticletitle{Automatically Identifying Bug Reports with Tactical Vulnerabilities by Deep Feature Learning}. In \bibinfo{booktitle}{\emph{32nd {IEEE} {ISSRE} 2021, Wuhan, China, October 25-28, 2021}}. \bibinfo{pages}{333--344}.
\newblock


\bibitem[Zhou et~al\mbox{.}(2015)]%
        {DBLP:conf/ease/ZhouNG15}
\bibfield{author}{\bibinfo{person}{Bo Zhou}, \bibinfo{person}{Iulian Neamtiu}, {and} \bibinfo{person}{Rajiv Gupta}.} \bibinfo{year}{2015}\natexlab{}.
\newblock \showarticletitle{Predicting concurrency bugs: how many, what kind and where are they?}. In \bibinfo{booktitle}{\emph{{EASE} 2015, Nanjing, China, April 27-29, 2015}}. \bibinfo{pages}{6:1--6:10}.
\newblock


\bibitem[Zhou et~al\mbox{.}(2020)]%
        {DBLP:journals/jss/ZhouLS20}
\bibfield{author}{\bibinfo{person}{Cheng Zhou}, \bibinfo{person}{Bin Li}, {and} \bibinfo{person}{Xiaobing Sun}.} \bibinfo{year}{2020}\natexlab{}.
\newblock \showarticletitle{Improving software bug-specific named entity recognition with deep neural network}.
\newblock \bibinfo{journal}{\emph{Journal of Systems and Software}}  \bibinfo{volume}{165} (\bibinfo{year}{2020}), \bibinfo{pages}{110572}.
\newblock


\bibitem[Zhu et~al\mbox{.}(2022)]%
        {DBLP:journals/kbs/ZhuTWL22}
\bibfield{author}{\bibinfo{person}{Ziye Zhu}, \bibinfo{person}{Hanghang Tong}, \bibinfo{person}{Yu Wang}, {and} \bibinfo{person}{Yun Li}.} \bibinfo{year}{2022}\natexlab{}.
\newblock \showarticletitle{Enhancing bug localization with bug report decomposition and code hierarchical network}.
\newblock \bibinfo{journal}{\emph{Knowledge-Based Systems}}  \bibinfo{volume}{248} (\bibinfo{year}{2022}), \bibinfo{pages}{108741}.
\newblock


\bibitem[Zou et~al\mbox{.}(2016)]%
        {DBLP:journals/ieicet/ZouXYZZH16}
\bibfield{author}{\bibinfo{person}{Jie Zou}, \bibinfo{person}{Ling Xu}, \bibinfo{person}{Mengning Yang}, \bibinfo{person}{Xiaohong Zhang}, \bibinfo{person}{Jun Zeng}, {and} \bibinfo{person}{Sachio Hirokawa}.} \bibinfo{year}{2016}\natexlab{}.
\newblock \showarticletitle{Automated Duplicate Bug Report Detection Using Multi-Factor Analysis}.
\newblock \bibinfo{journal}{\emph{{IEICE} Trans. Inf. Syst}}  \bibinfo{volume}{99-D} (\bibinfo{year}{2016}), \bibinfo{pages}{1762--1775}.
\newblock


\bibitem[Zou et~al\mbox{.}(2018)]%
        {8466000}
\bibfield{author}{\bibinfo{person}{Weiqin Zou}, \bibinfo{person}{David Lo}, \bibinfo{person}{Zhenyu Chen}, \bibinfo{person}{Xin Xia}, \bibinfo{person}{Yang Feng}, {and} \bibinfo{person}{Baowen Xu}.} \bibinfo{year}{2018}\natexlab{}.
\newblock \showarticletitle{How practitioners perceive automated bug report management techniques}.
\newblock \bibinfo{journal}{\emph{IEEE Transactions on Software Engineering}} \bibinfo{volume}{46}, \bibinfo{number}{8} (\bibinfo{year}{2018}), \bibinfo{pages}{836--862}.
\newblock


\end{thebibliography}

\end{document}